\documentclass[12pt,preprint]{aastex}
\usepackage{txfonts}
\usepackage{graphicx}

\newcommand{\cxo}{{\it Chandra}}
\newcommand{\oversim}[2]{\lower0.5ex\vbox{\baselineskip=0pt\lineskip=0.2ex
     \ialign{$\mathsurround=0pt #1\hfil##\hfil$\crcr#2\crcr\sim\crcr}}} 
\newcommand{\simless} {\mbox{$\mathrel{\mathpalette\oversim<}$}} 
\shorttitle{COUP census of X-ray stars in BN-KL and OMC-1S}
\shortauthors{Grosso et al.}
\begin{document}
\title{Chandra Orion Ultradeep Project census of X-ray stars \\ 
in the BN-KL and OMC-1S regions}
\author{%
N. Grosso\altaffilmark{1}, 
E.~D. Feigelson\altaffilmark{2}, 
K.~V. Getman\altaffilmark{2}, 
L. Townsley\altaffilmark{2}, 
P. Broos\altaffilmark{2}, 
E. Flaccomio\altaffilmark{3}, 
M.~J.~McCaughrean\altaffilmark{4,5}, 
G. Micela\altaffilmark{3}, 
S. Sciortino\altaffilmark{3}, 
J. Bally\altaffilmark{6}, 
N. Smith\altaffilmark{6}, 
A.~A. Muench\altaffilmark{7}, 
G.~P.~Garmire\altaffilmark{2}, 
F.\ Palla\altaffilmark{8}}


\altaffiltext{1}{Laboratoire d'Astrophysique de Grenoble,
       	         Universit{\'e} Joseph-Fourier,
                 F-38041 Grenoble cedex 9, France; \\ 
                 {\tt Nicolas.Grosso@obs.ujf-grenoble.fr}}
\altaffiltext{2}{Department of Astronomy and Astrophysics,
                 Pennsylvania State University, 525 Davey Laboratory,
                 University Park, PA 16802}
\altaffiltext{3}{INAF, Osservatorio Astronomico di Palermo, 
                 Piazza del Parlamento 1, 90134 Palermo, Italy}
\altaffiltext{4}{School of Physics, University of Exeter, Stocker Road,
                 Exeter EX4 4QL, Devon, UK}
\altaffiltext{5}{Astrophysikalisches Institut Potsdam,
                 An der Sternwarte 16,
                 D-14482 Potsdam, Germany}
\altaffiltext{6}{Center for Astrophysics and Space Astronomy, 
		 University of Colorado at Boulder, CB 389, Boulder, CO 80309}
\altaffiltext{7}{Harvard-Smithsonian Center for Astrophysics, 
                 60 Garden Street, Cambridge, MA 02138, USA}
\altaffiltext{8}{Osservatorio Astrofisico di Arcetri, Largo Enrico Fermi 5, 
                 50125 Firenze, Italy}
\begin{abstract}

We present a study of the X-ray sources detected in the vicinity of two
density peaks in the Orion Molecular Cloud~1 (OMC-1) behind the Orion
Nebula Cluster (ONC), as seen in the exceptionally deep ($\sim$\,10
days) exposure of \dataset[ADS/Sa.CXO#obs/COUP]{the \cxo{} Orion Ultradeep Project} 
(\dataset[ADS/Sa.CXO#obs/COUP]{COUP}).  We focus
on a $40\arcsec\times50\arcsec$ region around the \object[NAME BN Object]{Becklin-Neugebauer
object} and Kleinmann-Low nebula (collectively BN-KL) and a
$60\arcsec\times75\arcsec$ region around OMC-1S, a secondary
star-forming peak some $90\arcsec$ south of BN-KL\@.  Forty-three and
sixty X-ray sources were detected in BN-KL and OMC-1S, respectively, of
which half and one-third, respectively, were found to be foreground
members of the ONC, while the remaining sources are identified as
obscured X-ray sources with column densities $22\,\simless\,\log(N_{\rm
H}/{\rm cm^{-2}})\,\simless\,24$.  All but 5 and 18 of these sources
have near-infrared stellar counterparts, and 22 of these appear to be
young stellar objects (YSOs).  X-ray sources are found close to four
luminous mid-IR sources namely \object[NAME BN Object]{BN}, IRc3-i2, IRc2-C, and Source n; their
X-ray variability and spectral properties are typical of coronal
activity in low-mass stars rather than wind emission from massive stars,
suggesting that the X-ray emission may be arising from companions.  The
X-ray light curve of the X-ray source close to \object[NAME BN Object]{BN} shows a periodicity of
$\sim$8.3\,days and from an X-ray image deconvolution of the region around
\object[NAME BN Object]{BN}, we conclude that either \object[NAME BN Object]{BN} itself or a low mass companion with a
projected separation of $\simeq 200$\,AU was detected.  On the other
hand, no emission is seen from the bright radio Source I, held by some
to be the main source of luminosity in BN-KL\@.  In OMC-1S, \cxo{}
unveils a new subcluster of seven YSOs without infrared counterparts.
We compare the hard band X-ray luminosity functions of obscured X-ray
sources in BN-KL and OMC-1S with unobscured X-ray sources in the ONC,
and estimate that the true population of obscured sources in BN-KL and
OMC-1S is $\simeq$\,46 and 57 stars, with 90\% confidence intervals of
34--71 and 42--82 stars, respectively.

\end{abstract}


\section{Introduction} %

The nearest ($d\sim450$\,pc) massive molecular
cloud to the Sun is the Orion Molecular Cloud~1 \citep[henceforth
OMC-1;][]{genzel89}.  OMC-1 lies immediately behind the young (1\,Myr)
Orion Nebula Cluster (ONC) centered on the bright Trapezium OB stars, and
the Orion Nebula H\,{\small II} region is a thin concave blister on the
facing side of the cloud (O'Dell 2001).

At its highest density peak, just $\sim$\,0.1\,pc northwest of the
Trapezium, OMC-1 hosts the closest known region of high-mass star
formation, with a total infrared luminosity of $\sim 10^5$\,L$_\odot$
\citep[][ and references therein]{werner76,genzel89} reprocessed and
emitted by dust grains.  The dominant underlying source of this large
luminosity remains the subject of debate \citep{greenhill04}, but the
region also includes a number of embedded luminous young stars such as
the $\sim 10^4$\,L$_\odot$ \object[NAME BN Object]{Becklin-Neugebauer object} 
\citep[ henceforth \protect{\object[NAME BN Object]{BN}}]{becklin67} and several $\sim 10^3$\,L$_\odot$ stars \citep[ and
references therein]{gezari98}.  Three of these luminous infrared sources
(\object[NAME BN Object]{BN}, IRc2, and Source n) are associated with compact thermal radio
sources indicative of ultracompact H\,{\small II} regions or stellar
winds, and/or maser outflows \citep[][ and references
therein]{greenhill04}.  The cluster also illuminates the extended
Kleinmann-Low nebula \citep[ henceforth KL]{kleinmann67} and powers
strong shocked molecular outflows emanating from the region
\citep{allen93}.  Combined, BN-KL is the densest and most
chemically-rich region in the entire Orion giant molecular cloud complex
and is the nearest and best-studied molecular hot cores.  Nevertheless,
despite decades of multiwavelength studies, the region is far from fully
understood due to its extreme complexity.

The second density peak in OMC-1 of interest here is OMC-1S, located about 
$90\arcsec$ south of BN-KL\@. OMC-1S has roughly 10\% of the luminosity of
BN-KL, with $L_{\rm bol}\sim10^4$\,L$_\odot$ \citep{mezger90}. It has long
been known that the region hosts a number of deeply embedded near- and 
mid-infrared sources \citep[][]{mccaughrean88}, and there are also several 
collimated molecular outflows and a number of optical Herbig-Haro 
objects seen in the region \citep[see, e.g.,][]{odell01}. Based on a proper
motion study of the HH objects, \citet{odell03} proposed them all have the 
same origin which he termed the Optical Outflow Source (OOS), although it 
seems more likely that there are several exciting sources, as bolstered
by recent mid-infrared imaging of the region \citep{smith04}.

As the nearest regions of ongoing massive star formation, these density peaks 
in OMC-1 are clear targets for detailed study at all wavelengths. The most
recent detailed X-ray study of OMC-1 is that of \citet{garmire00}, who 
observed the region using \cxo{} and ACIS-I with an exposure time of 47.8\,ks 
(0.6\,days). In BN-KL, the luminous infrared Source~n was seen, which may 
constitute the first unambiguous (i.e.\ spatially resolved from other young 
stars) X-ray detection of a massive YSO\@. Faint X-ray emission was also seen 
close to, but apparently not coincident with \object[NAME BN Object]{BN}, and it remained unclear
whether or not the emission was coming from a separate, nearby source, or was
being produced in the outflow from \object[NAME BN Object]{BN} \citep{garmire00,tan04}.

In January 2003, \cxo{} and ACIS-I were again used to observe the region,
but this time for a total of 838\,ksecs ($\sim$\,10 days), yielding an
exceptionally deep study known as 
\dataset[ADS/Sa.CXO#obs/COUP]{the \cxo{} Orion Ultradeep Project} 
or \dataset[ADS/Sa.CXO#obs/COUP]{COUP} \citep[see ][]{getman05b}. 
Here we concentrate on the X-ray sources
detected along the line-of-sight towards BN-KL and OMC-1S with the
following main goals: to complete the census of obscured YSOs in these
regions as traced through their X-ray emission; to investigate the nature
of the X-ray sources associated with luminous mid-infrared stars in the 
BN-KL region; and to study the X-ray luminosity function of obscured sources
in OMC-1.

After briefly reviewing the observations and data analysis in
\S\ref{obs}, we present an overview of the X-ray sources and their
counterparts in BN-KL in \S\ref{source_bn}, before discussing X-ray
emission from \object[NAME BN Object]{BN} and the rotational modulation of the X-ray source close
to \object[NAME BN Object]{BN} in \S\ref{BN}, and then examining other X-ray sources close to
luminous mid-infrared stars in BN-KL in \S\ref{other_sources}.  The
X-ray sources and their counterparts in OMC-1S are reported in
\S\ref{source_south}, before a broader view is taken of the disk
fraction in OMC-1 in \S\ref{disk} and of the embedded stellar population
of OMC-1 in \S\ref{xlf}. Our conclusions are presented in \S\ref{conclusions}.

\section{Observations and data analysis} 
\label{obs}

COUP combines six near-consecutive exposures at a single pointing ($05^{\rm h}\,35^{\rm
m}\,17.0^{\rm s}$, $-05^\circ\,23\arcmin\,40.0\arcsec$, J2000) over
13.2\,days in January 2003, yielding a total on-source exposure time of
838\,ks (9.7\,days).  The observations were obtained using ACIS-I
\citep{garmire03} on-board the \cxo~X-ray Observatory
\citep{weisskopf02}.  \citet{getman05b} present a detailed discussion of
the COUP data, encompassing source detection, photon extraction,
spectral analysis, and variability analysis.  Here, we focus a subsample
of the 1616 COUP source catalog and data products of \citet{getman05b},
limited to X-ray sources along the line-of-sight towards two key regions
of the OMC-1 molecular cloud, namely BN-KL and OMC-1S, both located at
about 1.2$\arcmin$ off-axis in the COUP pointing.

\begin{figure*}[!h]
\centering
\includegraphics[width=\columnwidth]{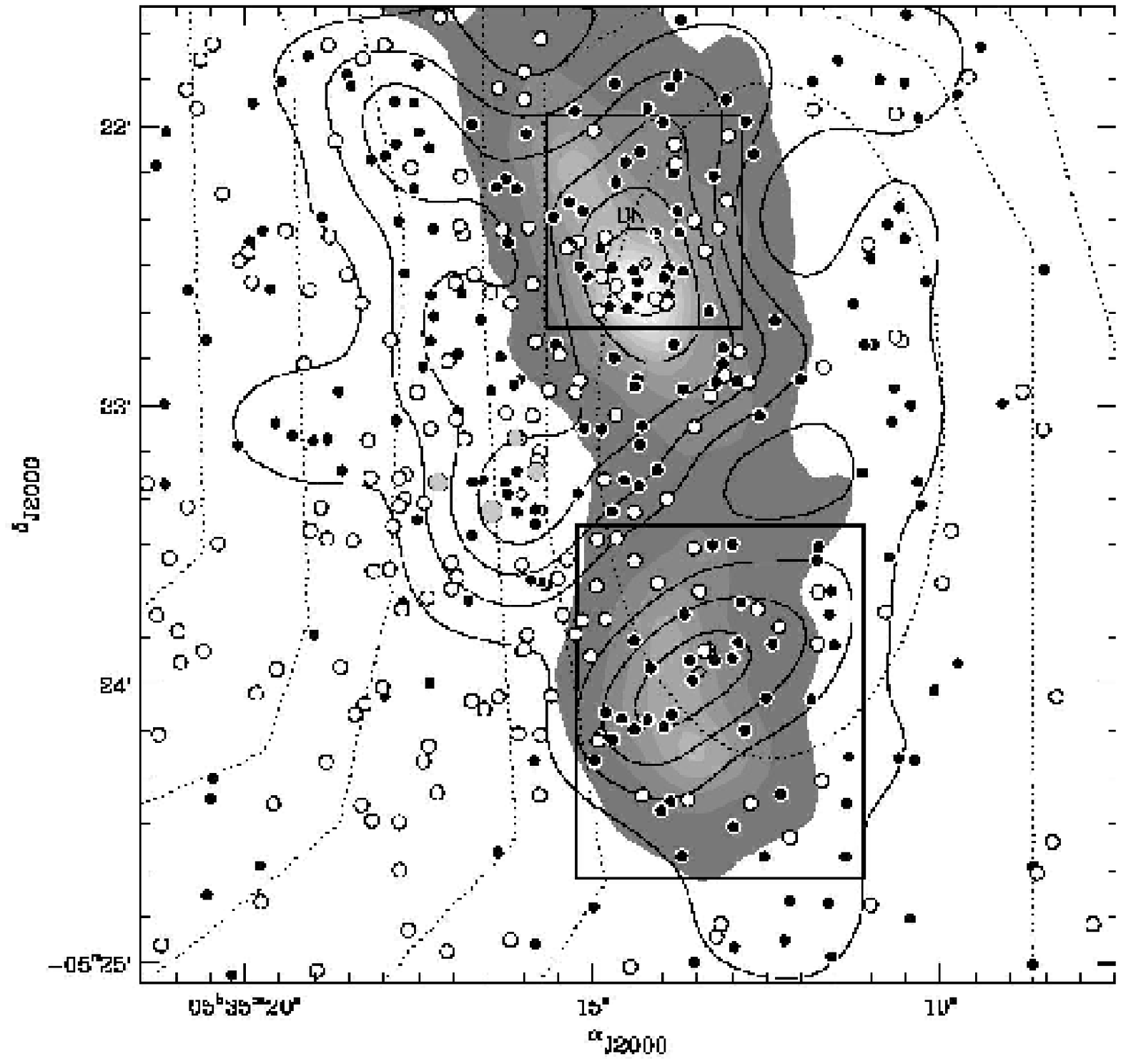}
 \caption{Map showing the OMC-1 areas studied in this work. The shaded contours 
show the SCUBA 450\,$\micron$ map of \citet{johnstone99}, with fluxes from 45 
to 450\,Jy\,beam$^{-1}$ in greyscale steps of 45\,Jy\,beam$^{-1}$. Black and 
white dots mark obscured ($N_{\rm H} > 10^{22}$\,cm$^{-2}$) and 
lightly-absorbed ($< 10^{22}$\,cm$^{-2}$) COUP X-ray sources, respectively. 
The four brightest Trapezium OB stars, $\theta^1$\,Ori\,A--D, are marked by 
larger grey dots. Continuous contours indicate the Gaussian-smoothed surface 
density of the obscured COUP X-ray sources from 1000--3500 stars\,pc$^{-2}$ 
in steps of 500 stars\,pc$^{-2}$, computed using a smoothing length of 
0.025\,pc (i.e.\ 11.5\arcsec{} at 450\,pc). Diamonds mark the three main 
density peaks: BN-KL, OMC-1S, and the Trapezium, with 3800, 2800, and 2900 
stars\,pc$^{-2}$, respectively. The dotted contours show the visual 
extinction derived from the $^{13}$CO map of \citet{bally87} from 20--45\,mag 
in steps of 5\,mag. The upper and lower rectangles correspond to BN-KL and 
OMC-1S, respectively, as studied here.
}
\label{map}
\end{figure*}

Figure~\ref{map} shows a 450\,$\micron$ wavelength image of the part of OMC-1 
behind the core of the ONC \citep{johnstone99}, where the two main 
submillimetre peaks delineate the BN-KL and OMC-1S regions: COUP sources 
are plotted for comparison. Lightly obscured COUP sources 
($N_{\rm H} < 10^{22}$\,cm$^{-2}$) are seen to be clustered around the 
Trapezium OB stars and can be classified as ONC members. By contrast, we can
analyse the spatial distribution of the more obscured
($N_{\rm H} \ge 10^{22}$\,cm$^{-2}$) COUP sources. Their surface density is
calculated using the kernel method and adopting a Gaussian shape kernel 
\citep[][ and references therein]{gomez93}, with a smoothing 
length\footnote{
The kernel density estimator is defined at each point ($\alpha,\delta$) of 
the computation grid by: \\
$D(\alpha,\delta)=\frac{1}{h^2} \Sigma_{i=1}^{n} 
K(\alpha_i,\delta_i,\alpha,\delta)$, where $h$ is the smoothing length, 
$(\alpha_i,\delta_i)$ are the positions of the $n$ sources, and $K$ is 
the weight function. Here we adopt a Gaussian shape for the kernel: 
$K(\alpha_i,\delta_i,\alpha,\delta)=\frac{1}{2 \pi} \exp{(-r^2/2h^2)}$, 
where $r^2=(\alpha-\alpha_i)^2 \cos^2 \delta_i+(\delta-\delta_i)^2$.} 
of $11.5\arcsec$, i.e.\ 0.025\,pc at 450\,pc. The obscured X-ray sources 
are seen to concentrate in three clusters associated with BN-KL, OMC-1S, and
the Trapezium OB stars, with peak surface densities of 3800, 2800, and 2900 
sources\,pc$^{-2}$, respectively. \citet{lada04} found a similar result by 
selecting all sources with infrared $K-L$ colors greater than 1.5 mag, roughly 
corresponding to $A_{\rm V} > 15$--26\,mag. 

In order to proceed further with our analysis, we adopt semi-arbitrary
definitions of BN-KL as the region shown in Fig.~6 of \citet{garmire00}, 
i.e.\ a $40\arcsec\times50\arcsec$ area centered on \object[NAME BN Object]{BN}, and for OMC-1S, a 
$60\arcsec\times75\arcsec$ area centered roughly on the corresponding 
submillimetre peak, covering the clustering of obscured X-ray sources.

\begin{figure*}[!h]
\centering
\includegraphics[width=\columnwidth]{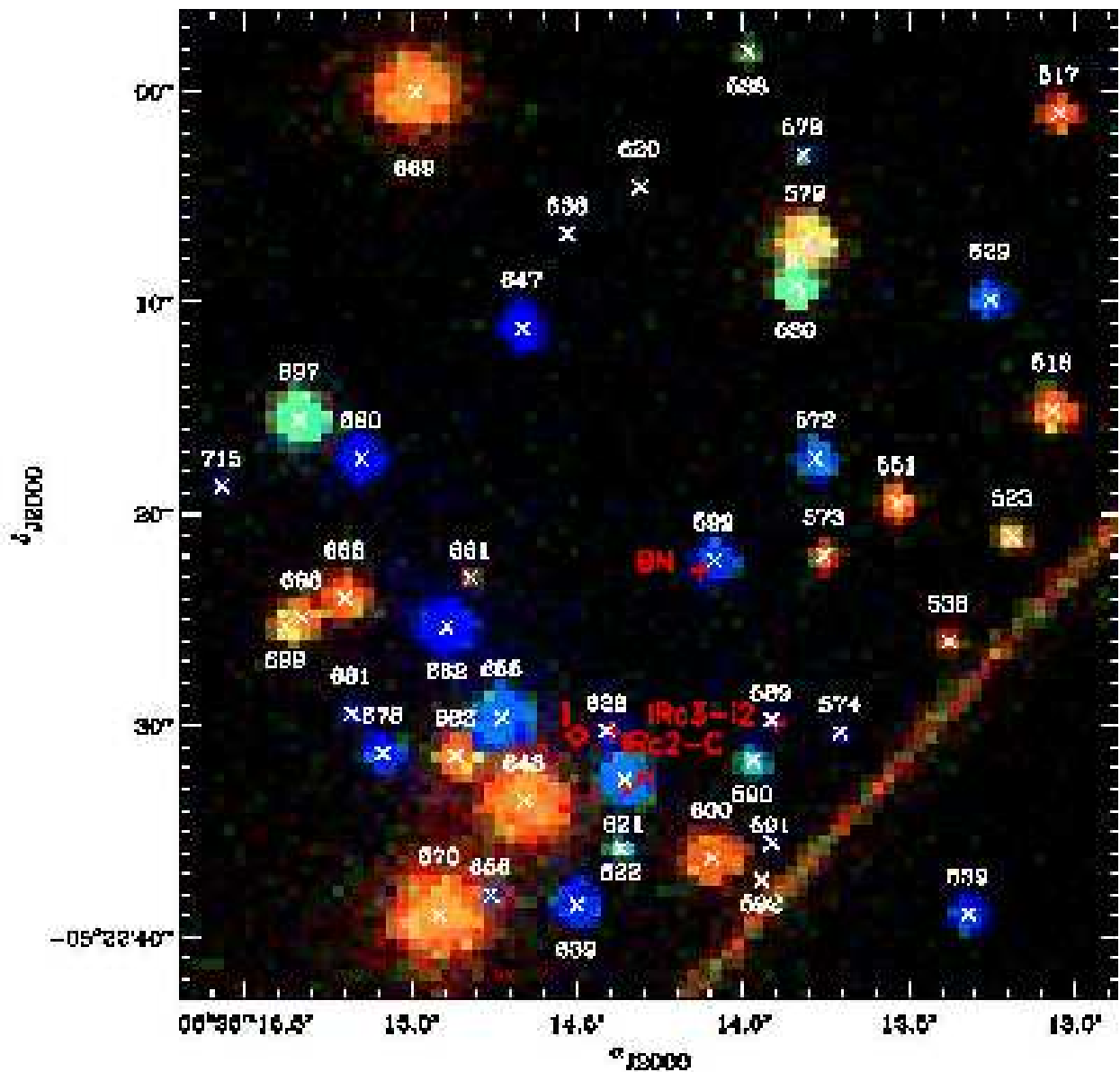}
 \caption{COUP view of BN-KL\@. Red, green, and blue represent photons in the 
0.5--1.7\,keV, 1.7--2.8\,keV, and 2.8--8.0\,keV bands, 
respectively, using the color-encoding scheme of \citet{lupton04}. Red crosses 
mark the positions of luminous mid-infrared sources with closely associated 
COUP X-ray sources: BN \citep[using the VLT $J_{\rm S}$ band position epoch 2002.0, 
adjusted to the mean COUP epoch of 2003.04 using the radio proper motions of ][]{tan04,menten95}; 
source {\it n} \citep[][ 7.8\,$\mu$m]{gezari98};
IRc3-i2 and IRc2-C \citep[][ 3.6\,$\mu$m]{dougados93}. The red diamond shows 
the radio position of source~I \citep{churchwell87}, which has no X-ray 
counterpart. The diagonal line across the lower right corner is the readout 
trail produced by bright X-ray source $\theta^1$\,Ori\,C collected during the 
readout of the CCD frame (so-called `out-of-time events'). COUP catalog 
numbers are shown in white. Table~\ref{bn_cxo_sources} and 
Table~\ref{bn_cxo_properties} describe the counterparts and X-ray source 
properties, respectively. Comments on individual COUP sources are given 
in Appendix~\ref{individual_bn}.
}
\label{bn_trichro}
\end{figure*}

\begin{figure*}[!h]
\centering
\includegraphics[width=\columnwidth]{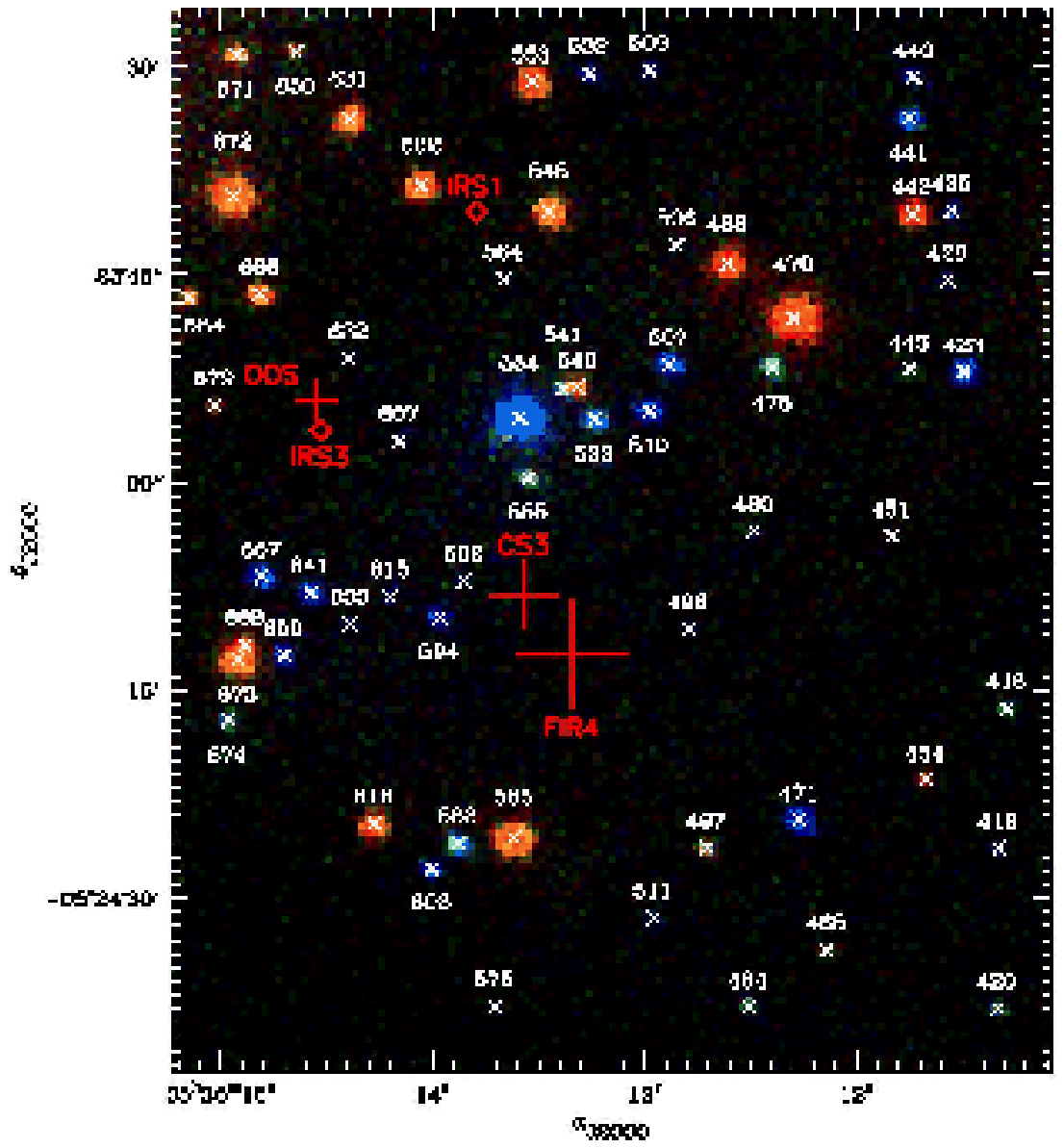}
 \caption{COUP view of OMC-1S, with the same color coding as for 
Fig.~\ref{bn_trichro}. Red diamonds mark the positions of IRS1 and IRS3, two 
bright mid-infrared sources in the region \citep{smith04} without X-ray 
counterparts. The luminous far-infrared/sub-millimetre source FIR4 
\citep{mezger90}, the dense molecular condensation CS3 \citep{mundy86}, and 
the hypothetical Optical Outflow Source \citep{odell03} are marked with red 
crosses. Table~\ref{omc1s_cxo_sources} and Table~\ref{omc1s_cxo_properties} 
describe the counterparts and X-ray source properties, respectively. 
Comments on individual COUP sources are given in Appendix~\ref{individual_omc1s}
}
\label{omc1s_trichro}
\end{figure*}

Figure~\ref{bn_trichro} shows a full resolution 
($0.492\arcsec\times0.492\arcsec$ pixel) view of our BN-KL region in the 
0.5--8\,keV band. Colors represent the photon energy where red, green, 
and blue show soft unabsorbed (0.5--1.7\,keV), medium 
(1.7--2.8\,keV), and hard (2.8--8.0\,keV) energy photons, 
respectively. This composite image was produced using the algorithm recently 
proposed by \citet{lupton04}, which ensures that a given X-ray photon color 
is mapped to a unique color in the RGB image, and that the color is correct 
even when the intensity is clipped at unity, compared to the burnt-out white 
produced by the standard RGB method. The 43 X-ray sources detected by COUP 
in this region above a limit of 3 net photons are plotted. 
Figure~\ref{omc1s_trichro} shows the equivalent COUP 0.5--8\,keV 
color composite image of the selected OMC-1S region, along with the positions 
of the 60 X-ray sources detected above the 3 net photons limit.

Table~\ref{bn_cxo_sources} and Table~\ref{omc1s_cxo_sources} list the X-ray
sources seen along the line-of-sight towards BN-KL and OMC-1S, respectively,
with their basic X-ray properties as adapted from \citet[][]{getman05b}. Also
given are associations with previously catalogued sources at optical, near- 
and mid-infrared wavelengths, along with other identifications with, for
example, radio sources. Of note in this latter regard are the spatial offsets 
between the COUP source and the nearest near-infrared source from the unified 
VLT $JHK_{\rm S}$ catalog of McCaughrean et al.\ (2005, in preparation): the RMS
offset between the X-ray and near-infrared sources is only 0.2\arcsec,
demonstrating the certainty to which such associations can be made. Finally,
an indication is given of the membership status of each X-ray source with
respect to the region, as determined by \citet[][]{getman05a}. Comments 
on individual COUP sources are given in Appendix~\ref{individual}.

The X-ray properties of the BN-KL and OMC-1S X-ray sources are described in 
more detail in Table~\ref{bn_cxo_properties} and 
Table~\ref{omc1s_cxo_properties}, respectively \citep[adapted from][]{getman05b}.
These include measures of the hardness ratio, column density, plasma
temperatures, emission measures, observed and extinction corrected 
luminosities, and an assessment of the source variability.

\section{X-ray sources along the line-of-sight towards BN-KL}
\label{source_bn}

Despite a total exposure time some 17.5 times longer than that of
\citet{garmire00}, COUP revealed just 17 new X-ray sources along the
line-of-sight towards the BN-KL field, an increase of just 65\%.  As
discussed in \S\ref{xlf}, this is due to a combination of factors,
including the underlying stellar mass function in the region and the
correlation between $L_{\rm X}$ and $L_{\rm bol}$, which is roughly
linear in the stellar regime \citep[see Figs.\ 1--5 of][]{preibisch05a}
and possibly steeper in the brown dwarf regime \citep[see Fig.~5
of][]{preibisch05b}.

The left panel of Fig.~\ref{nh} shows the distribution of column densities
for sources seen along the line-of-sight towards the region we have selected 
centered on BN-KL\@. A bimodal distribution is clearly seen. Half of the X-ray
sources have optical counterparts: we find 21 X-ray emitting optical stars 
compared to the 14 detected previously. 
The distribution shows that these sources have a median extinction of 
$\sim 4\times10^{21}$\,cm$^{-2}$, i.e.\ $A_\mathrm{V}\sim 2.5$\,mag only,
and it seems entirely likely that these unobscured sources are members of
the foreground ONC population centered on the Trapezium OB stars, just 
0.1\,pc away. Conversely, the second peak in the column density distribution 
are obscured by median column density of $\sim 10^{23}$\,cm$^{-2}$, i.e.\ 
$A_\mathrm{V}\sim70$\,mag; these sources are embedded in the background OMC-1
molecular cloud and while none have optical counterparts, all but five do
have infrared stellar counterparts. 

One concern is that some of the heavily embedded sources with or without
infrared counterparts may be background AGNs seen in X-rays through the cloud. 
\citet[][]{getman05a} have investigated the anticipated level of such 
contamination in the COUP sample by making Monte-Carlo simulations of COUP 
detected extragalactic background sources, using the expected 
$\log N$--$\log S$ for X-ray background AGNs \citep{giacconi01} convolved 
with the optical depth map for the OMC 
\citep[derived from $^{13}$CO map of][]{bally87}, and the COUP background map. 
\citet{getman05a} estimate that $\le 1$ background AGNs should be seen
towards our BN-KL region, since the considerable extinction associated with
the OMC shields us from such contamination.

In qualitative support of this finding, we see that the most deeply embedded 
X-ray sources in the BN-KL region, i.e.\ those without near-infrared
counterparts, all display time variability characteristic of young low-mass 
stars with X-ray flares (see Fig.~\ref{lc_source_x_bn}). The same is true of
many other X-ray sources in the region: Fig.~\ref{lc_multi_flares} shows the 
most variable X-ray source in the region, the obscured 
\object[COUP 0580]{COUP\,580}
($\log N_{\rm H} \simeq 22.2$, i.e.\ $A_\mathrm{V}\sim10$\,mag), which 
displayed a remarkable series of six impulsive X-ray flares.

\subsection{X-rays from BN and its vicinity} 
\label{BN}

It has long been unclear whether \object[NAME BN Object]{BN} is an X-ray emitter. 
Although there is considerable X-ray emission from the vicinity, earlier X-ray surveys with
{\it Einstein} \citep{ku79} and {\it ROSAT/HRI\/} \citep{gagne95}, for example, 
did not have sufficient spatial resolution and sensitivity to hard X-rays, respectively, 
to know whether any of this X-ray emission came from \object[NAME BN Object]{BN} itself.

In the first 1999 \cxo{} 
observations of \citet{garmire00}, a source with just 18 photons was seen 
very close to \object[NAME BN Object]{BN}, but with a measured positional offset of $1.1\arcsec$, 
too far from \object[NAME BN Object]{BN} to be considered as a true coincidence. The analysis of the 
combined 1999--2000 data by \citet{feigelson02} found a total of 42 photons,
no variability, a very hard spectrum with plasma energy $>10$\,keV and 
absorption $\log N_{\rm H} \simeq 22.6$, and a hard band X-ray luminosity 
of $\log L_{\rm h} \simeq 29.4$. They found a positional offset of 
$0.6\arcsec$ from the infrared source which, they stated in a footnote, 
was consistent with a true coincidence. 

More recently, \citet{tan04} showed that the X-ray emission lies roughly
north-west of \object[NAME BN Object]{BN}, along the projected proper motion vector of this
rapidly moving source.  As a consequence, \citet{tan04} proposed that
the X-ray emission does not come from \object[NAME BN Object]{BN} itself, but rather from
bowshock preceding the source in its passage through OMC-1.  With the
substantially deeper COUP data, we are in a better position to assess
this suggestion and reanalyse the question of whether or not the X-ray
emission is indeed coincident with the near-infrared source.

\subsubsection{COUP\,599a: An X-ray bright companion to BN}
\label{BN_companion}

In the COUP data, the nearest source to the nominal near-infrared position
of \object[NAME BN Object]{BN} as listed in the unified VLT catalog of McCaughrean et al.\ (2005, in
preparation) is \object[COUP 0599a]{COUP\,599a}, which shows a total of 847 counts which can be
successfully modeled by an isothermal plasma with $kT = 2.6$\,keV, a column 
density of $\log N_{\rm H} = 23.0$ (Fig.~\ref{Coup_599}), a luminosity of 
$\log L_{\rm h} = 30.2$ after correction for absorption, and variability 
over the 13 days of COUP observations by a factor of five in amplitude 
(see Fig.~\ref{Coup_599} and \S\ref{bn_rotational_modulation}). 

But is \object[COUP 0599a]{COUP\,599a} in fact \object[NAME BN Object]{BN}? 
Attempting to answer this question definitively, we have carried out a careful astrometric check. For overall consistency, the 
astrometry in the VLT catalog was derived from the $K_{\rm S}$ band mosaic image of 
the region. However, \object[NAME BN Object]{BN} is heavily saturated at $K_{\rm S}\sim 5$\,mag and due to 
the peculiar way in which infrared arrays saturate, its pixel position in 
that mosaic had to be defined by eye rather than via a centroid. 

Given the present detailed interest in \object[NAME BN Object]{BN}, we can do a better job using the
VLT $J_{\rm S}$ band data, in which \object[NAME BN Object]{BN} is much fainter ($\sim 15$\,mag) but still 
readily visible with high signal-to-noise, making an accurate determination
possible. We took a single, non-mosaiced $J_{\rm S}$ band image covering \object[NAME BN Object]{BN} from 
January 2002, one year prior to the COUP exposure. In this image, we 
established a local reference frame based on 
$\sim$\,10--15 nearby sources which are unsaturated in both the $J_{\rm S}$ and 
$K_{\rm S}$ data, and for which accurate sky coordinates are available from the VLT 
catalog, where the overall astrometric error with respect to the 2MASS (FK5) 
reference frame is 0.15\arcsec{} RMS (McCaughrean et al.\ 2005, in 
preparation).

Doing so, we obtained an epoch 2002.0 position for \object[NAME BN Object]{BN} of $05^{\rm h}35^{\rm m}
14.130^{\rm s}$, $-05^\circ22\arcmin22.72\arcsec$ (J2000.0), with an
astrometric error with respect to the local reference frame of 0.084\arcsec{}
RMS\@. In order to make a consistency check, we also repeated the same
process for $J_{\rm S}$ VLT frames covering \object[NAME BN Object]{BN} from December 1999 and January 2001,
spanning the full length of the VLT project. We find essentially the same
positions for \object[NAME BN Object]{BN} in both years as for January 2002, giving us confidence
in the accuracy of the position in the 2MASS global frame. 

We have also pushed the VLT data a little further and searched for evidence 
of the proper motion in \object[NAME BN Object]{BN} with respect to the surrounding stars in a local 
$x,y$ pixel coordinate reference frame (as opposed to the global 
$\alpha,\delta$ frame), under the assumption that these sources are relatively 
fixed over a two year period (the ONC 2D velocity dispersion of 
$\sim$\,3.3\,km\,s$^{-1}$ corresponds to 1.5 millarcsec per year at Orion).
We do see a slight shift in \object[NAME BN Object]{BN} from year to year, on the order of 
0.01--0.03\arcsec{} to the NW, although the fit errors are on the same order 
and thus the result is marginal at best. 

Therefore, we have simply taken the VLT-determined position from January
2002 and applied the radio-derived proper motion for \object[NAME BN Object]{BN} of
$0.0181\pm 0.0022$\arcsec\,yr$^{-1}$ at position angle $-37.7\pm5\degr$ 
\citep{tan04} to yield a projected epoch 2003.04 (i.e.\ at the time of
the COUP observation) position for \object[NAME BN Object]{BN} of $05^{\rm h}35^{\rm m}14.129^{\rm s}$, 
$-05^\circ22\arcmin22.71\arcsec$ (J2000.0).

This position lies 0.895\arcsec{} ($\sim$\,400\,AU) south-east of \object[COUP 0599a]{COUP\,599a}. 
The RMS astrometric error between the COUP and
VLT frames is 0.25\arcsec{} \citep{getman05b} and locally the alignment is
often better: no other source in the vicinity of \object[NAME BN Object]{BN} has an X-ray/infrared 
offset larger than 0.25\arcsec. Thus, we can conclude with confidence that 
\object[COUP 0599a]{COUP\,599a} is not \object[NAME BN Object]{BN}\@. 

What is \object[COUP 0599a]{COUP\,599a} then? \citet{tan04} proposed that the X-ray emission
seen just to the northwest of \object[NAME BN Object]{BN}, i.e.\ \object[COUP 0599a]{COUP\,599a}, originated in a wind 
bow shock as \object[NAME BN Object]{BN} moves rapidly through OMC-1. However, we now see that 
\object[COUP 0599a]{COUP\,599a} has strong, rapid variations in its flux and spectrum, and 
such variability is incompatible with the relatively constant X-ray emission 
that would be produced by such a shock. In fact, these properties make it 
much more likely that \object[COUP 0599a]{COUP\,599a} is simply a deeply embedded lower-mass source 
in OMC-1, plausibly a companion to \object[NAME BN Object]{BN} with a $\sim$\,400\,AU separation,
or otherwise just an unrelated cluster member along the same line-of-sight.

The column density $\log N_{\rm H} = 23.0$ inferred from the X-ray
spectrum of \object[COUP 0599a]{COUP\,599a} corresponds to an extinction of $A_{\rm V}\sim
50$\,mag, or $\sim$\,14\,mag in the $J_{\rm S}$ band.  Since there no
source detected at the position of \object[COUP 0599a]{COUP\,599a} in the VLT $J_{\rm S}$
band data to a limiting magnitude of $\sim 21$\,mag, \object[COUP 0599a]{COUP\,599a} must
have an intrinsic magnitude fainter than $J_{\rm S} \geq 7$\,mag at
Orion, or an absolute magnitude of $M_{J_{\rm S}} \geq -1.3$.  From the
1\,Myr isochrone of \citet{siess00}, this corresponds to an upper limit
mass of $\geq 4\,M_\odot$, i.e.\ \object[COUP 0599a]{COUP\,599a} could be a relatively
massive star.  In principle, stricter upper limits could be set on the
mass of \object[COUP 0599a]{COUP\,599a} by using the VLT $H$ and $K_{\rm S}$ band data, but
since \object[NAME BN Object]{BN} is already quite bright by 1.5\,$\mu$m and less than 1\arcsec{}
away, the VLT data are saturated at the location of \object[COUP 0599a]{COUP\,599a} and
thus it is impossible to claim any meaningful limits on its
non-detection at the longer wavelengths.  In any case, it seems quite
plausible that \object[COUP 0599a]{COUP\,599a} is a deeply embedded lower-mass star, either
unrelated or a companion to \object[NAME BN Object]{BN}\@.  Companions to higher-mass stars are
commonly found to be more X-ray luminous than the massive primary, both
in field and in unabsorbed ONC intermediate-mass stars
\citep{stelzer03,stelzer05}.

\subsubsection{COUP\,599b: A second companion or BN itself?}

Having established that \object[COUP 0599a]{COUP\,599a} is not \object[NAME BN Object]{BN}, the question remains whether
or not any fainter X-ray emission has been detected from \object[NAME BN Object]{BN} itself in the 
extremely deep COUP exposure. In order to pursue this question further,
we used the IDL Astrolib {\tt MAX\_LIKELIHOOD} routine to carry out an image 
deconvolution over a $43\arcsec \times 47\arcsec$ region around \object[NAME BN Object]{BN}\@. The 
routine iteratively seeks the Maximum Likelihood (ML) restoration of a blurred 
image with additive Poisson noise \citep{richardson72,lucy74}. The input point 
spread function was that obtained from the \cxo{} simulation program {\tt MARX} 
assuming the appropriate spectral parameters for \object[COUP 0599a]{COUP\,599a} and the 
deconvolution was stopped after 500 iterations.

All but one of the new features revealed in the reconstructed image are
consistent with noise unrelated to any known source at other wavelengths. The 
exception is a faint point source, \object[COUP 0599b]{COUP\,599b}, which lies $\sim 0.82\arcsec$ 
southeast of \object[COUP 0599a]{COUP\,599a}. Figure~\ref{bn_reconstructed} shows a small portion 
of the original COUP image, the corresponding ML reconstruction and a closeup 
of a $\pm 1.5\arcsec$ region around \object[NAME BN Object]{BN}\@. 

\object[COUP 0599b]{COUP\,599b} lies 0.4\arcsec{} from the VLT position for \object[NAME BN Object]{BN}: as described above,
this is a significantly larger offset than seen between VLT and COUP 
associations in the immediate vicinity. Similar offsets (0.3--0.4\arcsec) are 
seen between \object[COUP 0599b]{COUP\,599b} and the VLA radio positions for \object[NAME BN Object]{BN} as given by 
\citet{menten95}, \citet{tan04}, and \citet{zapata04a}. We found that the 
\citet{menten95} radio positions for Source~n and Source~I matched those for 
Sources~17 and 19 of \citet{zapata04a}, and that then slight adjustments 
($\Delta\alpha = -0.080\arcsec$ and $\Delta\delta = +0.038\arcsec$) were 
necessary to bring the \citet{zapata04a} frame into alignment with the 
COUP/VLT frame. After doing so, the \citet{zapata04a} position for \object[NAME BN Object]{BN} was 
also 0.4\arcsec{} from \object[COUP 0599b]{COUP\,599b}.

Thus it appears as though \object[COUP 0599b]{COUP\,599b} is not \object[NAME BN Object]{BN} either. 
However, \object[COUP 0599b]{COUP\,599b}
was detected with only $\simeq$\,50--70 photons in the PSF wings of 
\object[COUP 0599a]{COUP\,599a} and it is difficult to assess the positional errors after the
iterative ML deconvolution, and thus we cannot completely exclude the
possibility that \object[COUP 0599b]{COUP\,599b} coincides with \object[NAME BN Object]{BN}: further work will be needed.
For reference, a qualitative examination of likely \object[COUP 0599b]{COUP\,599b} photons suggests 
a heavily absorbed source with a low-energy cutoff similar to that seen in
\object[COUP 0599a]{COUP\,599a} ($\log N_{\rm H} = 23.0$), an estimated luminosity of 
$L_{\rm h} \sim 10^{29}$\,erg\,s$^{-1}$, and no dramatic variability during 
the 13-day COUP observation.

Both possibilities are consistent with the COUP results concerning
B~stars in the unobscured Orion Nebula Cluster \citep{stelzer05}.  Some
of these show X-rays coincident with the massive component, which may
arise from the massive star or from an unresolved close binary, while
others are resolved from the massive star as wide-binary companions.
The case of $\theta^1$\,Ori\,B is particularly instructive.  It is known
from infrared imaging to be at least a quintet and has two associated
COUP sources separated by $0.9\arcsec$.  It is not possible to
unambigously assign the X-rays to the components of this hierarchical
system, but one X-ray component has a softer spectrum and slow,
low-amplitude variability which may arise from the B2.5 primary, while
the other has a hard spectrum and high-amplitude flaring likely
associated with a binary low-mass subsystem.

\subsubsection{Periodic X-ray modulation of COUP\,599a}
\label{bn_rotational_modulation}

During a month-long (March--April 2000) near-infrared monitoring campaign,
\citet{hillenbrand01} found a puzzling nearly sinusoidal periodic variability 
in \object[NAME BN Object]{BN} in the $H$ and $K_{\rm S}$ bands. The period was 8.28 days and the peak-to-peak 
amplitude was $\sim$\,0.2\,mag, as measured in a $4\arcsec$ radius aperture, 
i.e.\ nominally covering both \object[COUP 0599a]{COUP\,599a} 
and \object[COUP 0599b]{COUP\,599b}.

Although our astrometric study suggests quite strongly that neither of these
two X-ray sources is in fact \object[NAME BN Object]{BN}, we have nevertheless made a detailed study 
of variability in the X-ray emission from \object[COUP 0599a]{COUP\,599a}, 
as there are sufficient 
photons to do so. There are several potential sources of variability in 
main-sequence and pre main-sequence stars: flares, the evolution of active 
regions, time-variable mass accretion \citep{flaccomio99}, modulation due to 
the inhomogeneity of the emitting region combined with stellar rotation,
and so on. 

Thanks to the 13\,day length of the COUP observation, we are able to search
for variability on similar timescales or less. The discovery of rotational 
modulation would be of particular interest as it would reveal the degree 
of inhomogeneity of the X-ray emitting plasma at the stellar surface. For
our analysis of \object[COUP 0599a]{COUP\,599a}, we closely followed the methods used by 
\citet{flaccomio05} for their extensive study of rotational modulation in the 
COUP sample. In brief, we computed the Lomb-Scargle Normalized Periodogram 
\citep[hereafter LNP;][]{lomb76,scargle82} on the binned light curve. Peaks
in the LNP were identified and False Alarm Probabilities (FAP) computed for
each of them using Monte Carlo techniques assuming correlated noise. The 
correlation timescale employed was on the order of the length of flares 
seen in \object[COUP 0599a]{COUP\,599a} as described below.

Figure~\ref{lc_bn} shows the light curve for \object[COUP 0599a]{COUP\,599a}. 
There is an underlying
gradual variability, with the suggestion of periodicity: there is also a flare
lasting $\sim$0.5\,days 2003 January 12 which was detected using the algorithm 
described by \citet{wolk05}. We have performed the LNP analysis on the whole
light curve minus this flare and a short time afterwards, as indicated in 
Fig.~\ref{lc_bn} by a horizontal bar, while Fig.~\ref{periodogram}a shows the 
corresponding periodogram. The largest peak appears at a frequency of 
0.121\,days$^{-1}$, corresponding to a period 
$P_{\rm rot,X}=8.27\pm0.10$\,days.

Dashed lines in the figure indicate LNP power thresholds of 1\% and 0.1\%, 
computed using 10000 Monte Carlo simulations and assuming a correlation 
time-scale, $\tau_{corr}$, of 15\,hrs. The simulations show that the 8.27\,day
peak has an FAP of $\sim$\,0.1\%, while changing $\tau_{corr}$ across the range 
5--25\,hrs yields an FAP from $<0.01$\% to 2\%. Figure~\ref{periodogram}b shows 
the \object[COUP 0599a]{COUP\,599a} light curve folded with the 8.27\,day period and repeated 
twice to show the periodicity more clearly. Different symbols indicate time 
bins belonging to different rotation cycles. We have also repeated the LNP 
analysis on the whole light curve, i.e.\ retaining the flare on 2003 
January~12, which yields a similar period ($P_{\rm rot,X}$=8.08\,days) but 
with a somewhat lower significance (FAP\,=\,0.4\%).

The apparent $\sim$8.3\,days X-ray periodicity in \object[COUP 0599a]{COUP\,599a} is so similar to 
the infrared period measured by \citet{hillenbrand01} for \object[NAME BN Object]{BN} that, at first
sight, it seems it cannot be coincidental. However, it is extremely hard to
understand this result, given that the X-ray variability is associated with
a source that is offset by $\sim$\,0.9\arcsec{} the infrared-variable \object[NAME BN Object]{BN}\@. 
First, even though \object[COUP 0599a]{COUP\,599a} lies within the aperture monitored by
\citet{hillenbrand01}, there is no evidence for an infrared source there
at the magnitudes that would be necessary to induce an apparent variability
in \object[NAME BN Object]{BN}\@. Imagine that \object[NAME BN Object]{BN} is in fact constant at $H = 9.2$\,mag, but that it 
appears to be variable by 0.2\,mag in a large aperture due to the proximity 
of another source which {\em is\/} varying. The most extreme situation would 
be if the adjacent source is varying by 100\,\%, in which case, in its full-on 
state it would be adding 0.2\,mag to the brightness of \object[NAME BN Object]{BN}, i.e.\ it would have 
a magnitude of $H = 10.9$\,mag in this state. It is straightforward to show
that in all other scenarios, in which the adjacent star varies by less than 
100\%, it would need to be brighter than $H = 10.9$\,mag in its full-on state,
and the VLT images emphatically rule out the presence of a source with 
anything like that brightness within 1\arcsec{} of \object[NAME BN Object]{BN}\@. 

The next explanation would be that both \object[NAME BN Object]{BN} and the \object[COUP 0599a]{COUP\,599a} exhibit near
identical $\sim$\,8\,day periodicities because they are physically linked,
but this seems equally unpalatable, as we cannot imagine a physical effect 
able to link the rotations, orbits, or magnetic activities of stars separated 
by $>400$\,AU\@. 

The third option is that the X-ray variations of \object[COUP 0599a]{COUP\,599a} do not in
fact represent a true periodicity, despite the apparent statistical
significance of the periodogram peak.  Such features can arise in time
series from aperiodic autoregressive (i.e.\ $1/f$-type noise) processes.
The non-sinusoidal shape of the folded lightcurve
(Fig.~\ref{periodogram}b) supports the idea that the variations are
non-rotational in origin and thus possibly aperiodic.  In this case, the
apparent similarity of the near-infrared and X-ray variations would
arise purely by chance and with no astrophysical meaning.

\subsection{Other X-ray sources close to luminous mid-infrared stars in BN-KL}
\label{other_sources}

\object[NAME BN Object]{BN} is not the only luminous source in the BN-KL region: Sources~I 
\citep{churchwell87} and~L \citep{menten95} are radio continuum sources
associated with compact H\,{\small II} regions, where radio Source~L is
also coincident with the luminous infrared Source~n \citep{lonsdale82}. 

COUP detected no X-ray emission from radio Source~I\@. On the other hand, radio
Source~L (infrared Source~n) has been previously detected by \cxo, both by 
\citet[][ their source 13]{garmire00} at a level of 61 counts using ACIS-I 
and by \citet[][ their source 272]{flaccomio03} using the HRC-I, at a level 
$\ga 3$ times than that measured by the first ACIS-I observation. Our longer 
\cxo{} exposure detected a total of 3779 counts from \object[COUP 0621]{COUP\,621}, coincident 
with Source~n, which makes a more detailed study of its X-ray properties 
possible. Figure~\ref{lc_luminous_stars} shows the light curve of \object[COUP 0621]{COUP\,621}, 
displaying several impulsive X-ray flares typical of the activity of low-mass 
stars. The suggestion then is that \object[COUP 0621]{COUP\,621} is a lower-mass companion to the 
luminous and higher-mass Source~n. 

In passing, it must be noted that \object[COUP 0621]{COUP\,621} is not coincident with the 
possible low-mass companion to Source~n recently detected 0.6\arcsec{} to its 
northwest using the infrared adaptive optics system NACO on the VLT 
\citep{lagrange04}. Indeed, the $\log L_{\rm h,c}=30.7$ determined for 
\object[COUP 0621]{COUP\,621} implies a relatively high mass, $\sim$\,1--3\,$M_\odot$ 
\citep{wolk05,preibisch05a}, but is also consistent with intermediate-mass 
B~star multiple systems \citep{stelzer05}. 

Finally, X-ray emission is also seen associated with two other luminous
mid-infrared sources in BN-KL:  \object[COUP 0628]{COUP\,628} 
and \object[COUP 0589]{COUP\,589} are X-ray
counterparts to component~C of IRc2 and component~i2 of IRc3
\citep{dougados93}, respectively.  Figure~\ref{lc_luminous_stars} shows
the corresponding X-ray light curves.  A plausible interpretation is
that we are again seeing lower-mass components of the higher-mass
multiple systems IRc2 and IRc3.

\section{X-ray sources along the line-of-sight towards OMC-1S}
\label{source_south}

As with BN-KL, the number of new sources (31) detected in the OMC-1S region
is relatively small, despite the 17.5 times longer exposure of COUP relative
to the original ACIS-I observations of \citet{garmire00}, and also as with
BN-KL, the most likely explanation lies in the shape of the X-ray luminosity 
function (see \S\ref{xlf}).

\begin{figure*}
\centering
\begin{tabular}{@{}c@{}c@{}}
\includegraphics[angle=0,width=0.5\columnwidth]{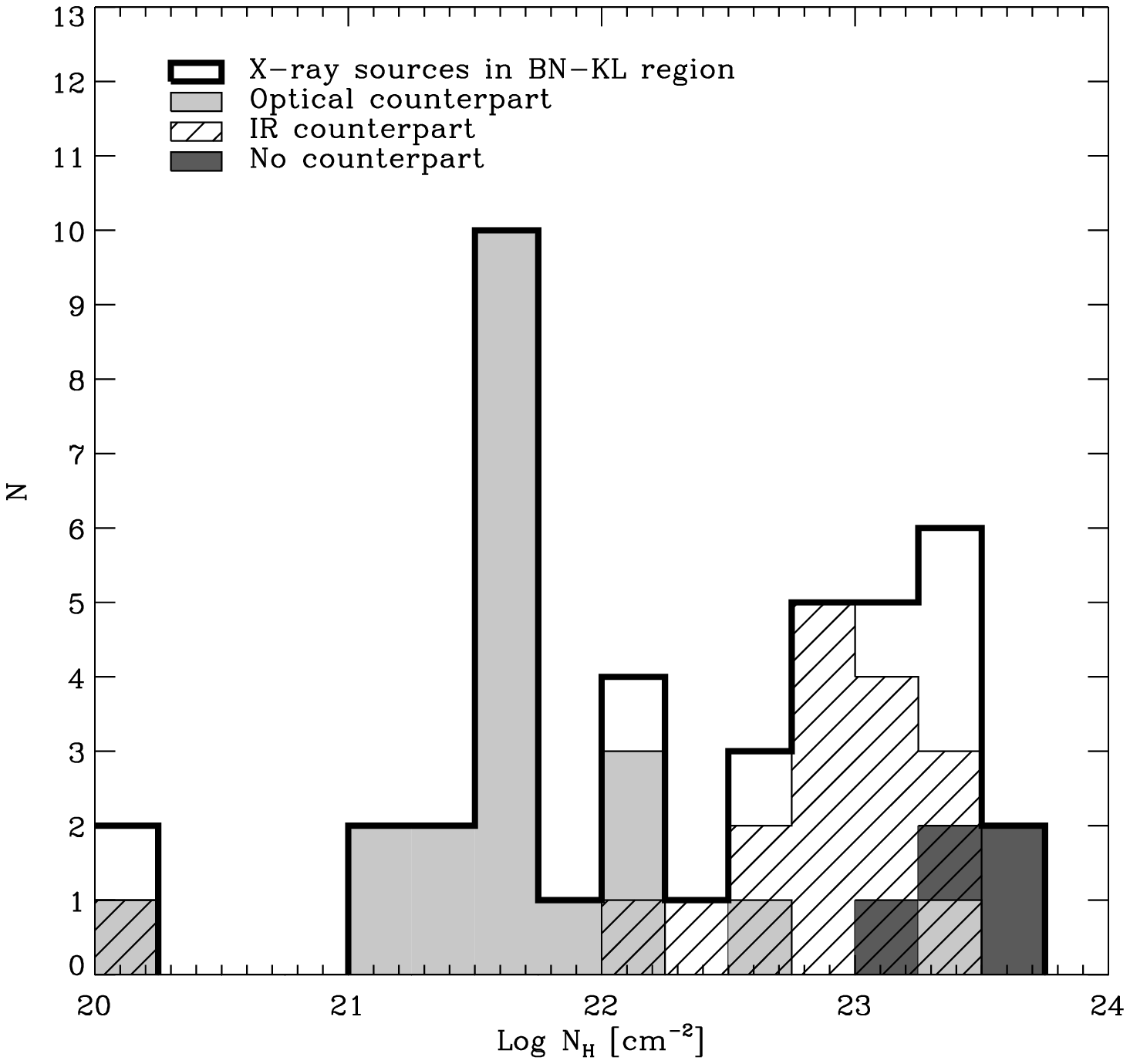} & \includegraphics[angle=0,width=0.5\columnwidth]{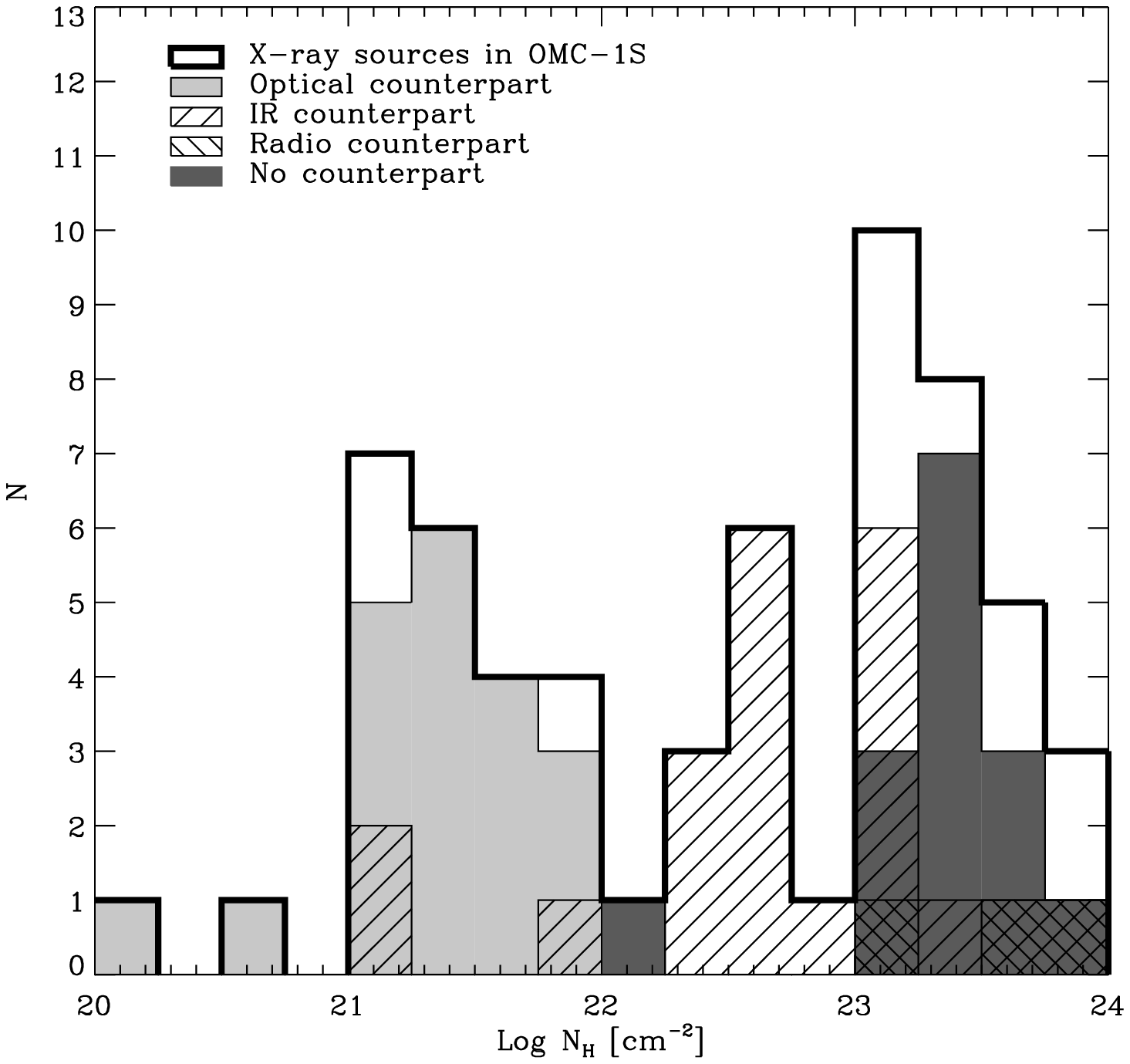}\\
\end{tabular}
 \caption{Distribution of column densities for COUP sources in BN-KL (left) and 
OMC-1S (right). X-ray sources with optical counterparts have the lowest column 
densities, while X-ray sources with near-infrared counterparts and no
counterparts at all have progressively higher column densities.
Two population of X-ray sources are visible: the unobscured stars of the 
Orion Nebula Cluster and obscured sources in the OMC-1 cores BN-KL and
OMC-1S, which have column densities roughly 30 and 70 times higher, 
respectively. 
}
\label{nh}
\end{figure*}

\begin{figure*}
\centering
\begin{tabular}{cc}
\includegraphics[width=0.5\columnwidth]{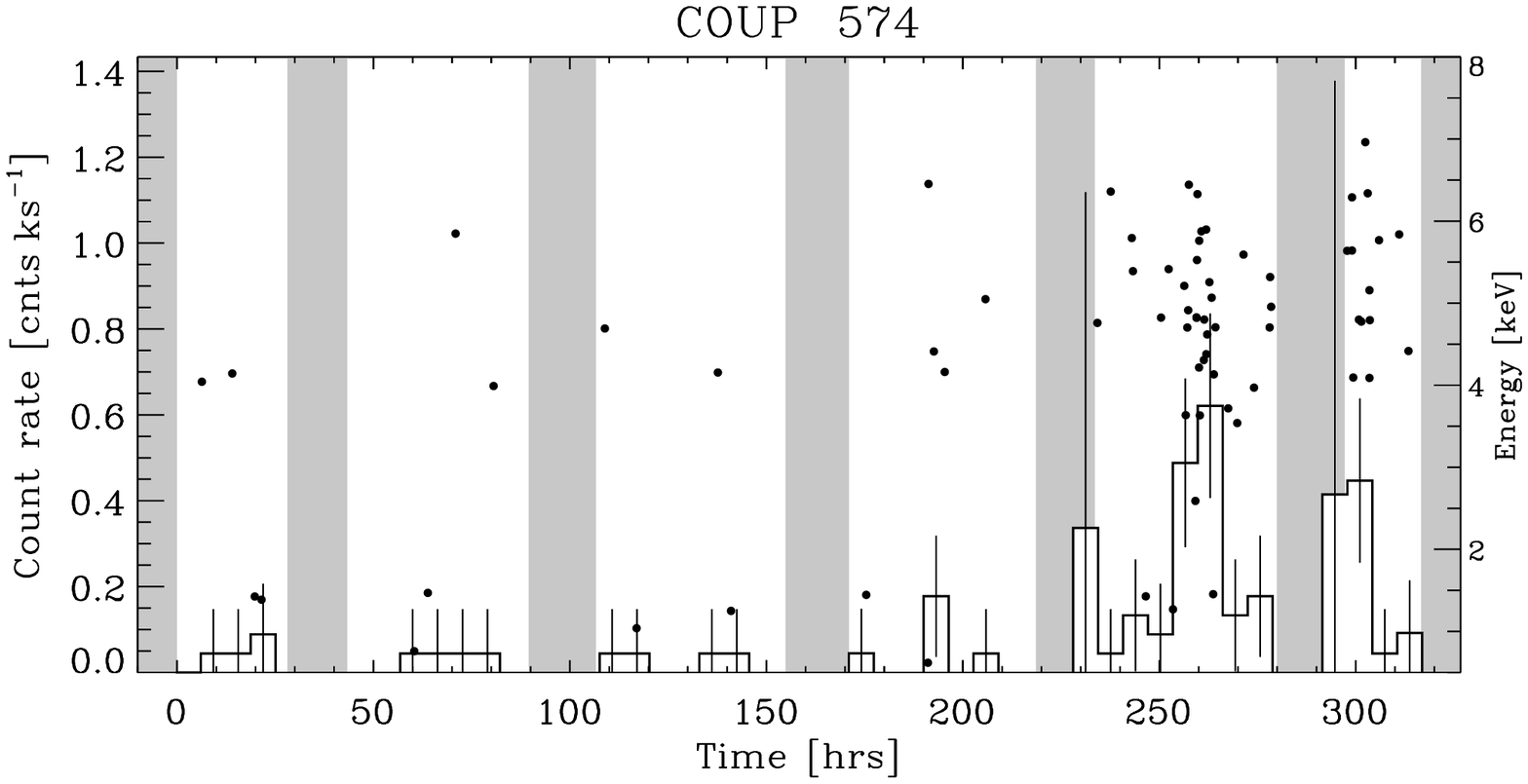} & \includegraphics[width=0.5\columnwidth]{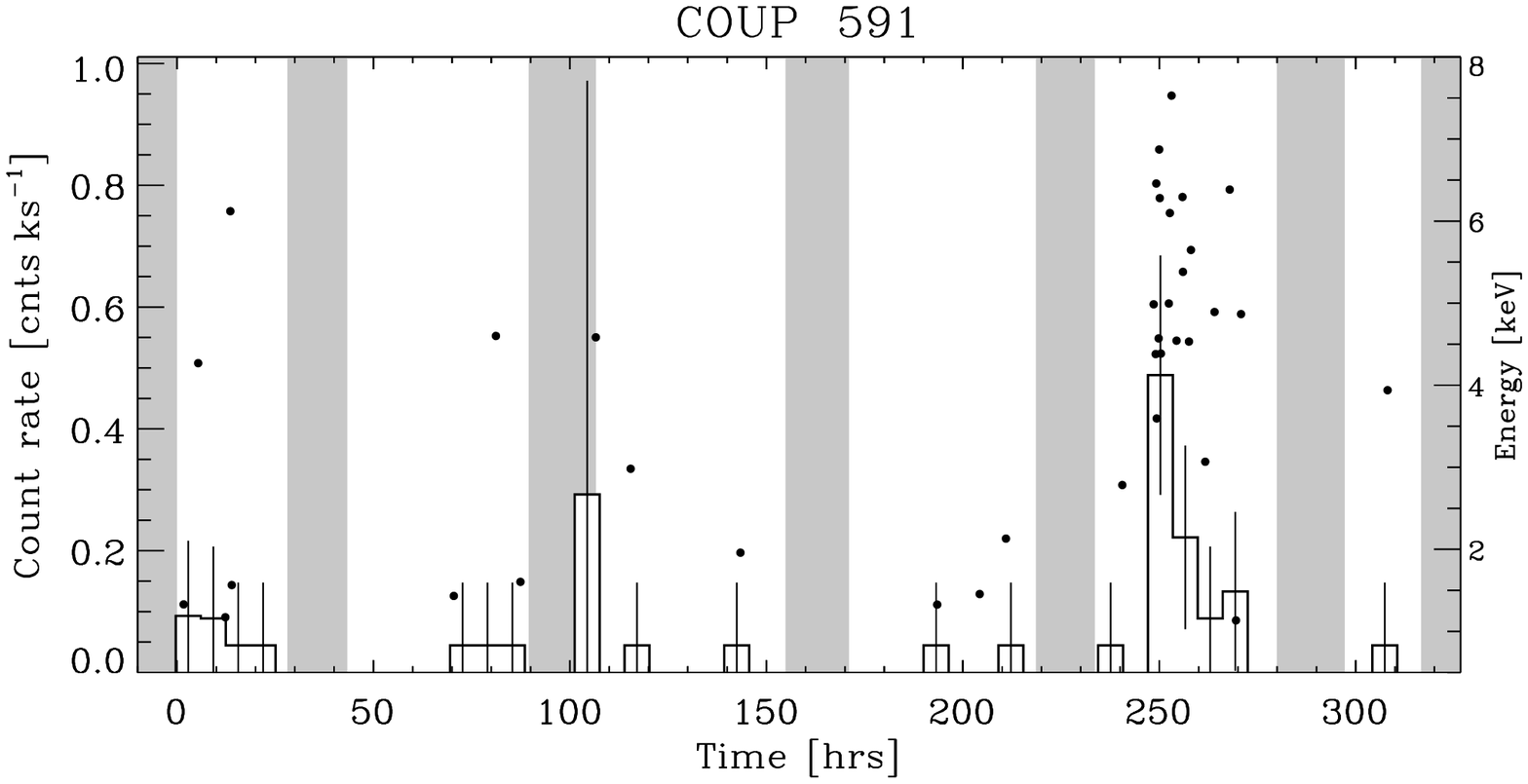}\\
\includegraphics[width=0.5\columnwidth]{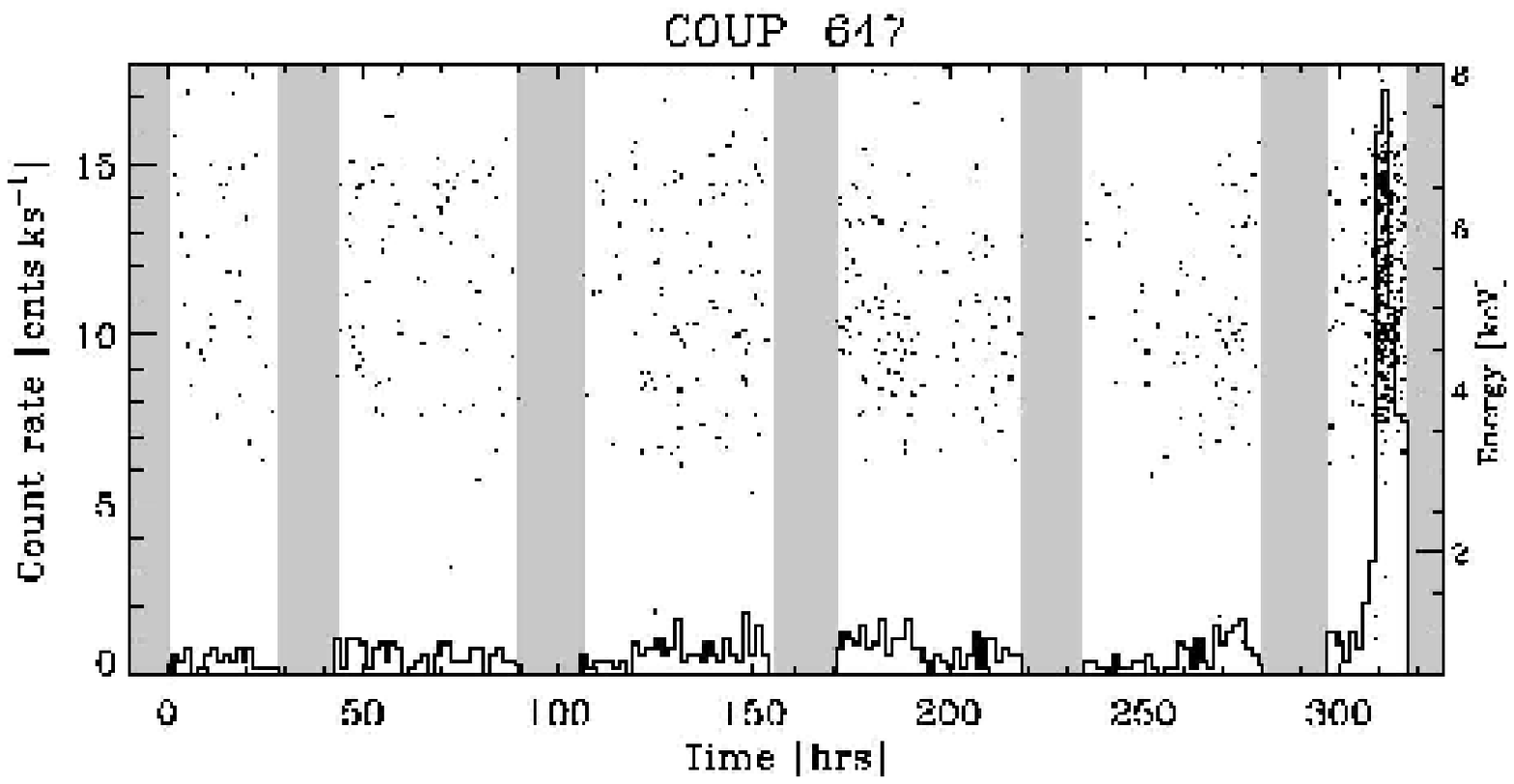} & \includegraphics[width=0.5\columnwidth]{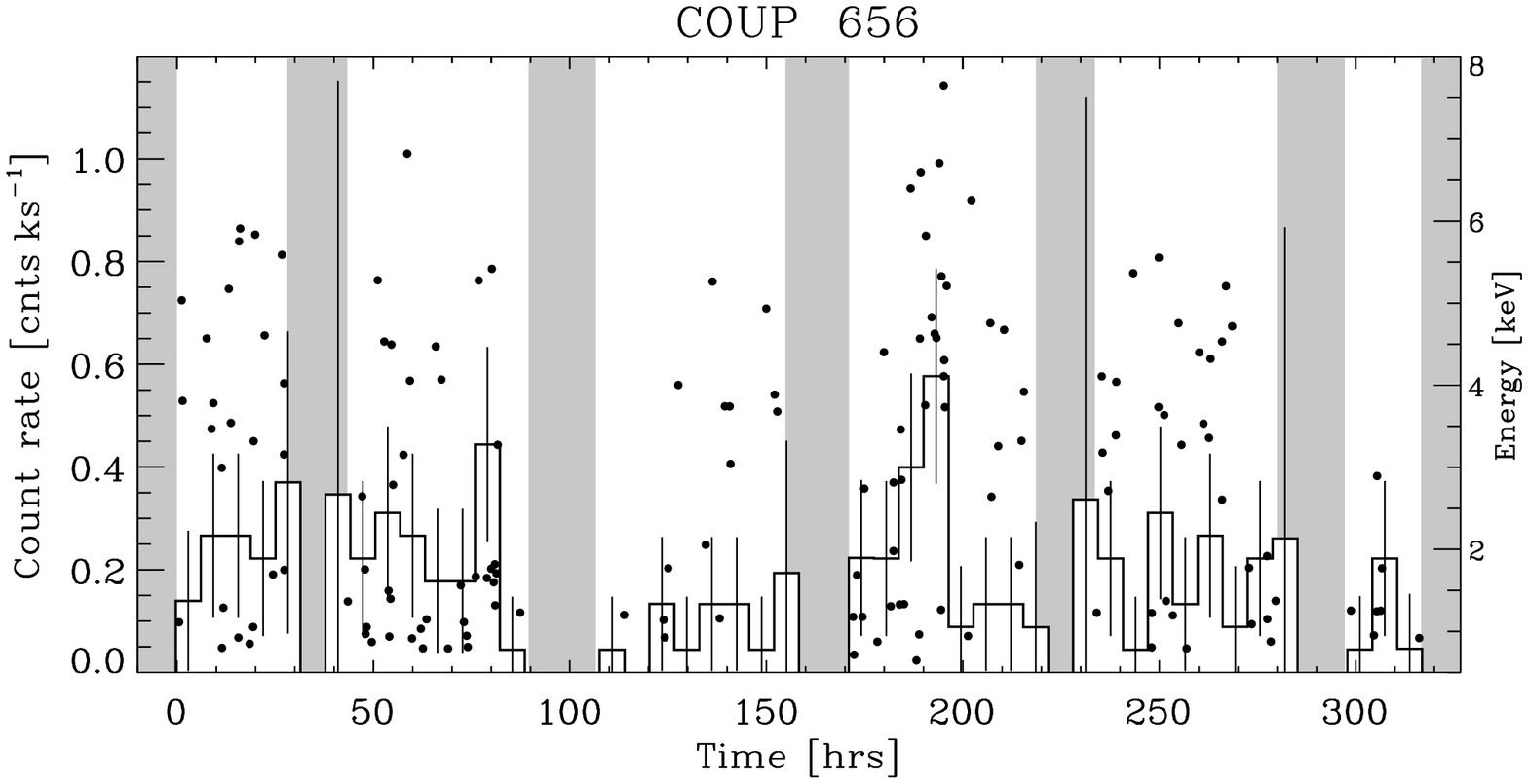}\\
\includegraphics[width=0.5\columnwidth]{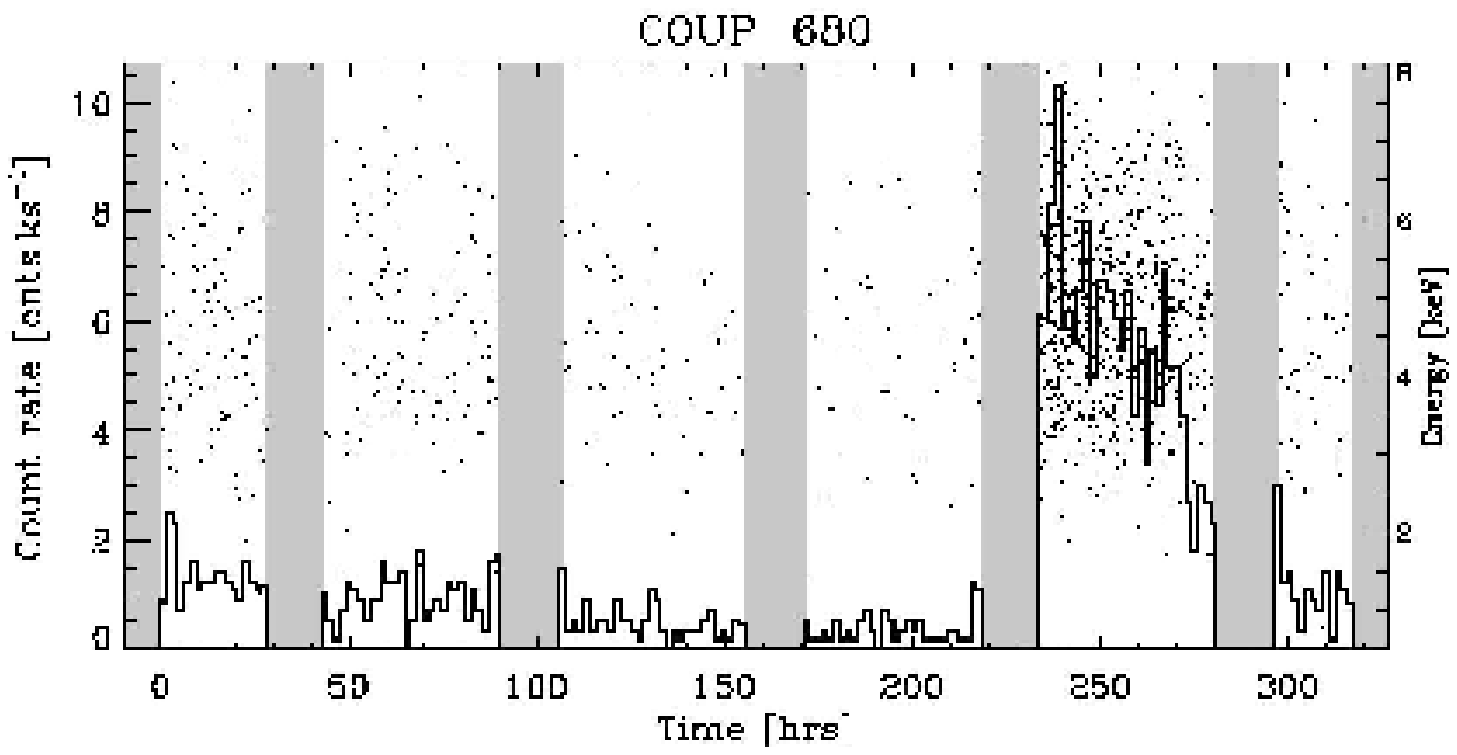} \\
\end{tabular}
 \caption{0.5--8.0\,keV light curves for COUP sources in BN-KL without 
infrared counterparts. Vertical grey stripes indicate the five passages of \cxo{}
through the Van Allen belts where ACIS was taken out of the focal plane, 
and thus was not observing Orion. Dots mark the arrival times of individual X-ray photons 
with their corresponding energies given on the righthand axis. These obscured X-ray sources 
display the typical variability of young low-mass stars with X-ray flares. 
}
\label{lc_source_x_bn}
\end{figure*}

\begin{figure*}
\centering
\includegraphics[width=0.5\columnwidth]{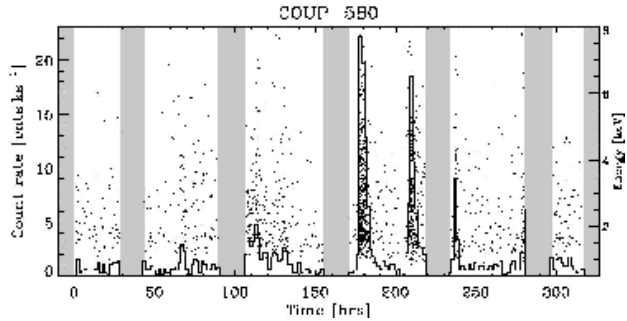}\\
 \caption{The X-ray light curve for COUP\,580, highly obscured and the most 
variable X-ray source in BN-KL\@. The source displayed a remarkable series of 
6 impulsive flares (peaking at 69, 113, 180, 210, 237, and 279\,h during the
COUP observation). The separation between each consecutive flares is 2.0, 
2.7, 1.3, 1.2, and 1.7 days.
}
\label{lc_multi_flares}
\end{figure*}

\begin{figure*}[h]
\begin{minipage}{0.5\columnwidth}
\centering
 \includegraphics[width=\columnwidth]{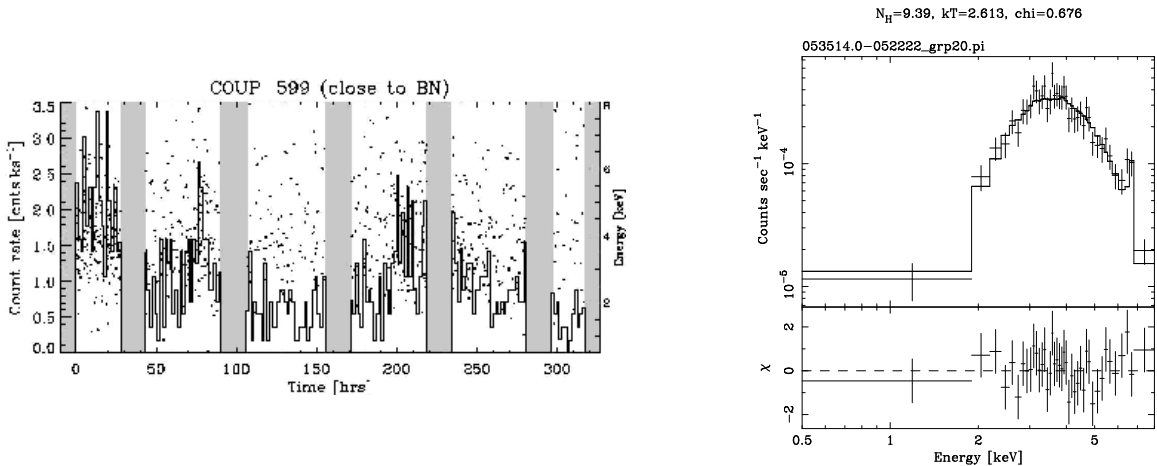}
\end{minipage}
\ \
\begin{minipage}{0.5\columnwidth}
\centering
\includegraphics[angle=-90,width=0.65\columnwidth]{f7b.ps}
\end{minipage}
\caption{The X-ray light curve and X-ray spectrum for COUP\,599a, an X-ray 
bright source close to (but not coincident with) BN, without an infrared
counterpart.}
\label{Coup_599}
\end{figure*}

\begin{figure*}
\centering
\begin{minipage}[t]{1.0\textwidth}
  \centering
  \includegraphics[angle=0.,scale=0.33]{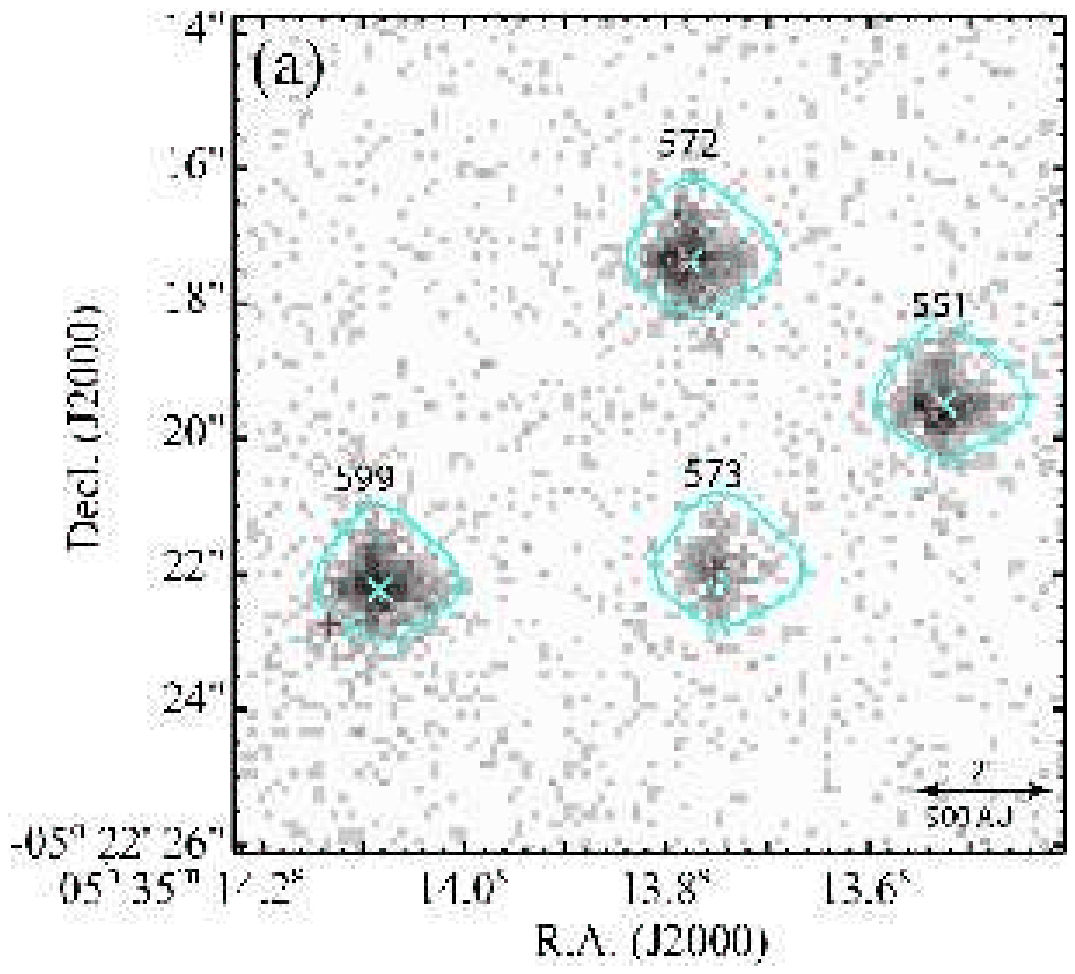} \hspace{0.05in}
  \includegraphics[angle=0.,scale=0.33]{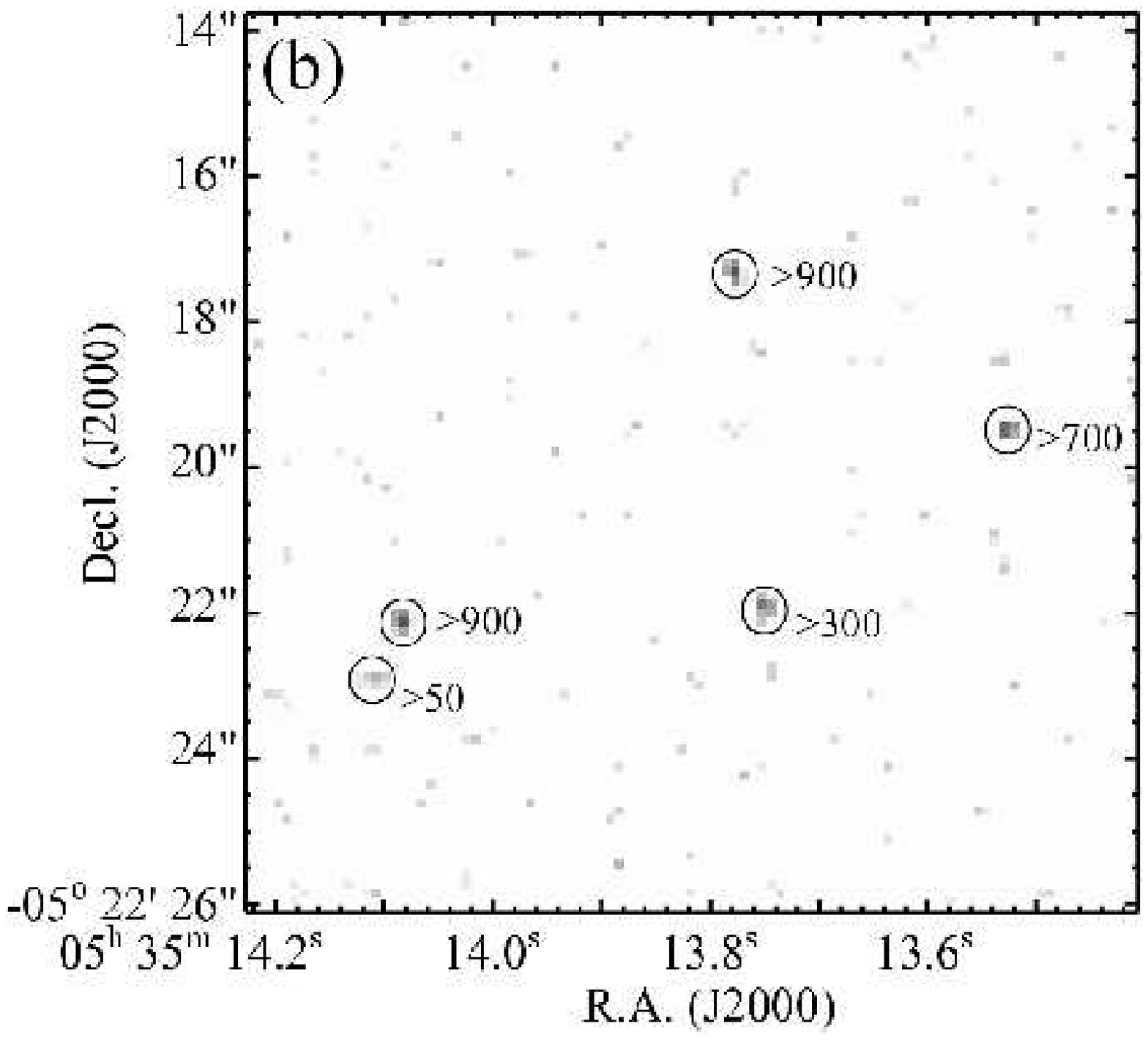} \hspace{0.2in}
\end{minipage}
  \begin{minipage}[t]{1.0\textwidth}
  \centering
  \includegraphics[angle=0.,scale=0.45]{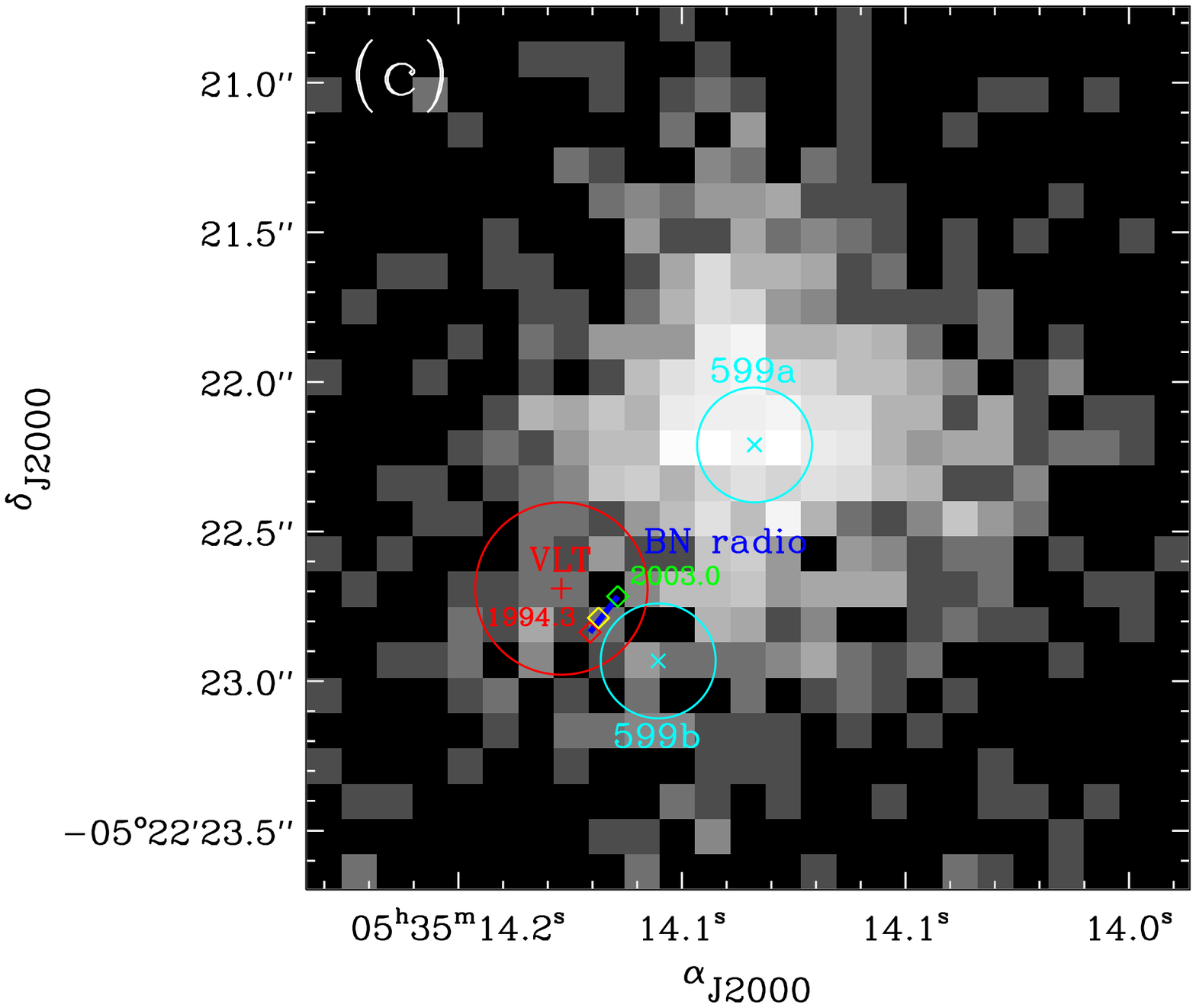} \hspace{0.0in}
\end{minipage}
\caption{COUP images in the full 0.5--8.0\,keV band covering the 
immediate neighborhood of BN at a finer scale of 0.125\arcsec/pixel. 
The image intensities are scaled logarithmically. 
(a) A $12.3\arcsec\times 12.3$\arcsec{} region of the original image; cyan
crosses mark the COUP source centroids and the cyan polygons delineate the
86\% enclosed energy regions, while the red crosses mark VLT $JHK_{\rm S}$
stellar positions.  (b) The Maximum Likelihood reconstruction of the same
field, showing the new source, COUP\,599b, 
to the southeast of COUP\,599a; 
the number of X-ray counts detected for each source is marked. (c) A further 
magnified view of the original image covering $3\arcsec\times 3$\arcsec{}
around COUP\,599a and BN.\@ COUP\,599a and COUP\,599b are marked with cyan
crosses and 0.2\arcsec{} error circles, while the VLT $J_{\rm S}$ band position
for BN is marked with a red cross and a 0.3\arcsec{} error circle. The blue 
line shows the proper motion of the radio position of BN adjusted to the 
VLT frame from epoch 1994.3 \citep{menten95} to epoch 2003.04, computed from 
\citet{tan04}. The VLA positions of BN adjusted to the VLT frame are marked 
with red, yellow, and green diamonds, corresponding to epoch 1994.3 
\citep{menten95}, epoch 1994--1997 \citep[ Z12=B=BN]{zapata04a}, and COUP 
epoch 2003.0 \citep{menten95,tan04}, respectively.}
\label{bn_reconstructed}
\end{figure*}

\begin{figure*}[t]
\centering
\includegraphics[clip=true,width=0.8\columnwidth]{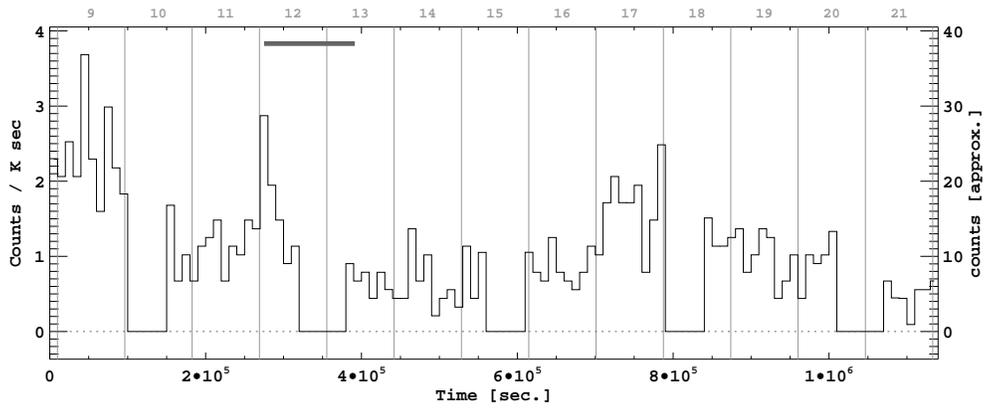}
\caption{The X-ray light curve for COUP\,599a in the 0.5--8.0\,keV 
band with a coarser bin width of 10000\,s. Boundaries between days in 2003 
January are marked by vertical lines and the dates are displayed above the 
upper axis. The horizontal bar indicates the time interval that has been 
excluded in the Lomb-Scargle Normalized Periodogram (LNP) analysis.
}
\label{lc_bn}
\end{figure*}

\begin{figure*}[h]
\centering
\includegraphics[clip=true,width=0.8\columnwidth]{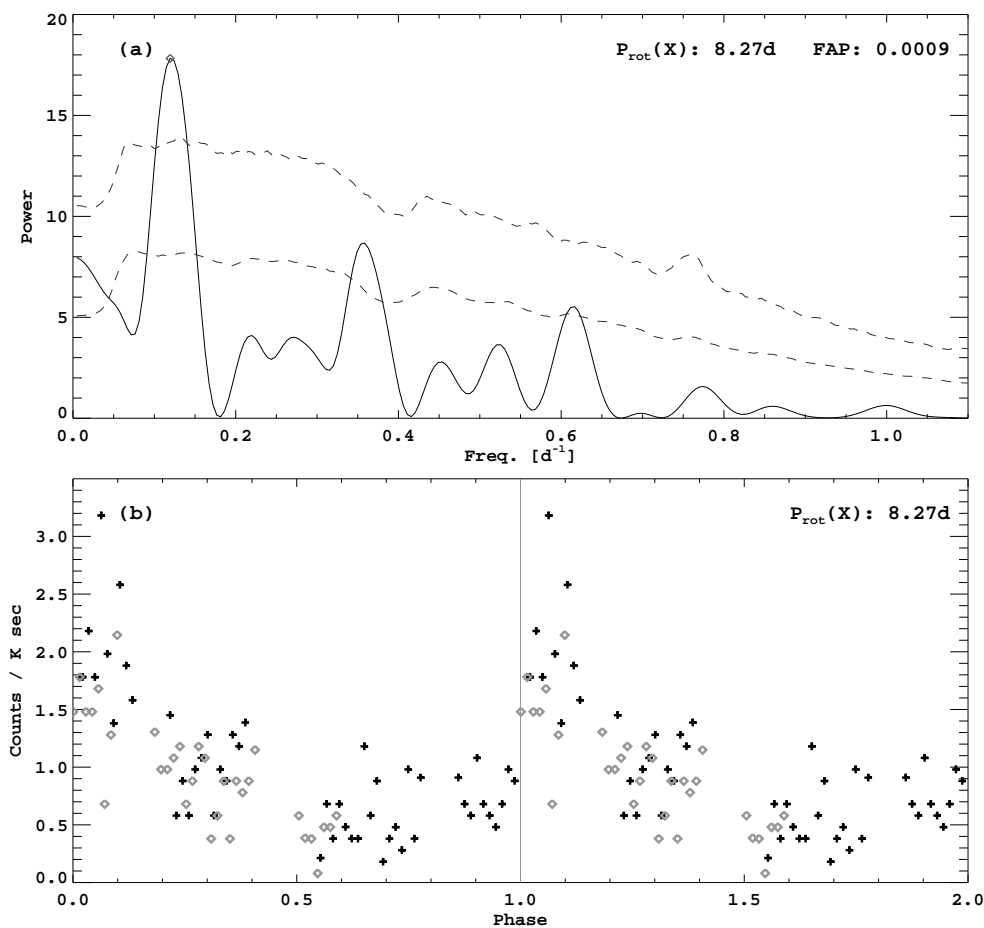}
\caption{(a) Lomb-Scargle Normalized Periodogram for COUP\,599a (solid line), 
computed from the light curve in Fig.~\ref{lc_bn} after removing the flare 
on 2003 January 12. Dashed lines indicate False Alarm Probability thresholds 
of 1\% and 0.1\%, calculated assuming noise with a correlation timescale of 
15\,h. (b) Light curve of COUP\,599a folded with the period corresponding to 
the LNP peak, 8.27\,days. Black pluses and grey diamonds indicate data points 
belonging to the first and the second rotational periods, respectively.
}
\label{periodogram}
\end{figure*}

\begin{figure*}[!h]
\centering
\begin{tabular}{cc}
\includegraphics[width=0.5\columnwidth]{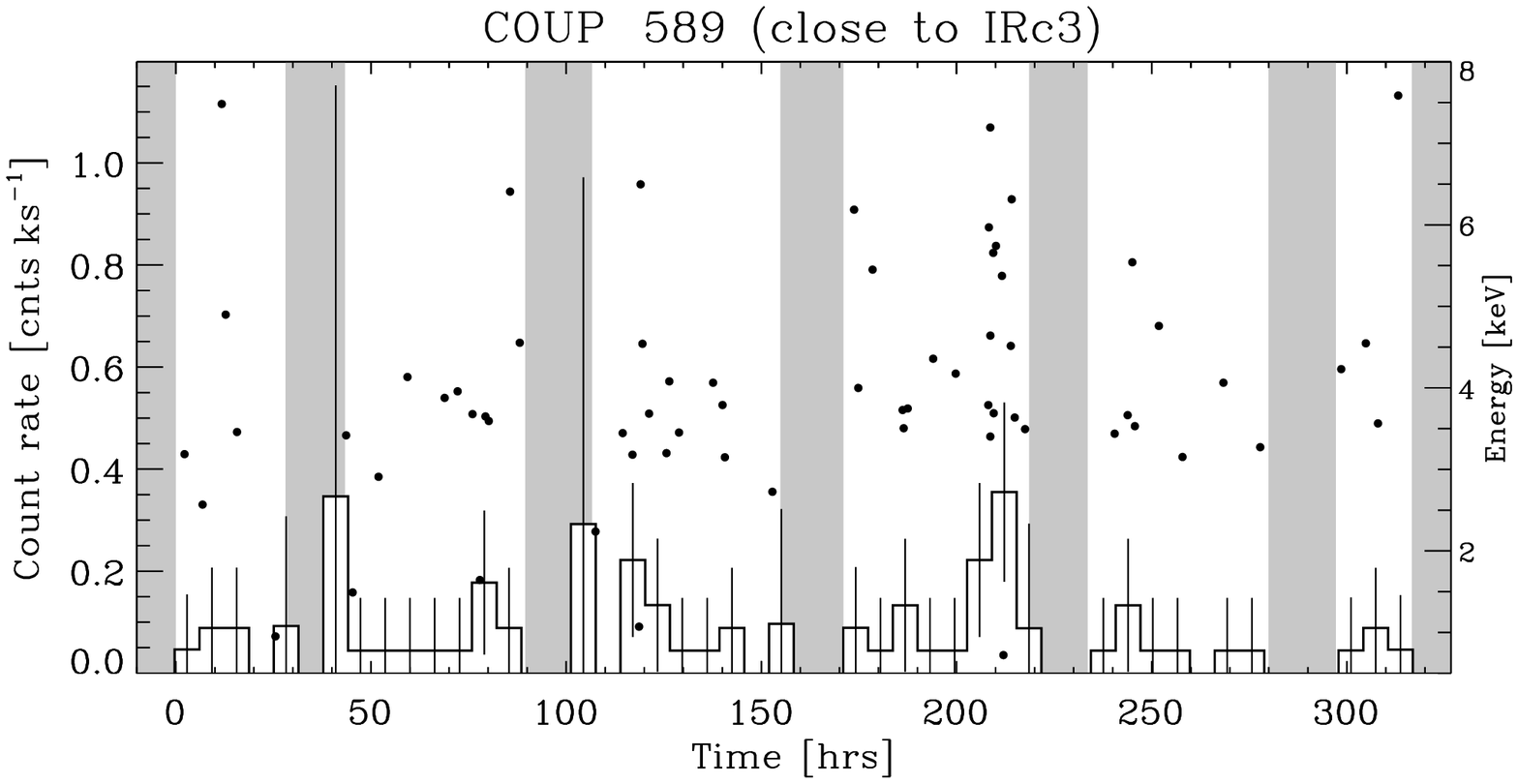} & \includegraphics[width=0.5\columnwidth]{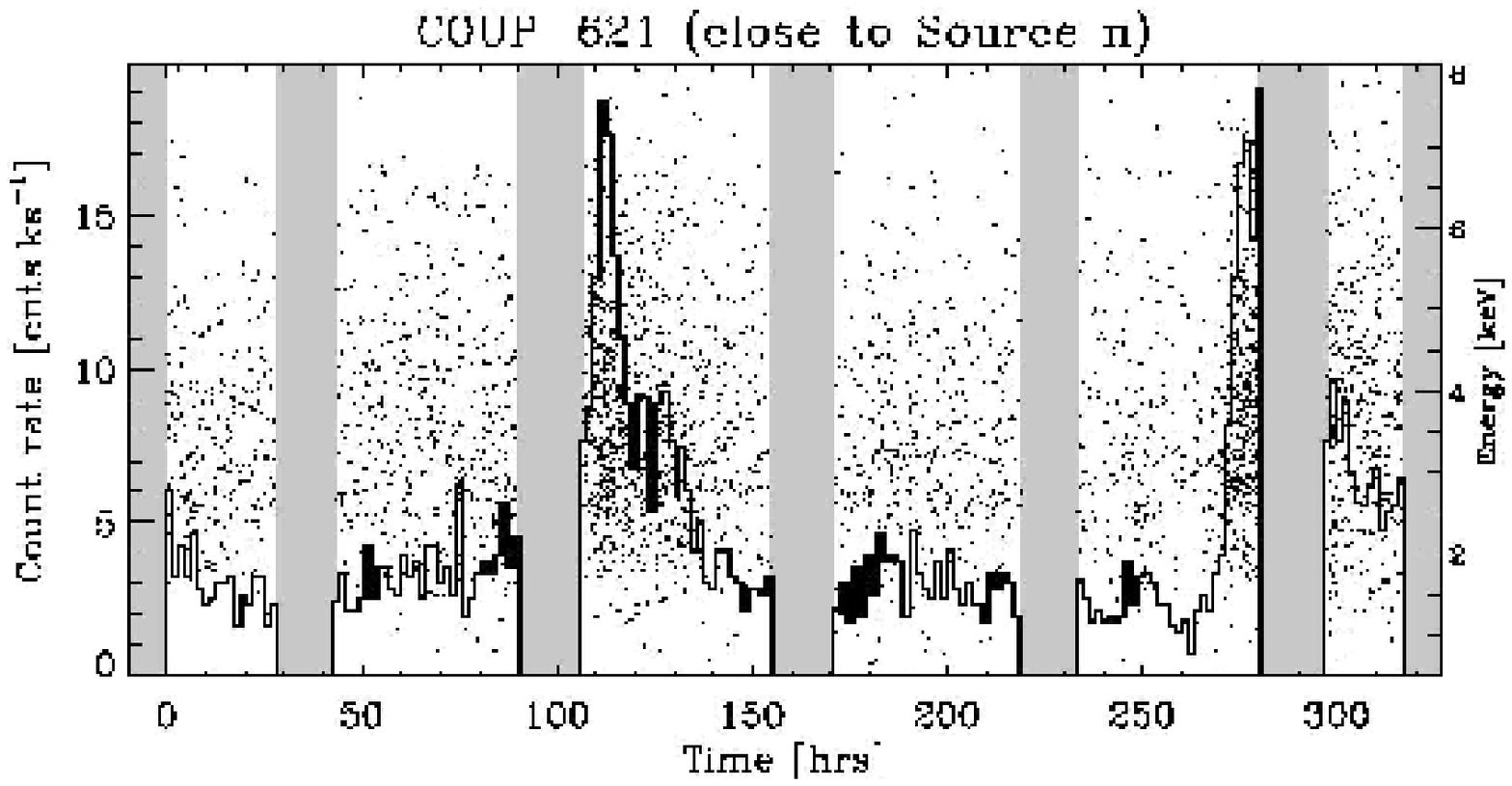} \\
\includegraphics[width=0.5\columnwidth]{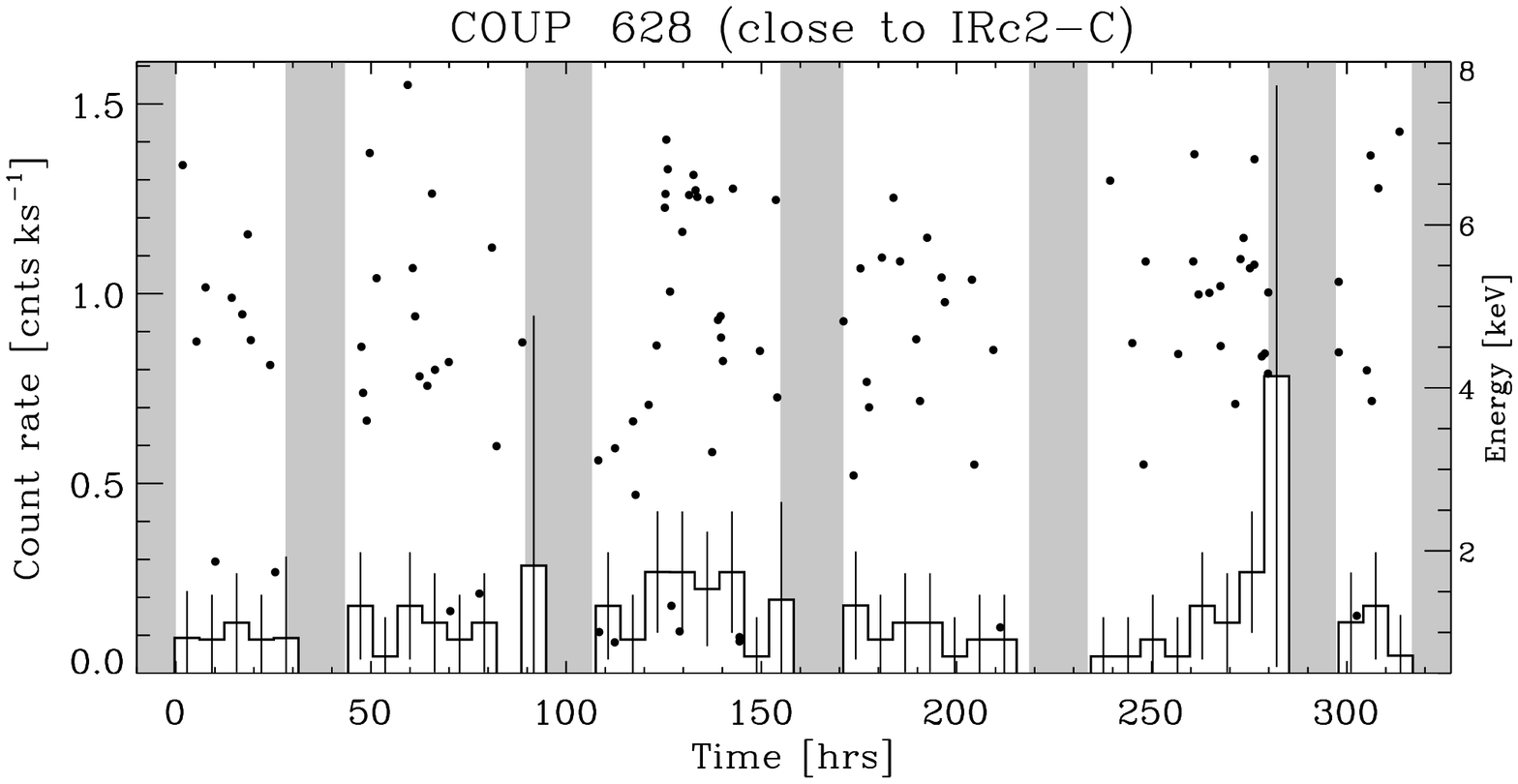}
\end{tabular}
 \caption{0.5--8.0\,keV band X-ray light curves for other COUP sources 
close to luminous mid-infrared stars in BN-KL\@. See Fig.~\ref{bn_trichro} 
for the location of these sources. 
}
\label{lc_luminous_stars}
\end{figure*}

One-third of the COUP sources within the region we have defined as OMC-1S 
have optical counterparts. The right panel of Fig.~\ref{nh} shows the 
distribution of column densities for these sources, with a median value of 
$\sim 2.3\times10^{21}$\,cm$^{-2}$, i.e.\ $A_{\rm V}\sim 1.5$\,mag. These 
relatively unobscured sources are thus likely associated with ONC, since
OMC-1S lies within 1\arcmin{} in projection from the centre of the cluster.
The other COUP sources along the line-of-sight towards OMC-1S are obscured 
by a median column density of $\sim 1.6 \times 10^{23}$\,cm$^{-2}$ or 
$A_{\rm V}\sim 100$\,mag (Fig.~\ref{nh}); all but 18 of these sources have 
near-infrared stellar counterparts. 

By design, the VLT $JHK_{\rm S}$ catalog of McCaughrean et al.\ (2005, in 
preparation) lists only point sources, but the VLT images show a number of
extended nebula sources which also exhibit X-ray emission \citep[see also][]{kastner05}. 
Of particular note in OMC-1S is
\object[COUP 0554]{COUP\,554}, which is seen in the near-infrared as a red nebulosity with a 
FWHM of $\sim$\,1.5\arcsec{} or $\sim$\,700\,AU, with a somewhat more 
extended low-level extension to the southwest. It appears to be a small 
reflection nebula at the opening of a cavity surrounding a deeply embedded 
YSO and has long been implicated as signposting one of the key embedded
sources in the OMC-1S region \citep[source S1 of][]{mccaughrean88}. 

\begin{figure*}[h]
\begin{minipage}{0.5\columnwidth}
\centering
 \includegraphics[width=\columnwidth]{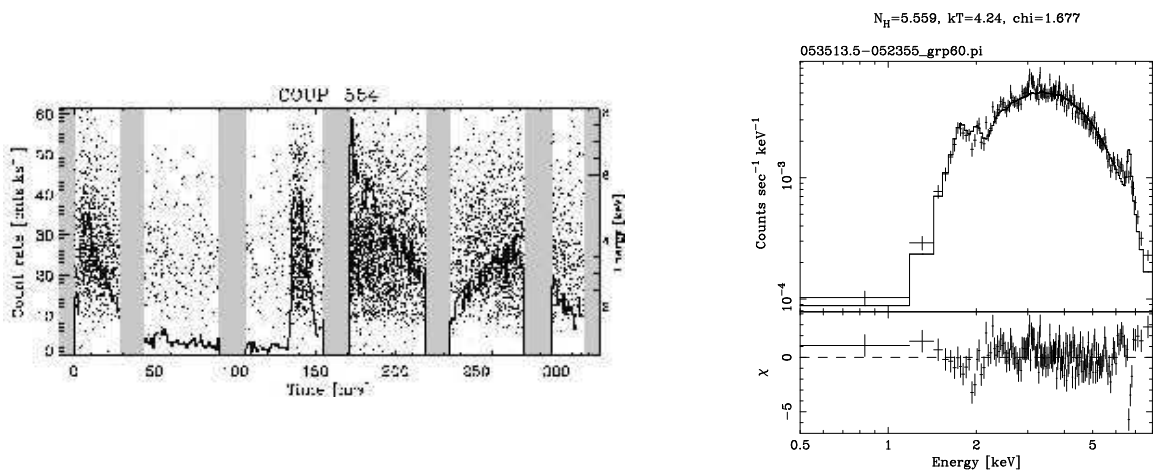}
\end{minipage}
\ \
\begin{minipage}{0.5\columnwidth}
\centering
\includegraphics[angle=-90,width=0.65\columnwidth]{f14b.ps}
\end{minipage}
\caption{0.5--8.0\,keV band X-ray light curve and X-ray spectrum for 
COUP\,554 in OMC-1S\@. This X-ray source is the brightest obscured 
($\log N_{\rm H}$=22.7, i.e.\ $A_{\rm V}\sim30$\,mag) COUP source in OMC-1S\@. The
near-infrared counterpart is a compact ($\sim$\,1.5\arcsec{} FWHM) extended 
reflection nebula, while the X-ray source appears to be a bona fide T\,Tauri 
star with enhanced magnetic activity.}
\label{coup_554}
\end{figure*}

The X-ray emission from \object[COUP 0554]{COUP\,554} is characteristic of a T~Tauri star:
its light curve and spectrum are shown in Fig.~\ref{coup_554}. It is 
slightly displaced to the northeast with respect to the centroid of the 
near-infrared emission, supporting the embedded YSO and reflection nebula 
scenario. \object[COUP 0554]{COUP\,554} is also detected at mid-infrared wavelengths 
\citep[IRS4 of][]{smith04} and at 1.3\,cm 
\citep[source 136-356 of][]{zapata04b},
but not at 3.6\,cm, again supporting the presence of an embedded YSO\@. 

Finally, according to \citet[][ component~B]{gaume98} this source
displays no near-infrared excess, whereas \citet[][ source
TPSC-46]{lada00} claim that there is evidence for near-infrared excess,
but given that the source is now clearly seen to be extended at those
wavelengths, it is not immediately obvious that this can be directly
related to the presence of a disk or protostellar status.

Just to the south of \object[COUP 0554]{COUP\,554}, \object[COUP 0555]{COUP\,555} 
is also associated with a
mid-infrared source \citep[ IRS5]{smith04} and 1.3\,cm radio emission
\citep[ source 136-359]{zapata04b}, but this time also with a near-infrared
point source \citep[ component~C; VLT catalog source 382]{gaume98}. It also has the characteristics
expected of an embedded T~Tauri star, rather than a protostar as suggested 
by \citet[][ source TPSC-16]{lada00}.

Few if any of the X-ray sources without near-infrared counterparts are
expected to be background AGN shining through the cloud, as was also the
case for BN-KL\@.  The \citet{getman05a} modelling of possible
background extragalactic contaminants predicts that only $\sim$\,1 AGN
should be seen over the OMC-1S area.  Thus most of the remaining 18
X-ray sources without near-infrared counterparts are expected to be
genuine embedded YSOs. Figure~\ref{lc_source_x_1s} shows their X-ray
light curves:  several X-ray flares are evident.  Seven of these sources
(namely \object[COUP 0582]{COUP\,582}, \object[COUP 0594]{594}, \object[COUP 0615]{615}, 
\object[COUP 0633]{633}, \object[COUP 0641]{641}, \object[COUP 0659]{659}, 
and \object[COUP 0667]{667}) are
seen to lie in a small region just to the southeast of the core of
OMC-1S and likely represent a new subcluster of embedded YSOs.  One of
these sources, \object[COUP 0594]{COUP\,594}, was recently detected in a sensitive VLA
1.3\,cm observation of the region where it displayed a large
($\sim$\,20\%) right circular polarization indicative of gyrosynchrotron
emission from the active magnetosphere of a young low-mass star
\citep[source 140-410 of][]{zapata04b}.

Conversely, it is worth looking at the most luminous mid- and far-infrared
sources in OMC-1S and examining their X-ray properties. No X-ray emission was
detected at the position of the luminous far-infrared/sub-millimetre source 
FIR4 \citep{mezger90} or at the dense molecular condensation CS3 
\citep{mundy86}, both of which were suggested to be the the exciting source 
of an embedded, highly collimated molecular outflow \citep{schmid-burgk90}.
Similarly, COUP did not detect the brightest mid-infrared source in OMC-1S,
IRS1 of \citet{smith04}, located at the base of the prominent jet that powers 
HH\,202, or IRS3 of \citep{smith04}, the third most luminous mid-infrared 
source in the region. Lastly, no X-ray emission was detected at the exact
location of `Optical Outflow Source' of \citet{odell03} (see Fig.~\ref{omc1s_trichro}).

The nearest X-ray source to the OOS is \object[COUP 0632]{COUP\,632}, 3.8\arcsec{} away from it.
\object[COUP 0632]{COUP\,632} is a faint source with just 16 counts collected, but is remarkably
the most embedded X-ray source in the entire COUP catalog, with 
$\log N_{\rm H}$=23.94, i.e.\ $A_{\rm V}\sim500$\,mag (see Fig.~\ref{most_embedded}
for its light curve and X-ray spectrum). This extreme level of extinction
could be produced by a circumstellar envelope plus a disk seen close to 
edge-on, and indeed, \object[COUP 0632]{COUP\,632} is associated with the second brightest 
mid-infrared source in the region, IRS2 of \citet{smith04}. The spectral 
energy distribution of IRS2 continues to rise to 20\,$\mu$m indicating a 
large infrared excess typical of a Class~I source, i.e.\ an evolved protostar 
\citep[see Fig.~1 of][]{smith04}. It was also suggested to be a protostellar
candidate by \citet[ source TPSC-1]{lada00}. 

\citet{smith04} proposed that \object[COUP 0632]{COUP\,632}/IRS2 was the most likely
driving source of the HH\,529 jet which extends east from OMC-1S at a
position angle of $\sim 100^\circ$; \citet{odell03} had previously
linked HH\,529 with their OOS\@.  The source was also recently detected
at 1.3\,cm \citep[source~144-351 of][]{zapata04b}, but not at 3.6\,cm
\citep{zapata04a}, and as its bolometric luminosity \citep[$\ga
7.9$\,L$_\odot$;][]{smith04} is too small to drive an H\,{\small II}
region, \citet{zapata04a} proposed that the radio emission emanates from
the magnetosphere (gyrosynchrotron emission) or the ionized outflow
(free-free emission) of a low-mass star.

\section{Near-infrared disk emission from X-ray sources in OMC-1}
\label{disk}

Approximately 70\% of the X-ray detected stars in BN-KL and OMC-1S have 
near-infrared counterparts and we have examined these sources in near-infrared
color-color diagrams (see Fig.~\ref{color_color}) to explore the fraction 
with excess emission due to disks \citep[e.g.,][]{mccaughrean96,
hillenbrand98,lada00,lada04}. A given source is said to have a excess
emission if it lies to the right of the main sequence colors with 
extended reddening vectors, after allowing for 0.1\,mag photometric
errors in all three bands used in a given color-color diagram.

We have used three different color-color diagrams, namely $H - K_{\rm S}$ vs.\
$J - H$, $K_{\rm S} - L^\prime$ vs.\ $J - H$, and $K_{\rm S} - L^\prime$ vs.\ $H - K_{\rm S}$,
combining the VLT data of McCaughrean et al.\ (2005, in preparation) with the
$L^\prime$ data of \citet{lada04}. There appears to be little difference 
between BN-KL and OMC-1S, with the latter perhaps including a handful more 
extremely red sources than the former, suggesting that OMC-1S may be younger
and its YSOs more deeply embedded. However, BN-KL is a much more confused
region than OMC-1S given the brightness of \object[NAME BN Object]{BN} itself and the surrounding
nebulosity, and thus faint, extremely red sources will be harder to see
there.

In both BN-KL and OMC-1S, the excess fraction increases from
$\sim$\,25\% in the first diagram to $\sim$\,60\% in the last, similar
to the results obtained by \citet{lada04} for stars in the core of the
ONC without any selection on X-ray properties.  Under the assumption
that excess emission is due to circumstellar disks, we therefore deduce
that these deeply embedded star formation sites in OMC-1 have disk
fractions indistinguishable from the larger cluster.

\section{Stellar population of OMC-1} \label{xlf} 

Having examined the two regions and individual sources within them in
detail, we can now look at the overall properties of the populations of
X-ray sources embedded in BN-KL and OMC-1S\@.  In order to differentiate
between embedded OMC-1 sources and unobscured foreground ONC stars along
the line-of-sight to the two regions, we consider only X-ray sources
without optical counterparts and with $\log N_{\rm H} \ge 22$ (see
Fig.~\ref{nh}).  As we continue, recall that BN-KL, with its
$\sim10^5$\,L$_\odot$ \citep[][ and references
therein]{werner76,genzel89} has ten times the luminosity of OMC-1S\@.

\subsection{X-ray luminosity function comparisons}
\label{xlf_comp.sec}

Different young stellar clusters, including the ONC, NGC\,1333, and
IC\,348, appear to have X-ray luminosity functions (XLFs) with a
universal lognormal shape \citep{feigelson05b}.  Once a
richness-dependent tail above $\log L_{\rm t} \leq 31.5$ is removed, the
XLF appears lognormal with a mean $<\log L_{\rm t}> = 29.3$ and standard
deviation $\sigma(\log L_{\rm t}) = 1.0$ \citep{feigelson05}.  The shape
of the XLF can be roughly understood as a convolution of the IMF, which
breaks from the Salpeter powerlaw below $\simeq 0.5$\,M$_\odot$, and the
correlation between X-ray luminosity and mass.  These two effects result
in a steep fall-off in the number of fainter X-ray stars in a young
stellar cluster, explaining why the factor of ten increase in the
limiting sensitivity of COUP over previous \cxo{} observations of the
region led to only a very small increase in the number of detected
lightly-absorbed ONC stars.  If we assume that the stellar populations
embedded in the OMC-1 cores have a similar average XLF shape, then we
can infer the total obscured population from a comparison of the
incomplete (due to obscuration) OMC-1 source counts with the more
complete ONC source counts.  We restrict our analysis to the hard X-ray
band as that is less affected by obscuration.

Following \citet{preibisch05a}, our ONC reference counts are taken from the
COUP `lightly absorbed optical sample', which comprises sources from 
\citet{hillenbrand97} which were classified as ONC members, for which
spectral types are known, and which have $A_{\rm V} \le 5$\,mag. 

The lightly absorbed optical sample consists of 584 stars, 551 of which
were detected as X-ray sources by COUP and have X-ray hard band
corrected luminosities.  For the 33 stars in the lightly absorbed
optical sample which were not detected by \cxo, we used PIMMS to
transform the count rate upper limits into the luminosity upper limits,
assuming a Raymond-Smith plasma spectrum with a temperature of
$kT=1$\,keV, 0.2 times solar abundance, and computing the absorbing
hydrogen column from the visual extinction, according to the relation
$N_{\rm H}=A_{\rm V} \times 1.6\,10^{21}$\,cm$^{-2}$\,mag$^{-1}$
\citep{vuong03}.

Table~\ref{statistics} summarizes the statistics of the BN-KL and OMC-1S samples.
The obscured samples of BN-KL and OMC-1S consist of 22 and 36 stars, respectively, 
but hard band luminosities are available for only 21 and 32 stars.

Figure~\ref{lh} shows histograms of the differential XLFs in the unobscured
ONC, and in the obscured samples of BN-KL, and OMC-1S\@. 
The observed (absorbed) hard band luminosities are 
in panels (a)--(c), while the intrinsic (corrected for absorption) hard band 
luminosities in panels (d)--(f). Quantitative analysis is more readily 
performed using the integral XLFs in Fig.~\ref{integral_xlf}, where ONC XLF 
is the maximum-likelihood Kaplan-Meier estimator\footnote{This estimator of 
the empirical distribution function takes into account the 
nondetections in ONC sample. The Kaplan-Meier and two-sample statistical 
methods used here -- the Gehan and Peto-Peto generalized Wilcoxon tests and 
the Logrank test -- are standard methods of univariate survival analysis as 
described by \citet{feigelson85}.}. 
We can see that the OMC-1 stellar populations have average luminosities 
systematically higher by $\Delta\! \log L_{\rm h,c} \sim 1$ than the
unobscured ONC population. This difference is present at a significance level 
of $P$=99.48--99.98\%. On the other hand, the BN-KL and OMC-1S XLFs are 
statistically indistinguishable.

The elevated XLFs of the OMC-1 obscured populations is consistent with
the reasonable presumption that more heavily absorbed fainter stars are
missing from the samples.  We therefore introduce into the OMC-1 samples
a distribution of missing stars mimicking the distribution of unobscured
ONC stars in the interval $26.5 \le \log L_{\rm h,c} \le 29.5$ (i.e.\
from the minimum $\log L_{\rm h,c}$ of the ONC sample to about 80\% of
the XLF of the OMC-1 samples).  Forty trial sets were computed for each
trial number of added stars.  In order to make the XLFs compatible with
that of the unobscured ONC sample at the $P$=10\,\% confidence level, 13
and 10 stars had to be added to the 21 and 32 stars with measured
$L_{\rm h,c}$ of BN-KL and OMC-1S, respectively.  There are 1 and 4
faint X-ray sources in BN-KL and OMC-1S, respectively, which have no
hard X-ray luminosities and were therefore not included in our XLF
analysis, leaving 18 of the 23 nominally `missing' X-ray stars in
OMC-still to be identified.

Twenty-five stars had to be added to each region to bring their mean and
median XLF values to the same levels as the unobscured ONC sample.  If
$>$50 stars were added to each region, their XLFs become too heavily
weighted at low luminosities at the $P<10$\% confidence level.  From
this analysis, we estimate that the true populations of obscured sources
in BN-KL and OMC-1S are $\simeq$46 and 57 stars, with 90\% confidence
intervals of 34--71 and 42--82 stars, respectively.

\subsection{Initial Mass Function comparisons}

We now extend this analysis to an examination of the stellar initial
mass function (IMF) for obscured X-ray sources in BN-KL and OMC-1S\@.
The procedure developed here is approximate and its reliability is
uncertain.  For example, we do not account for scatter or nonlinearity
in the ONC X-ray/mass relation based on the evolutionary tracks of
\citet{siess00}, nor for the inefficiency of COUP in detecting the
very-low-mass population of any Orion sample.  Nonetheless, this effort
gives the first estimate of the full IMF associated with the OMC-1
cores.  Previous studies of these populations were limited to luminous
OB stars or protostars with heavy disks and/or jets.

Starting with the observed $\log L_{\rm h,c}$ sample, a correction to
$L_{\rm h,c}$ values is applied to convert it to total X-ray luminosity
in the 0.5--8\,keV energy corrected from absorption, $L_{\rm t,c}$.
From the ONC optical sample, excluding stars with $T_{\rm eff}>6000$\,K, 
we obtain $\log L_{\rm t,c}= 0.81(\pm0.01) \times \log L_{\rm h,c} + 6.1(\pm 0.3)$.  
We then apply the X-ray luminosity-mass relation derived
by \citet{preibisch05a} for ONC optical sample (excluding the O-, B-,
and A-type stars), $\log L_{\rm t,c} = 30.43(\pm 0.04) + 1.56(\pm0.08)\times \log M$.  

Thirteen and ten stars, for BN-KL and OMC-1S respectively, are added 
to the observed $\log L_{\rm h,c}$ to account for the missing low X-ray
luminosity stars needed to match the OMC XLF (\S\ref{xlf_comp.sec}).  To
accomplish this, 10,000 IMF trial sets are generated where these stars
are randomly distributed in the luminosity interval $26.5 \le \log
L_{\rm h,c} \le 29.5$ using the formulae above in the mass interval
0.01\,M$_\odot$ $\le M \le$ 0.53\,M$_\odot$.  The IMF below
0.53\,M$_\odot$ is set at the average of these 10,000 trial sets.  Due
to uncertainties in the mass-X-ray luminosity relation, IMFs obtained in
this fashion may not accurately reflect the true IMFs.  However our aim
here is only to compare the X-ray IMFs of different sample.

Figure~\ref{bn_imf} shows the OMC-1 core IMFs resulting from these
calculations, together with the IMF of the lightly-absorbed ONC
population for comparison.  All three IMFs peak around 0.6\,M$_\odot$.
This similarity is not surprising, as it results from our matching their
XLFs and applying the same mass-X-ray luminosity relation.  One can see,
however, a small excess in stars with masses $M>1$ M$_\odot$ in the 
BN-KL region over both the ONC and OMC-1S samples.  

\subsection{Discussion}

BN-KL being ten times brighter than OMC-1S, one would predict that the low-mass population 
should be correspondingly ten times higher. However our census of the X-ray low-mass stars 
show that both regions have roughly the same number of stars, $N\sim50$. 
We discuss several possible explanations for this surprising result.

One possible cause could be a deficit on COUP sources in BN-KL produced
by confusion in the COUP image, particularly around IRc2.
Figure~\ref{bn_trichro} shows a compact cluster of about 10
X-ray sources in this area.  However, patchy diffuse emission is seen.
It may be due in part to extended point spread function winds of bright 
ONC X-ray sources, but some intrinsic emission from deeply embedded
stars, each emitting fewer counts than needed to trigger the COUP
source detection algorithms \citep{getman05b}, is likely present. 
However, it is hard to imagine that dozens of missing
X-ray sources are hidden in a so small region.

Another effect to explore combines two physical effects.  While $L_{\rm bol}$
increases faster than linearly with mass for OB stars, there is a
strongly stochastic variation in the mass of the single most massive
(hence highest $L_{\rm bol}$) member of sparse stellar clusters.  This
latter point is well-illustrated in Fig.~1 of \citet{bonnell99} which
shows a huge range in highest-mass member of sparse clusters.  For
example, if the mass of the most massive star of the cluster is 5 and
24\,M$_{\rm \odot}$, the model of Bonnell \& Clarke predicts that the
10\% quantile and the 50\% quantile (median) of the cluster members,
respectively, are both equal to 50.  Thus, integrated $L_{\rm bol}$ is
a poor predictor of cluster richness.  

However, again it is not clear this stochastic sampling of the upper IMF
can account for BN-KL and OMC-1S showing very similar stellar
populations.  The most massive star of OMC-1S is probably component~C
which exhibits colors consistent with a ZAMS A0 star \citep{gaume98}.
In contrast, the BN-KL region suggests a scenario where the cluster
luminosity is not dominated by a single massive source (once thought to
be IRc2), but by a cluster of luminous sources \citep{gezari98,
greenhill04}.  The most luminous may be Source~I with B3~V type or
earlier producing $\ga 3\times10^3$\,L$_\odot$.  While a single
massive star might emerge from stochastic sampling off a standard IMF, 
a cluster of intermediate-mass stars is less likely.

Thirdly, the differences in far-infrared (FIR) luminosities of these
sub-regions may not, by themselves, indicate strong differences in the
respective stellar populations.  \citet{carpenter00} derive FIR
estimates for a range of clusters in W3/W4/W5 along with estimates of
the spectral type of the cluster's most massive star; they find no
correlation between strength of FIR and the existence of massive stars.
This can be explained as an evolutionary effect where extremely young
hypercompact H\,{\small II} regions are quenched due to extinction in
the FIR and ``get brighter'' as they get older.  The BN-KL star cluster
may thus be older, but not significantly richer, than the OMC-1S star
cluster.

Finally, the most straightforward interpretation of Fig.~\ref{bn_imf}
may be correct: the BN-KL region has a top-heavy IMF compared to
both the OMC-1S and ONC populations.  Top-heavy IMFs have been proposed
for masses $M \ga 50$\,M$_\odot$ in certain starburst regions and
galaxies, but have not been reported for masses around 1--4\,M$_\odot$
as suggested by Fig.~\ref{bn_imf}.

\section{Conclusions}
\label{conclusions}

The exceptionally deep COUP exposure of the Orion Nebula has allowed us
to examine the embedded stellar populations associated with BN-KL and
OMC-1S in the background OMC-1 cloud with the following results:

\begin{list}{}{\listparindent 2cm}

\item[1.]  A total of 43 and 60 X-ray sources are detected in BN-KL and
OMC-1S, respectively, 26 and 29 of which were previously seen in a
shorter \cxo{} observation \citep{garmire00}.  Half and one-third of the
X-ray sources seen along the line-of-sight towards BN-KL and OMC-1S are
likely foreground members of the Orion Nebula Cluster.  Some 22 and
36 sources are obscured X-ray sources in BN-KL and OMC-1S, respectively;
all but 5 and 18 of these sources have infrared stellar counterparts.
Together, 22 of these new obscured X-ray sources without infrared
counterparts appear to be YSOs in OMC-1.  In particular, in OMC-1S,
\cxo{} reveals a compact subcluster of seven new YSOs without infrared
counterparts.

\item[2.]  X-ray sources are located close to four luminous mid-infrared
sources in BN-KL:  \object[NAME BN Object]{BN}, IRc3-i2, IRc2-C, and Source n.  Their X-ray
variability and spectral properties are typical of coronal activity of
low-mass companions rather than wind emission from massive stars, so we
may have detected companions rather than the massive stars themselves.
However, interpretation of X-ray emission even from unobscured OB
stars in the ONC is often difficult due to the frequent presence of
magnetic effects even in mass star winds \citep{schulz03, stelzer05}. 

\item[3.]  We have investigated the region immediately surrounding \object[NAME BN Object]{BN}
itself in detail, using an X-ray image deconvolution to reveal a new,
faint source which may be \object[NAME BN Object]{BN} itself, or a low-mass companion to \object[NAME BN Object]{BN} at
$\sim$\,200\,AU.  A nearby brighter X-ray source is definitively shown
not to be \object[NAME BN Object]{BN}, but it remains unexplained how a possible X-ray
periodicity of $\sim$\,8.3\,days in this source may be related to the
very similar periodicity seen for \object[NAME BN Object]{BN} in the near-infrared.  Finally, no
X-ray emission was seen from the bright radio Source~I, one of the
likely main sources of luminosity in the region.

\item[4.]  From a comparison of the hard band X-ray luminosity functions
of obscured X-ray sources in BN-KL and OMC-1S with that for unobscured
X-ray sources in the ONC, and applying a statistical relation linking
X-ray luminosities with stellar masses in the ONC, we estimate that the
true populations of obscured sources in BN-KL and OMC-1S are $\simeq$46
and 57 stars, with 90\% confidence intervals of 34--71 and 42--82 stars,
respectively.  We discuss, without clear conclusion, possible
explanations for their apparently similar populations in light of
the greatly enhanced FIR radiation of BN-KL.  

\end{list}

\acknowledgments COUP is supported by the \cxo{} Guest Observer grant SAO GO3-4009A 
(E.  Feigelson, PI). Further support was provided by the \cxo{} ACIS Team contract NAS8-38252.

Facility: \facility{CXO(ACIS)}

\appendix
\section{Comments on individual COUP sources in BN-KL and OMC-1S regions}
\label{individual}

	\subsection{Individual COUP sources in BN-KL}
	\label{individual_bn}

\begin{itemize}
\item[\protect{\object[COUP 0518]{COUP\,518}}] This X-ray source is not a close binary as proposed by 
     \citet{garmire00} in note {\it b} of their Table~2. Offset by
     1.1\arcsec{} from the UKIRT/MAX mid-infrared source MAX34 ($N$=7.4\,mag) of
     \citet{robberto05}.
\item[\protect{\object[COUP 0523]{COUP\,523}}] Proplyd 132-221 \citep{odell96}; see \citet{kastner05} for
     further details. Close binary source CB1 in the NICMOS 2.15\,$\mu$m image
     of \citet{stolovy98}. Offset by 0.8\arcsec{} from the UKIRT/MAX 
     mid-infrared source MAX35 ($N$=7.2\,mag) of \citet{robberto05}.
\item[\protect{\object[COUP 0551]{COUP\,551}}] Proplyd 135-220 \citep{odell96} with a possible embedded
     silhouette disk \citep{bally00}; see \citet{kastner05} for further 
     details. 
     Faint 8.4\,GHz radio source (Menten 2000, personal communication; see 
     Garmire et al.\ 2000, Table~2, note {\it d}). 
\item[\protect{\object[COUP 0572]{COUP\,572}}] Offset by 0.5\arcsec{} from the Gemini/TReCS mid-infrared 
     source S11\,=\,137-217 (0.1\,mJy at 11.7\,$\mu$m) of \citet{smith05}.
\item[\protect{\object[COUP 0573]{COUP\,573}}] Proplyd 137-222 \citep{odell96}; see \citet{kastner05} for
     further details. Protostar candidate TPSC-29 of \citet{lada00}.
\item[\protect{\object[COUP 0578]{COUP\,578}}] Close binary source CB2 in the NICMOS 2.15\,$\mu$m image
     of \citet{stolovy98}. Offset by 0.7\arcsec{} from the UKIRT/MAX 
     mid-infrared source MAX48 ($N$=6.3\,mag) of \citet{robberto05}.
\item[\protect{\object[COUP 0579]{COUP\,579}}] Proplyd 138-207 \citep{odell96}; see \citet{kastner05} for
     further details. The identification by \citet{garmire00} with the infrared 
     source {\it g} (see note {\it f} in their Table~2) is a typo. Offset by
     0.8\arcsec{} from the the UKIRT/MAX mid-infrared source MAX46 ($N$=4.8\,mag)
     of \citet{robberto05}.
\item[\protect{\object[COUP 0589]{COUP\,589}}] Associated with a methanol maser \citep{johnston92}. Peak
     brightness at 11.6\,$\mu$m of 4.0\,Jy\,arcsec$^{-2}$ \citep{gezari98}.
\item[\protect{\object[COUP 0592]{COUP\,592}}] Source near the readout trail, contaminated by soft photons 
     (see Table~\ref{bn_cxo_properties}).
\item[\protect{\object[COUP 0599b]{COUP\,599b}}] Peak brightness at 11.6\,$\mu$m of 121\,Jy\,arcsec$^{-2}$ 
     \citep{gezari98}.
\item[\protect{\object[COUP 0600]{COUP\,600}}] Associated with 6 water masers \citep{gaume98}.
\item[\protect{\object[COUP 0620]{COUP\,620}}] Close binary source CB3 in the NICMOS 2.15\,$\mu$m image
     of \citet{stolovy98}. Offset by 0.6\arcsec{} from the Gemini/TReCS 
     mid-infrared source S15\,=\,143-205 (0.1\,mJy at 11.7\,$\mu$m) of
     \citet{smith05}.
\item[\protect{\object[COUP 0621]{COUP\,621}}] \citet{tsujimoto05} detect a 6.4\,keV iron fluorescent line 
     in the COUP spectrum of this source. Protostar candidate TPSC-43 
     of \citet{lada00}.
\item[\protect{\object[COUP 0628]{COUP\,628}}] Peak brightness at 11.6\,$\mu$m of 7.5\,Jy\,arcsec$^{-2}$ 
     \citep{gezari98}.
\item[\protect{\object[COUP 0638]{COUP\,638}}] The hydrogen column density derived from the X-ray median 
     energy is too high to be consistent with the optical/infrared counterpart: 
     either the identification is wrong despite the perfect positional match 
     or the computed X-ray median energy is inaccurate due to the 
     faintness of the source. 
\item[\protect{\object[COUP 0639]{COUP\,639}}] This source is not the counterpart of the source IRc18 of 
     \citet{gezari98}, located 1.6\arcsec{} away, as proposed by 
     \citet{garmire00} in note {\it j} of their Table~2.
\item[\protect{\object[COUP 0647]{COUP\,647}}] \citet{tsujimoto05} detect a 6.4\,keV iron fluorescent line 
     in the COUP X-ray spectrum of this source.
\item[\protect{\object[COUP 0648]{COUP\,648}}] Faint 8.4\,GHz radio source (Menten 2000, personal
     communication; see Garmire et al.\ 2000, Table~2, note {\it k}).
\item[\protect{\object[COUP 0655]{COUP\,655}}] Close binary source CB4 in the NICMOS 2.15\,$\mu$m image
     of \citet{stolovy98}. Close double infrared source in \citet{muench02}. 
     Typo in \citet{garmire00}, Table~2, note {\it m}~: optical ID and 
     $K$ instead of optical ID and $V$, respectively. Protostar candidate 
     TPSC-77 of \citet{lada00}. 
\item[\protect{\object[COUP 0662]{COUP\,662}}] Coincident with a faint anonymous source in NICMOS 
     2.15\,$\mu$m image of \citet[][ see their Fig.~1 at offset
     11.6\arcsec, $-2.0$\arcsec]{stolovy98}. 
\item[\protect{\object[COUP 0670]{COUP\,670}}] Offset by 0.2\arcsec{} from the Gemini/TReCS mid-infrared 
     source S26\,=\,149-239 (0.4\,mJy at 11.7\,$\mu$m) of \citet{smith05}
     and by 0.5\arcsec{} from the UKIRT/MAX mid-infrared source MAX66 ($N$=4.8\,mag)
     of \citet{robberto05}. 
\item[\protect{\object[COUP 0678]{COUP\,678}}] Coincident with a faint anonymous source both in the NICMOS 
     2.15\,$\mu$m image of \citet[][ see their Fig.~1 at offsets 
     14.5\arcsec, $-9.0$\arcsec]{stolovy98}, and in the VLT/NACO 
     2.27\,$\mu$m image of \citet[][ see their Fig. 2 at offsets 
     $-2.5$\arcsec, 7.4\arcsec]{lacombe04}. This X-ray source is not the 
     counterpart of the optical source 9086 of \citet{hillenbrand97} 
     --corresponding to the near-infrared source 606 of \citet{muench02}, 
     and VLT source 476 of McCaughrean et al.\ (2005, in preparation)-- 
     as discussed in \citet{garmire00}. 
\item[\protect{\object[COUP 0681]{COUP\,681}}] Coincident with a faint anonymous source both in the NICMOS 
     2.15\,$\mu$m image of \citet[][ see target OMC-2C in the HST data 
     archives]{stolovy98} and in the VLT/NACO 2.27\,$\mu$m image of 
     \citet[][ see their Fig. 2 at offsets $-3.7$\arcsec, 4.5\arcsec]
     {lacombe04}. In the VLT/NACO image, this source illuminates a 
     1\arcsec-size fan-shape nebula oriented towards the south. 
\item[\protect{\object[COUP 0697]{COUP\,697}}] Offset by 0.4\arcsec{} from the Gemini/TReCS mid-infrared 
     source S30\,=\,153-216 (0.5\,mJy at 11.7\,$\mu$m) of \citet{smith05}
     and by 0.8\arcsec{} from the UKIRT/MAX mid-infrared source MAX71 ($N$=4.7\,mag)
     of \citet{robberto05}. 
\item[\protect{\object[COUP 0698]{COUP\,698}}] Candidate double X-ray source in \citet{garmire00}, Table~2, 
     note {\it w}, now resolved (\object[COUP 0699]{COUP\,699} at $1.7$\arcsec{} with PA 
     $119\degr$). Offset by 0.3\arcsec{} from the Gemini/TReCS mid-infrared 
     source S31\,=\,153-225 (0.1\,mJy at 11.7\,$\mu$m) of \citet{smith05}.
\item[\protect{\object[COUP 0699]{COUP\,699}}] Proplyd 154-225 \citep{odell96}; see \citet{kastner05} for
     further details. Offset by 0.7\arcsec{} from the UKIRT/MAX mid-infrared 
     source MAX70 ($N$=7.0\,mag) of \citet{robberto05}.
\end{itemize}

	\subsection{Individual COUP sources in OMC-1S}
	\label{individual_omc1s}

\begin{itemize}
\item[\protect{\object[COUP 0420]{COUP\,420}}] Offset by 0.2\arcsec{} from the UKIRT/MAX mid-infrared 
     source MAX18 ($N$=6.3\,mag) of \citet{robberto05}. Protostar candidate 
     TPSC-74 of \citet{lada00}.
\item[\protect{\object[COUP 0423]{COUP\,423}}] Offset by 0.5\arcsec{} from the UKIRT/MAX mid-infrared 
     source MAX19 ($N$=6.9\,mag) of \citet{robberto05}. Protostar candidate 
     TPSC-32 of \citet{lada00}. 
\item[\protect{\object[COUP 0434]{COUP\,434}}] Offset by 0.3\arcsec{} from the Gemini/TReCS mid-infrared 
     source S1\,=\,116-421 (0.02\,mJy at 11.7\,$\mu$m) of \citet{smith05} 
     and by 0.4\arcsec{} from the UKIRT/MAX mid-infrared source MAX20 ($N$=8.0\,mag) of 
     \citet{robberto05}.
\item[\protect{\object[COUP 0441]{COUP\,441}}] Protostar candidate TPSC-64 of \citet{lada00}. 
\item[\protect{\object[COUP 0443]{COUP\,443}}] Proplyd 117-352 \citep{odell96}; see \citet{kastner05} for
     further details. 
\item[\protect{\object[COUP 0465]{COUP\,465}}] Proplyd 121-434 \citep{odell96}; see \citet{kastner05} for
     further details. 
\item[\protect{\object[COUP 0470]{COUP\,470}}] Offset by 0.4\arcsec{} from the near-infrared source 302AB of \citet{muench02}. 
     Offset by 0.2\arcsec{} from the Gemini/TReCS mid-infrared 
     source IRS8 of \citet{smith04}, by 0.2\arcsec{} from Gemini/TReCS 
     mid-infrared source S3\,=\,123-348 (0.1\,mJy at 11.7\,$\mu$m) of
     \citet{smith05}, and by 0.4\arcsec{} from 
     the UKIRT/MAX mid-infrared source MAX28 ($N$=6.0\,mag) of 
     \citet{robberto05}.
\item[\protect{\object[COUP 0484]{COUP\,484}}] Protostar candidate TPSC-50 of \citet{lada00}.
\item[\protect{\object[COUP 0488]{COUP\,488}}] Offset by 0.3\arcsec{} from the near-infrared source 318 of \citet{muench02}. 
     Offset by 0.15\arcsec{} from the Gemini/TReCS mid-infrared 
     source IRS7 of \citet{smith04}, by 0.1\arcsec{} from the Gemini/TReCS 
     mid-infrared source S4\,=\,126-344 (0.2\,mJy at 11.7\,$\mu$m) of
     \citet{smith05}, and by 0.3\arcsec{} from
     the UKIRT/MAX mid-infrared source MAX31 ($N$=5.5\,mag) of 
     \citet{robberto05}.
\item[\protect{\object[COUP 0545]{COUP\,545}}] Offset by 0.1\arcsec{} from the near-infrared source 327 of \citet{muench02}. 
     Offset by 0.1\arcsec{} from Gemini/TReCS mid-infrared 
     source IRS9 of \citet{smith04}, by 0.1\arcsec{} from the Gemini/TReCS 
     mid-infrared source S8\,=\,134-340 (0.1\,mJy at 11.7\,$\mu$m) of
     \citet{smith05}, and by 0.2\arcsec{} from
     the UKIRT/MAX mid-infrared source MAX41 ($N$=6.3\,mag) of 
     \citet{robberto05}.
\item[\protect{\object[COUP 0554]{COUP\,554}}] Associated with a compact ($\sim$\,1.5\arcsec{} FWHM) 
     extended nebula in the VLT $JHK_{\rm S}$ images (McCaughrean et al.\ 2005, in 
     preparation). The offset and $K_{\rm S}$ band magnitude are from the 2MASS 
     counterpart 2MASSJ\,05351356$-$0523552. 
     Offset by 0.2\arcsec{} from the near-infrared source 276 of \citet{muench02}. 
     Offset by 0.2\arcsec{} from the
     Gemini/TReCS mid-infrared source IRS4 of \citet{smith04}, by 0.2\arcsec{}
     from the Gemini/TReCS mid-infrared source S10\,=\,136-356 (0.1\,mJy at 11.7\,$\mu$m) of
     \citet{smith05}, and by 0.4\arcsec{}
     from the UKIRT/MAX mid-infrared source MAX43 ($N$=6.3\,mag) of 
     \citet{robberto05}. Protostar candidate TPSC-46 
     of \citet{lada00}.
\item[\protect{\object[COUP 0555]{COUP\,555}}] Associated with water masers \citep{gaume98}. 
     Offset by 0.2\arcsec{} from the near-infrared source 263 of \citet{muench02}. 
     Offset by 0.3\arcsec{} from the Gemini/TReCS mid-infrared source IRS5 of
     \citet{smith04}, by 0.3\arcsec{} from the Gemini/TReCS mid-infrared 
     source S9\,=\,136-360 (0.1\,mJy at 11.7\,$\mu$m) of \citet{smith05}, 
     and by 0.5\arcsec{} from the UKIRT/MAX mid-infrared source MAX42 ($N$=6.8\,mag) of
     \citet{robberto05}. Protostar candidate TPSC-16 
     of \citet{lada00}.
\item[\protect{\object[COUP 0594]{COUP\,594}}] Associated with water masers \citep{gaume98}.
\item[\protect{\object[COUP 0607]{COUP\,607}}] Associated with water masers \citep{gaume98}. 
\item[\protect{\object[COUP 0616]{COUP\,616}}] Proplyd 143-425 \citep{odell96}; see \citet{kastner05} for
     further details. Offset by 0.1\arcsec{} from the UKIRT/MAX mid-infrared 
     source MAX54 ($N$=5.6\,mag) of \citet{robberto05}. 
\item[\protect{\object[COUP 0631]{COUP\,631}}] Proplyd 144-334 \citep{odell96}; see \citet{kastner05} for
     further details. Offset by 0.2\arcsec{} from the near-infrared source 350 of \citet{muench02}. 
     Offset by 0.2\arcsec{} from the Gemini/TReCS 
     mid-infrared counterpart IRS11 of \citet{smith04}, by 0.2\arcsec{} from
     the Gemini/TReCS mid-infrared source S19\,=\,144-334 (0.3\,mJy at 11.7\,$\mu$m) of
     \citet{smith05}, and by 0.2\arcsec{} from
     the UKIRT/MAX mid-infrared source MAX59 ($N$=5.9\,mag) of 
     \citet{robberto05}.
\item[\protect{\object[COUP 0632]{COUP\,632}}] Offset by 0.4\arcsec{} from the near-infrared source 293 of \citet{muench02}. 
     Offset by 0.2\arcsec{} from the Gemini/TReCS mid-infrared 
     source IRS2 of \citet{smith04}, by 0.2\arcsec{} from the Gemini/TReCS 
     mid-infrared source S20\,=\,144-351 (2.5\,mJy at 11.7\,$\mu$m) of
     \citet{smith05}, and by 0.5\arcsec{} from
     the UKIRT/MAX mid-infrared source MAX58 ($N$=4.0\,mag) of 
     \citet{robberto05}. 
     Protostar candidate TPSC-1 of \citet{lada00}.
\item[\protect{\object[COUP 0671]{COUP\,671}}] Proplyd 149-329 \citep{odell96}; see \citet{kastner05} for
     further details.
\end{itemize}


\begin{deluxetable}{@{}ccccrr@{}lr@{}lr@{}lccrlrccccc@{}}
\rotate
\tablecaption{\cxo/ACIS-I sources in BN-KL region.}
\tabletypesize{\scriptsize}
\tablewidth{0pt}
\tablehead{
& & & & & & & \multicolumn{13}{c}{Counterparts}  \\
& & & & & & & \multicolumn{13}{c}{\hrulefill} \\
\multicolumn{7}{c}{\cxo/ACIS-I sources} & & & \multicolumn{3}{c}{Optical} & \multicolumn{7}{c}{Infrared}  & \multicolumn{1}{c}{Radio}  \\
\multicolumn{7}{c}{\hrulefill}       & & &  \multicolumn{3}{c}{\hrulefill} &  \multicolumn{7}{c}{\hrulefill} &  \multicolumn{1}{c}{\hrulefill}  \\
\noalign{\smallskip}
COUP & G00 & $\alpha_{\rm J2000}$ & $\delta_{\rm J2000}$ & \multicolumn{1}{c}{Net}  & \multicolumn{2}{c}{$N_{\rm H}$}  & \multicolumn{2}{c}{$\theta$} &  \multicolumn{2}{c}{H97} & $I$ & \multicolumn{1}{c}{M05} & \multicolumn{1}{c}{$K_{\rm S}$} & \multicolumn{1}{c}{ID} & \multicolumn{1}{c}{$L$} & \multicolumn{1}{c}{$K-L$} & ID & MIR & ID & Class\\
\#  &	\#	&	$5^\mathrm{h}35^\mathrm{m}$ & $-5\degr$    &  \multicolumn{1}{c}{Cts} & \multicolumn{2}{c}{cm$^{-2}$}  & \multicolumn{2}{c}{\arcsec} & \multicolumn{2}{c}{\#}  & mag & \multicolumn{1}{c}{\#} & \multicolumn{1}{c}{mag} & & mag & \multicolumn{1}{c}{mag} & \\
(1) & (2) & (3) & (4) & \multicolumn{1}{c}{(5)} & \multicolumn{2}{c}{(6)} & \multicolumn{2}{c}{(7)} & \multicolumn{2}{c}{(8)} &  (9) & \multicolumn{1}{c}{(10)} &  \multicolumn{1}{c}{(11)} &  \multicolumn{1}{c}{(12)} &  (13) &  \multicolumn{1}{c}{(14)} & (15) & (16) & (17) & (18)
}
\startdata
\object[COUP 0517]{517~} &      ~1 & $13\fs03$ & $22\arcmin00\farcs9$ &   581 &       21.1 &                 &     0.2 &              &    \object[\protect{[H97b] 395}]{395}   &   &    15.3 &     350 &    12.1 & \nodata & \nodata & \nodata &    \nodata &    \nodata &    \nodata &             ONC \\
\object[COUP 0518]{518~} &      ~2 & $13\fs05$ & $22\arcmin15\farcs1$ &  1274 &       21.5 &                 &     0.2 &              &    \object[\protect{[H97b] 9029}]{9029} &   &    14.2 &     354 &    10.9 &    M673 &    10.1 &     0.6 &          a &      MAX34 &    \nodata &             ONC \\
\object[COUP 0523]{523~} &      ~3 & $13\fs18$ & $22\arcmin21\farcs1$ &   280 &       21.9 &                 &     0.2 &              &    \object[\protect{[H97b] 399b}]{399}  & b &    16.8 &     361 &    10.5 &   M651A &     9.0 &     1.1 &          b &      MAX35 &    \nodata &             ONC \\
\object[COUP 0529]{529~} &      ~4 & $13\fs24$ & $22\arcmin09\farcs8$ &   480 &       22.9 &                 &     0.1 &              &    \nodata                              &   & \nodata &     367 &    13.1 & \nodata & \nodata & \nodata &    \nodata &    \nodata &    \nodata &             OMC \\
\object[COUP 0538]{538~} & \nodata & $13\fs37$ & $22\arcmin26\farcs1$ &   123 &       21.5 &                 &     0.1 &              &    \object[\protect{[H97b] 404}]{404}   &   &    14.6 &     374 &    11.4 & \nodata & \nodata & \nodata &          c &    \nodata &    \nodata &             ONC \\
\object[COUP 0539]{539~} & \nodata & $13\fs31$ & $22\arcmin39\farcs1$ &   502 &       23.0 &                 &     0.3 &   $^\dagger$ &    \nodata                              &   & \nodata & \nodata & \nodata &    L129 &    10.0 & \nodata &    \nodata &    \nodata &    \nodata &             OMC \\
\enddata
\end{deluxetable}


\begin{deluxetable}{@{}cccrccccr@{}lcccccc@{}}
\rotate
\tablecaption{X-ray properties of \cxo/ACIS-I sources in BN-KL region.}
\tabletypesize{\scriptsize}
\tablewidth{0pt}
\tablehead{
\noalign{\smallskip}
   &   &    &    &         &                   &     &     & \multicolumn{8}{c}{X-ray spectral fitting} \\
\vspace{-0.6cm}\\
   &   &    &    &         &                   &     &     & \multicolumn{8}{c}{\hrulefill}  \\
   & &  & \multicolumn{5}{c}{X-ray photon statistics} &      &            & \multicolumn{2}{c}{Temperature} & \multicolumn{2}{c}{Emission Measure} & \multicolumn{2}{c}{Luminosity}  \\
\vspace{-0.6cm}\\
   & & & \multicolumn{5}{c}{\hrulefill}  &           &       & \multicolumn{2}{c}{\hrulefill} & \multicolumn{2}{c}{\hrulefill} & \multicolumn{2}{c}{\hrulefill}  \\
\noalign{\smallskip}
COUP & ID & Class   & \multicolumn{1}{c}{Net} & $\log P_{\rm KS}$ & $MedE $ & $HR2$ & $HR3$ & \multicolumn{2}{c}{$N_{\rm H}$} & $kT_1$ & $kT_2$ & \multicolumn{2}{c}{$\log EM_1, EM_2$}  & \multicolumn{2}{c}{$\log L_\mathrm{h}, L_\mathrm{h,c}$} \\
\#   &    &         & \multicolumn{1}{c}{Cts} &                   & keV            &       &       &    \multicolumn{2}{c}{cm$^{-2}$}  & \multicolumn{2}{c}{keV} & \multicolumn{2}{c}{cm$^{-3}$} & \multicolumn{2}{c}{erg\,s$^{-1}$} \\
(1)& (2)& (3)     & \multicolumn{1}{c}{(4)}    & (5)               & (6) & (7) & (8)  & \multicolumn{2}{c}{(9)}    & (10) & (11) & (12) & (13) & (14) & (15) 
}
\startdata
\object[COUP 0517]{517~} &       H395 &             ONC &   581 & -4.0 & 1.1 &    -0.7 &    -0.4 &            21.1 &                 &            ~0.8 &             2.0 &            51.9 &            52.0 &            28.6 &        28.6 \\
\object[COUP 0518]{518~} &          a &             ONC &  1274 & -4.0 & 1.3 &    -0.5 &    -0.4 &            21.5 &                 &            ~0.8 &             2.4 &            52.2 &            52.5 &            29.1 &        29.2 \\
\object[COUP 0523]{523~} &          b &             ONC &   280 & -2.9 & 1.5 &    -0.4 &    -0.3 &            21.9 &                 &            ~0.8 &             3.1 &            52.1 &            51.8 &            28.6 &        28.7 \\
\object[COUP 0529]{529~} &        367 &             OMC &   480 & -4.0 & 3.8 &    ~1.0 &    ~0.6 &            22.9 &                 &            ~2.9 &         \nodata &            53.2 &         \nodata &            29.6 &        29.9 \\
\object[COUP 0538]{538~} &          c &             ONC &   123 & -0.9 & 1.2 &    -0.6 &    -0.8 &            21.5 &                 &            ~1.0 &         \nodata &            51.6 &         \nodata &            27.7 &        27.7 \\
\object[COUP 0539]{539~} &       L129 &             OMC &   502 & -4.0 & 4.0 &    ~0.9 &    ~0.8 &            23.0 &                 &            ~3.1 &         \nodata &            53.3 &         \nodata &            29.6 &        30.0 \\
\object[COUP 0551]{551~} &          e &             ONC &   654 & -4.0 & 1.3 &    -0.6 &    -0.4 &            21.6 &                 &            ~0.6 &             2.4 &            52.2 &            52.2 &            28.8 &        28.8 \\
\object[COUP 0572]{572~} &         aa &             OMC &   794 & -4.0 & 3.7 &    ~0.9 &    ~0.6 &            22.8 &                 &            ~3.0 &         \nodata &            53.3 &         \nodata &            29.8 &        30.1 \\
\enddata
\tablecomments{~ \\
Col.\ 1: COUP identifier. \\
Col.\ 2: Name of optical/near-infrared/mid-infrared/radio counterpart. \\
Cols.\ 3--4: Source class based on counterparts and net counts, respectively. \\ 
Col.\ 5: Logarithm of the probability that the source is constant, as measured
         using a Kolmogorov-Smirnov 1-sample test (i.e.\ a value of $-4.0$
         indicates a probablity of less than $10^{-4}$ that the source is
         constant. \\
Col.\ 6: Median energy of the source photons. \\
Cols.\ 7--8: Hardness ratios in 0.5--1.7\,keV vs.\ 1.7--2.8\,keV
         and [1.7--2.8]\,keV vs.\ 2.8--8.0\,keV, respectively.\\ 
Col.\ 9: Logarithm of the hydrogen column density (see notes to Col.~6 in
         Table~\ref{bn_cxo_sources}). \\
Cols.\ 10--11: Plasma temperatures .\\
Cols.\ 12--13: Emission measures. \\
Col.\ 14: Observed X-ray luminosity in the 2.8--8.0\,keV band. \\
Col.\ 15: 2.8--8.0\,keV band luminosity corrected for absorption.
\\
Comments on individual sources:\\
\object[COUP 0574]{COUP\,574}.\ Fitting procedure failed to constrain $kT$, that was fixed to 15\,keV. \\
\object[COUP 0592]{COUP\,592}.\ Source near the readout trail, contaminated by soft photons. 
Fitting procedure failed to constrain $N_\mathrm{H}$, that was fixed to 20. 
This low value of extinction is not consistent with the fact that this X-ray source has no 
optical counterpart but an IR counterpart (see Fig.~\ref{nh}). Luminosity values are considered 
not reliable, and won't be used here. \\
\object[COUP 0661]{COUP\,661}.\ Fitting procedure failed to constrain $N_\mathrm{H}$, that was fixed to 20. 
}
\label{bn_cxo_properties}
\end{deluxetable}


\begin{deluxetable}{@{}ccccrr@{}lr@{}lr@{}lccrlrccccc@{}}
\rotate
\tablecaption{\cxo/ACIS-I sources in OMC-1S.}
\tabletypesize{\scriptsize}
\tablewidth{0pt}
\tablehead{
& & & & & & & \multicolumn{13}{c}{Counterparts}  \\
& & & & & & & \multicolumn{13}{c}{\hrulefill} \\
\multicolumn{7}{c}{\cxo/ACIS-I sources} & & & \multicolumn{3}{c}{Optical} & \multicolumn{7}{c}{Infrared}  & \multicolumn{1}{c}{Radio}  \\
\multicolumn{7}{c}{\hrulefill}       & & &  \multicolumn{3}{c}{\hrulefill} &  \multicolumn{7}{c}{\hrulefill} &  \multicolumn{1}{c}{\hrulefill}  \\
\noalign{\smallskip}
COUP & G00 & $\alpha_{\rm J2000}$ & $\delta_{\rm J2000}$ & \multicolumn{1}{c}{Net}  & \multicolumn{2}{c}{$N_{\rm H}$}  & \multicolumn{2}{c}{$\theta$} &  \multicolumn{2}{c}{H97} & $I$ & \multicolumn{1}{c}{M05} & \multicolumn{1}{c}{$K_{\rm S}$} & \multicolumn{1}{c}{ID} & \multicolumn{1}{c}{$L$} & \multicolumn{1}{c}{$K-L$} & ID & MIR & ID & Class\\
\#  &	\#	&	$5^\mathrm{h}35^\mathrm{m}$ & $-5\degr$    &  \multicolumn{1}{c}{Cts} & \multicolumn{2}{c}{cm$^{-2}$}  & \multicolumn{2}{c}{\arcsec} & \multicolumn{2}{c}{\#}  & mag & \multicolumn{1}{c}{\#} & \multicolumn{1}{c}{mag} & \multicolumn{1}{c}{\#} & mag & \multicolumn{1}{c}{mag}  \\
(1) & (2) & (3) & (4) & \multicolumn{1}{c}{(5)} & \multicolumn{2}{c}{(6)} & \multicolumn{2}{c}{(7)} & \multicolumn{2}{c}{(8)} &  (9) & \multicolumn{1}{c}{(10)} &  \multicolumn{1}{c}{(11)} &  \multicolumn{1}{c}{(12)} &  (13) &  \multicolumn{1}{c}{(14)} & (15) & (16) & (17) & (18)
}
\startdata
\object[COUP 0416]{416} & \nodata & $11\fs28$ & $24\arcmin16\farcs4$ &    50 &       22.6 &                 &     0.1 &              & \nodata                              &   & \nodata &     274 &    15.6 &     L76 &    12.5 &     1.5 &    \nodata &    \nodata &    \nodata &             OMC \\
\object[COUP 0419]{419} & \nodata & $11\fs31$ & $24\arcmin26\farcs5$ &    24 &       23.7 &                 &     0.2 &              & \nodata                              &   & \nodata &     277 &    14.6 &     L77 &    14.3 &     0.8 &    \nodata &    \nodata &    \nodata &             OMC \\
\object[COUP 0420]{420} & \nodata & $11\fs32$ & $24\arcmin38\farcs2$ &    65 &       22.3 &                 &     0.1 &              & \nodata                              &   & \nodata &     276 &    12.6 &     L78 &     9.6 &     3.0 &    \nodata &      MAX18 &    \nodata &             OMC \\
\object[COUP 0423]{423} &     211 & $11\fs48$ & $23\arcmin52\farcs0$ &   429 &       23.2 &                 &     0.2 &              & \nodata                              &   & \nodata &     282 &    13.0 &     L80 &    10.8 &     1.6 &    \nodata &      MAX19 &    \nodata &             OMC \\
\object[COUP 0425]{425} & \nodata & $11\fs54$ & $23\arcmin40\farcs2$ &    62 &       23.1 &                 & \nodata &              & \nodata                              &   & \nodata & \nodata & \nodata & \nodata & \nodata & \nodata &    \nodata &    \nodata &    \nodata &             OMC \\
\object[COUP 0426]{426} & \nodata & $11\fs56$ & $23\arcmin45\farcs3$ &    24 &       23.5 &                 &     0.2 &   $^\dagger$ & \nodata                              &   & \nodata & \nodata & \nodata &     L82 &    12.9 & \nodata &    \nodata &    \nodata &    \nodata &             OMC \\
\enddata
\tablecomments{ \\
Cols.\ 1--18 are as for Table~\ref{bn_cxo_sources}, with the following
differences and/or additional notes: \\
Col.\ 2: All 31 ACIS-I sources of \citet{garmire00} were recovered in the COUP 
     observation except their sources 330 and 874 (CXOU\,J053514.9$-$052411 and 
     CXOU\,J053514.6$-$052350, respectively). \\
Cols.\ 12--14: $L$ band magnitudes (sources with `M') from \citet{muench02}
     and $L^\prime$ band magnitudes (sources with `L') from \citet{lada04}. \\
Col.\ 15: $K$ band components from \citet{gaume98}. \\
Col.\ 16: Mid-infrared counterparts MAX and S refer to \citet{robberto05}
     and \citet{smith05}, respectively. \\
Col.\ 17: New 3.6\,cm radio sources from \citet{zapata04a} are marked with `Z';
     new 1.3\,cm radio sources from \citet{zapata04b} are labelled with the
     format `xxx-yyy'. \\
Col.\ 18: Membership classification from \citet{getman05a}, where `Unc' stands for uncertain case.
}
\label{omc1s_cxo_sources}
\end{deluxetable}


\begin{deluxetable}{@{}cccrccccr@{}lcccccc@{}}
\rotate
\tablecaption{X-ray properties of \cxo/ACIS-I sources in OMC-1S.}
\tabletypesize{\scriptsize}
\tablewidth{0pt}
\tablehead{
\noalign{\smallskip}
   &   &    &    &         &                   &     &     & \multicolumn{8}{c}{X-ray spectral fitting} \\
\vspace{-0.6cm}\\
   &   &    &    &         &                   &     &     & \multicolumn{8}{c}{\hrulefill}  \\
   & &  & \multicolumn{5}{c}{X-ray photon statistics} &        &          & \multicolumn{2}{c}{Temperature} & \multicolumn{2}{c}{Emission Measure} & \multicolumn{2}{c}{Luminosity}  \\
\vspace{-0.6cm}\\
   & & & \multicolumn{5}{c}{\hrulefill}  &         &         & \multicolumn{2}{c}{\hrulefill} & \multicolumn{2}{c}{\hrulefill} & \multicolumn{2}{c}{\hrulefill}  \\
\noalign{\smallskip}
COUP & ID & Class   & \multicolumn{1}{c}{Net} & $\log P_{\rm KS}$ & $MedE$ & $HR2$ & $HR3$ & \multicolumn{2}{c}{$N_{\rm H}$} & $kT_1$ & $kT_2$ & \multicolumn{2}{c}{$\log EM_1, EM_2$}  & \multicolumn{2}{c}{$\log L_\mathrm{h}, L_\mathrm{h,c}$} \\
\#   &    &         &  \multicolumn{1}{c}{Cts}    &                   & keV            &       &       &    \multicolumn{2}{c}{cm$^{-2}$}   & \multicolumn{2}{c}{[keV]} & \multicolumn{2}{c}{cm$^{-3}$} & \multicolumn{2}{c}{erg\,s$^{-1}$} \\
(1)& (2)& (3)     & \multicolumn{1}{c}{(4)}    & (5)               & (6) & (7) & (8)  & \multicolumn{2}{c}{(9)}    & (10) & (11) & (12) & (13) & (14) & (15) 
}
\startdata
\object[COUP 0416]{416} &        274 &             OMC &    50 & -2.1 & 3.1 &    ~0.6 &    ~0.3 &            22.6 &                 &            ~2.3 &         \nodata &            52.1 &         \nodata &            28.4 &            28.7 \\
\object[COUP 0419]{419} &        277 &             OMC &    24 & -1.0 & 5.2 & \nodata &    ~1.0 &            23.7 &                 &            ~1.9 &         \nodata &            53.2 &         \nodata &            28.5 &            29.7 \\
\object[COUP 0420]{420} &        276 &             OMC &    65 & -1.0 & 2.8 &    ~0.3 &    ~0.2 &            22.3 &                 &            ~3.5 &         \nodata &            51.7 &         \nodata &            28.4 &            28.5 \\
\object[COUP 0423]{423} &        282 &             OMC &   429 & -4.0 & 4.7 &    ~0.8 &    ~0.9 &            23.2 &                 &            ~5.8 &         \nodata &            53.2 &         \nodata &            29.7 &            30.2 \\
\object[COUP 0425]{425} &    \nodata &             OMC &    62 & -4.0 & 4.6 & \nodata &    ~0.9 &            23.1 &                 &            ~6.4 &         \nodata &            52.2 &         \nodata &            28.8 &            29.1 \\
\object[COUP 0426]{426} &        L82 &             OMC &    24 & -2.2 & 4.3 & \nodata &    ~0.9 &            23.5 &                 &            ~1.2 &         \nodata &            53.4 &         \nodata &            28.4 &            29.5 \\
\object[COUP 0434]{434} &       H366 &             ONC &    64 & -0.6 & 1.1 &    -0.7 &    -0.9 &            21.5 &                 &            ~0.8 &         \nodata &            51.4 &         \nodata &            27.2 &            27.2 \\
\object[COUP 0440]{440} &        297 &             OMC &    57 & -3.3 & 4.8 &    -0.5 &    ~1.0 &            23.2 &                 &            15.0 &         \nodata &            52.1 &         \nodata &            28.8 &            29.1 \\
\enddata
\tablecomments{~ \\
Cols.\ 1--15 are as for Table~\ref{bn_cxo_properties}.
\\
Comments on individual sources:\\
\object[COUP 0440]{COUP\,440}.\ Fitting procedure failed to constrain $kT$, that was fixed to 15\,keV. \\
\object[COUP 0442]{COUP\,442}.\ Fitting procedure failed to constrain $N_\mathrm{H}$, that was fixed to 20. \\
\object[COUP 0443]{COUP\,443}.\ Fitting procedure failed to constrain $kT$, that was fixed to 15\,keV. \\
\object[COUP 0496]{COUP\,496}.\ Fitting procedure failed to constrain $kT$, that was fixed to 15\,keV. \\
\object[COUP 0509]{COUP\,509}.\ Fitting procedure failed to constrain $kT$, that was fixed to 15\,keV. \\
\object[COUP 0582]{COUP\,582}.\ Fitting procedure failed to constrain $kT$, that was fixed to 15\,keV. \\
\object[COUP 0632]{COUP\,632}.\ Fitting procedure failed to constrain $kT$, that was fixed to 15\,keV.
}
\label{omc1s_cxo_properties}
\end{deluxetable}


\begin{deluxetable}{ccc}
\tablecaption{Statistics of the membership classification of COUP sources in BN-KL and OMC-1S regions.}
\tablewidth{0pt}
\tablehead{
Membership classification & BN-KL   & OMC-1S \\
(1)       & (2)     & (3) }
\startdata
ONC       & 21      & 23 \\
OMC       & 22      & 29 \\
OMC or EG? & \nodata & 7 \\
Unc.      & \nodata & 1 \\
\multicolumn{3}{c}{\hrulefill}\\
Total     & 43      & 60\\
OMC\,+\,OMC or EG? & 22      & 36 \\
OMC\,+\,OMC or EG? with $L_{\rm h,c}$ & 21     & 32
\enddata
\tablecomments{~\\
Column (1) provides the membership classification established by \citet{getman05a}, 
where `ONC', `OMC', `EG', and `Unc', stand for ONC source, OMC source, extragalactic source, 
and uncertain case. We extend here the `OMC' classification to sources without optical 
counterpart and with $\log N_{\rm H} \ge 22$, which are not classified 'EG' by \citep{getman05a}.\\
Columns (2) and (3) give the corresponding number of COUP sources with this membership classification for BN-KL and OMC-1S, respectively.}
\label{statistics}
\end{deluxetable}


\clearpage

\clearpage
\begin{figure*}[!h]
\centering
\begin{tabular}{cc}
\includegraphics[width=0.5\columnwidth]{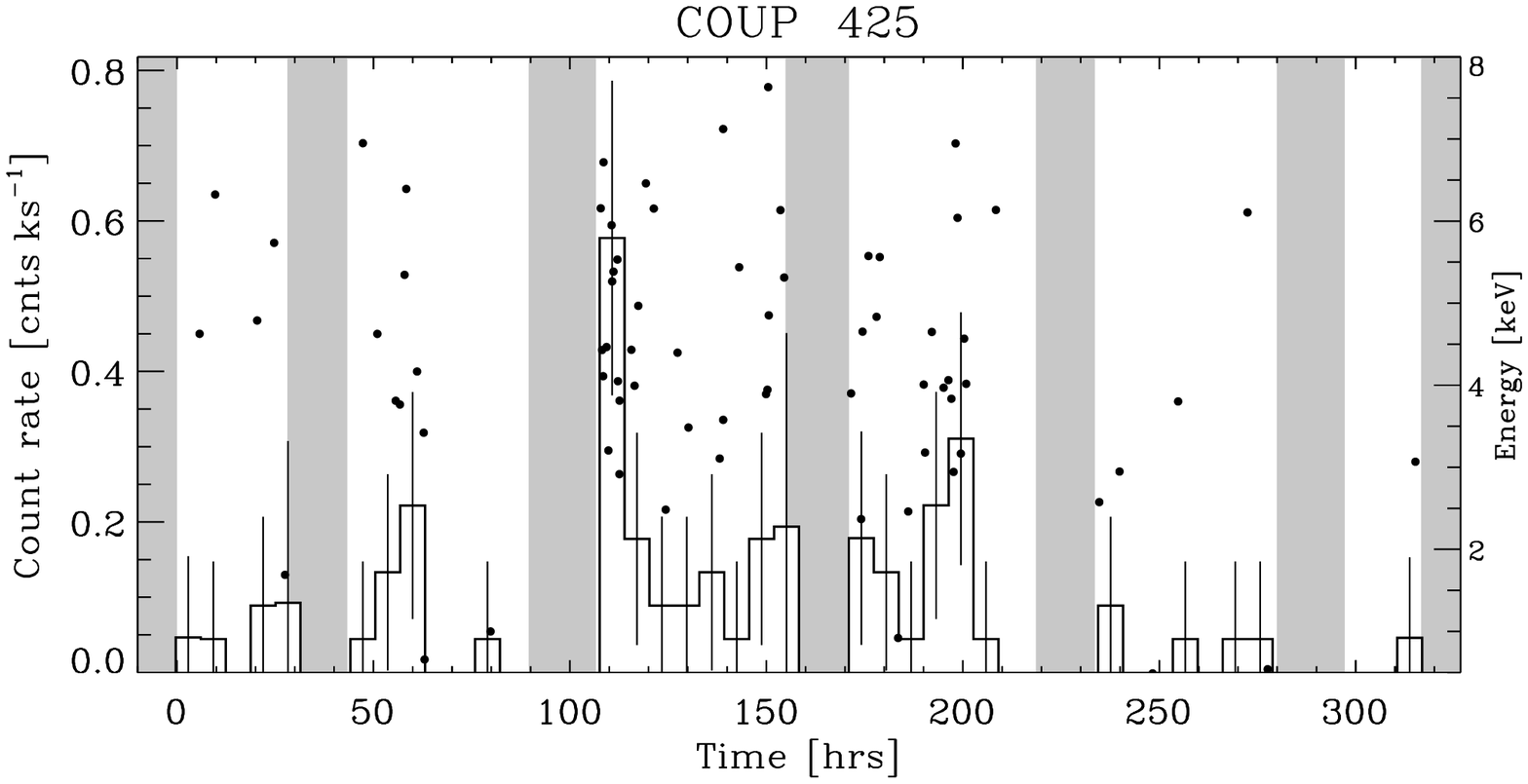} & \includegraphics[width=0.5\columnwidth]{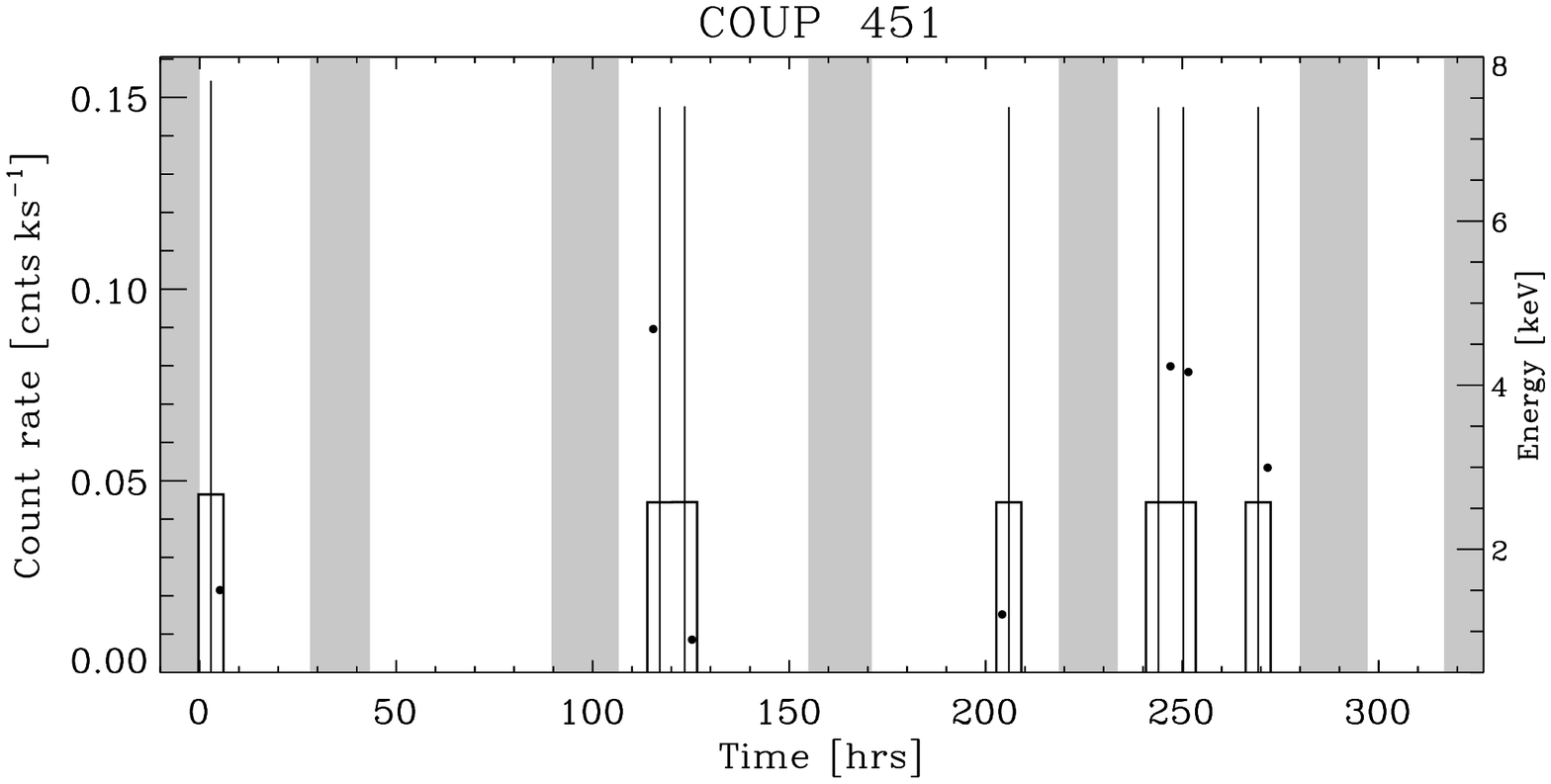}\\
\includegraphics[width=0.5\columnwidth]{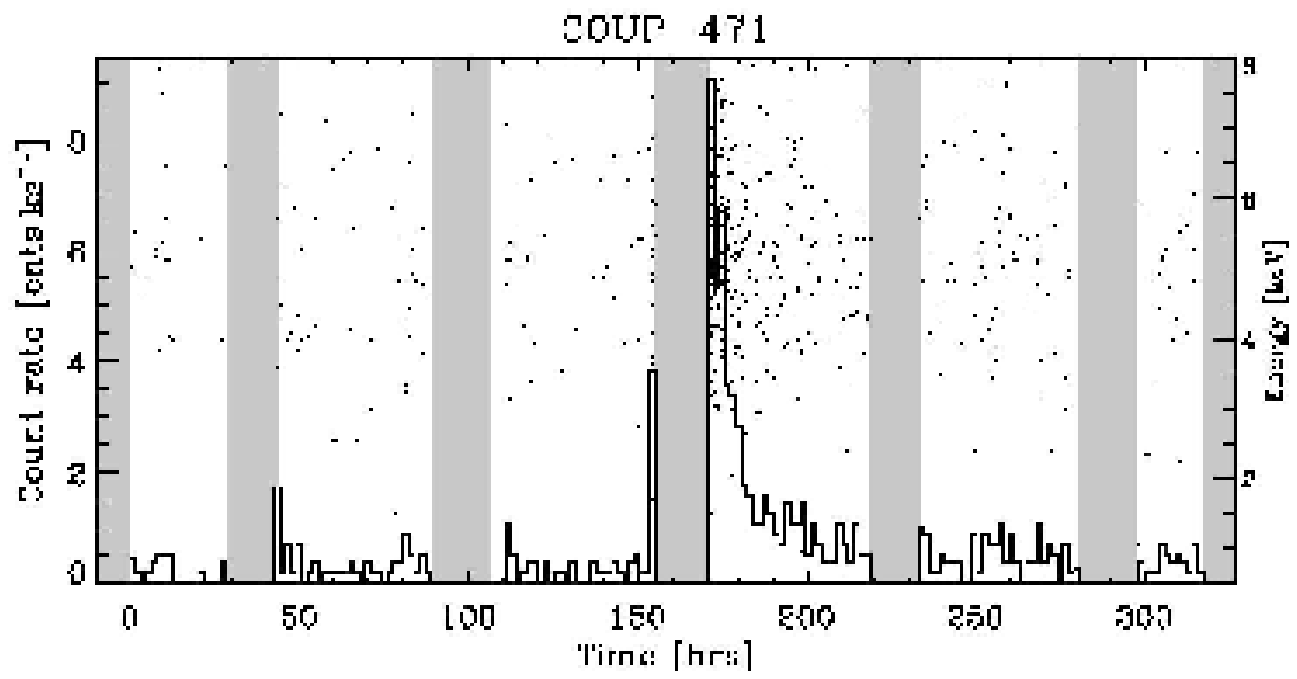} & \includegraphics[width=0.5\columnwidth]{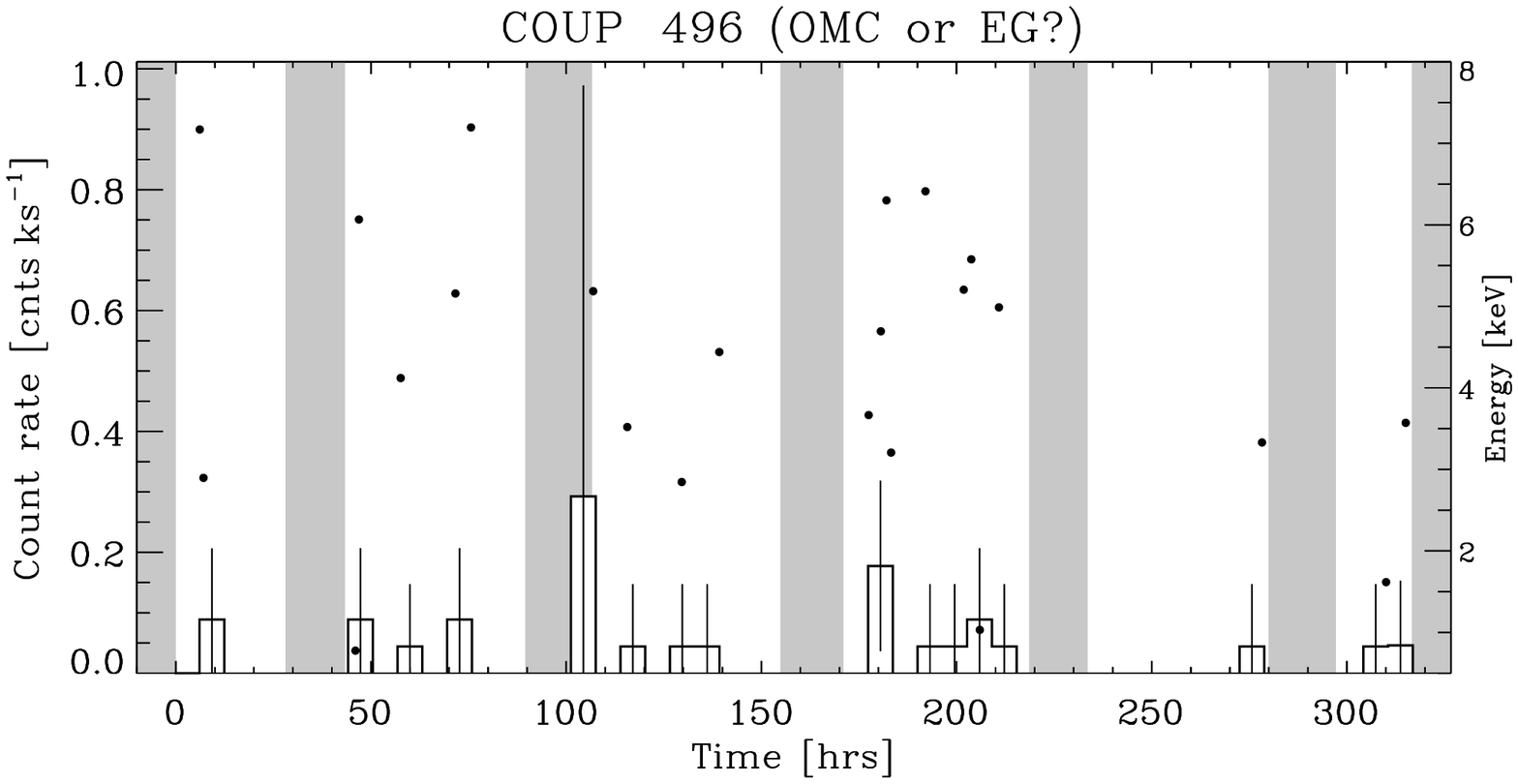}\\
\includegraphics[width=0.5\columnwidth]{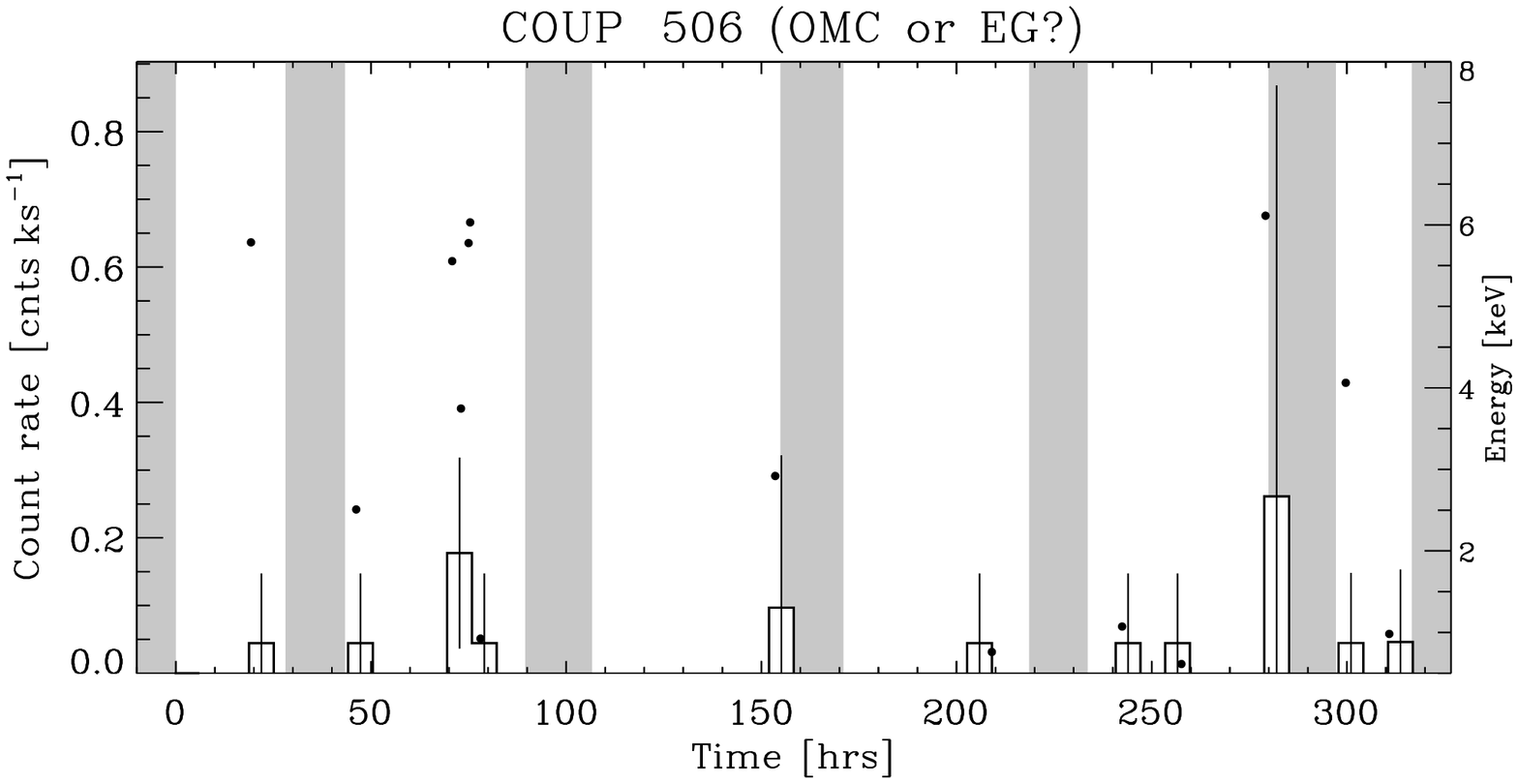} & \includegraphics[width=0.5\columnwidth]{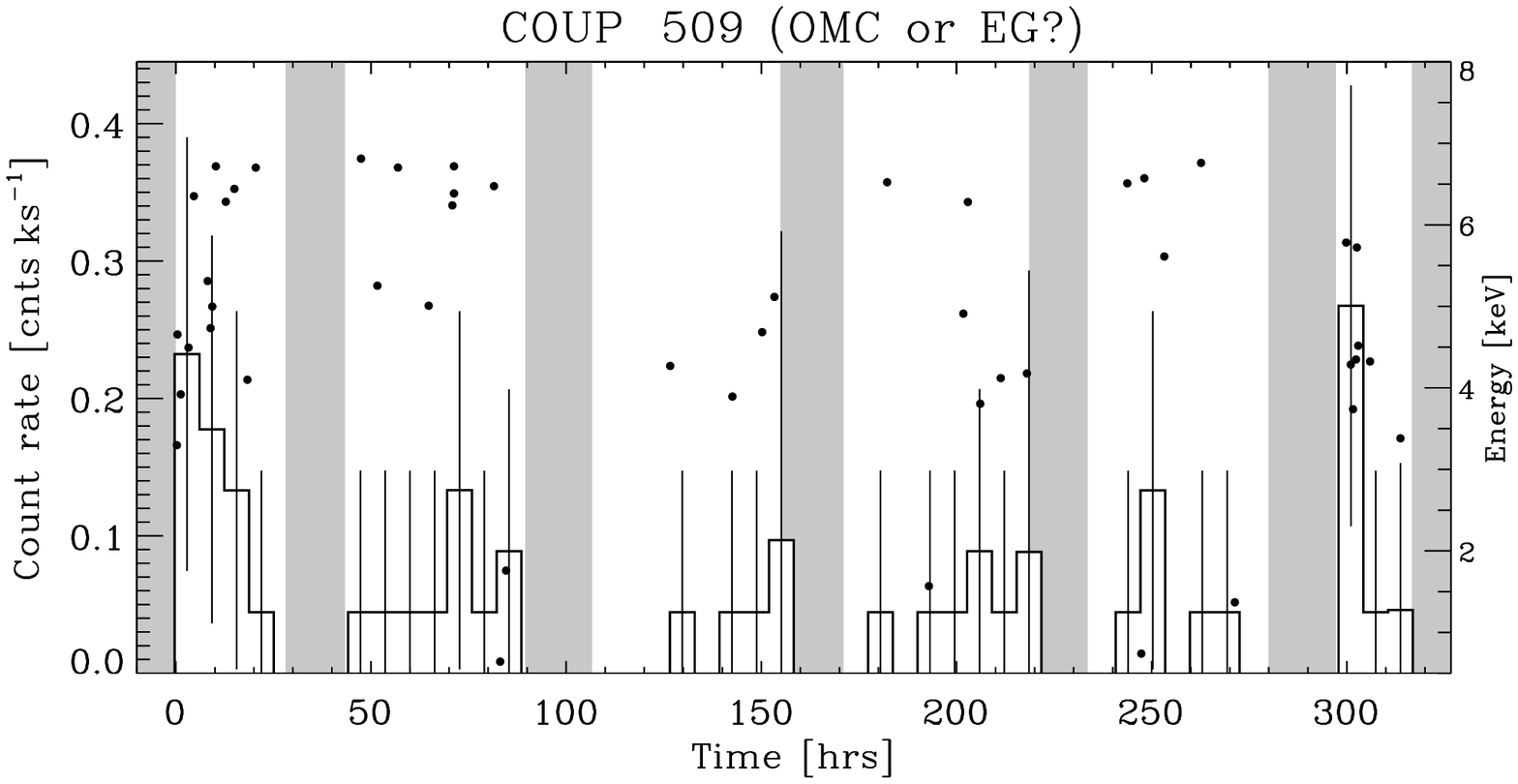}\\
\includegraphics[width=0.5\columnwidth]{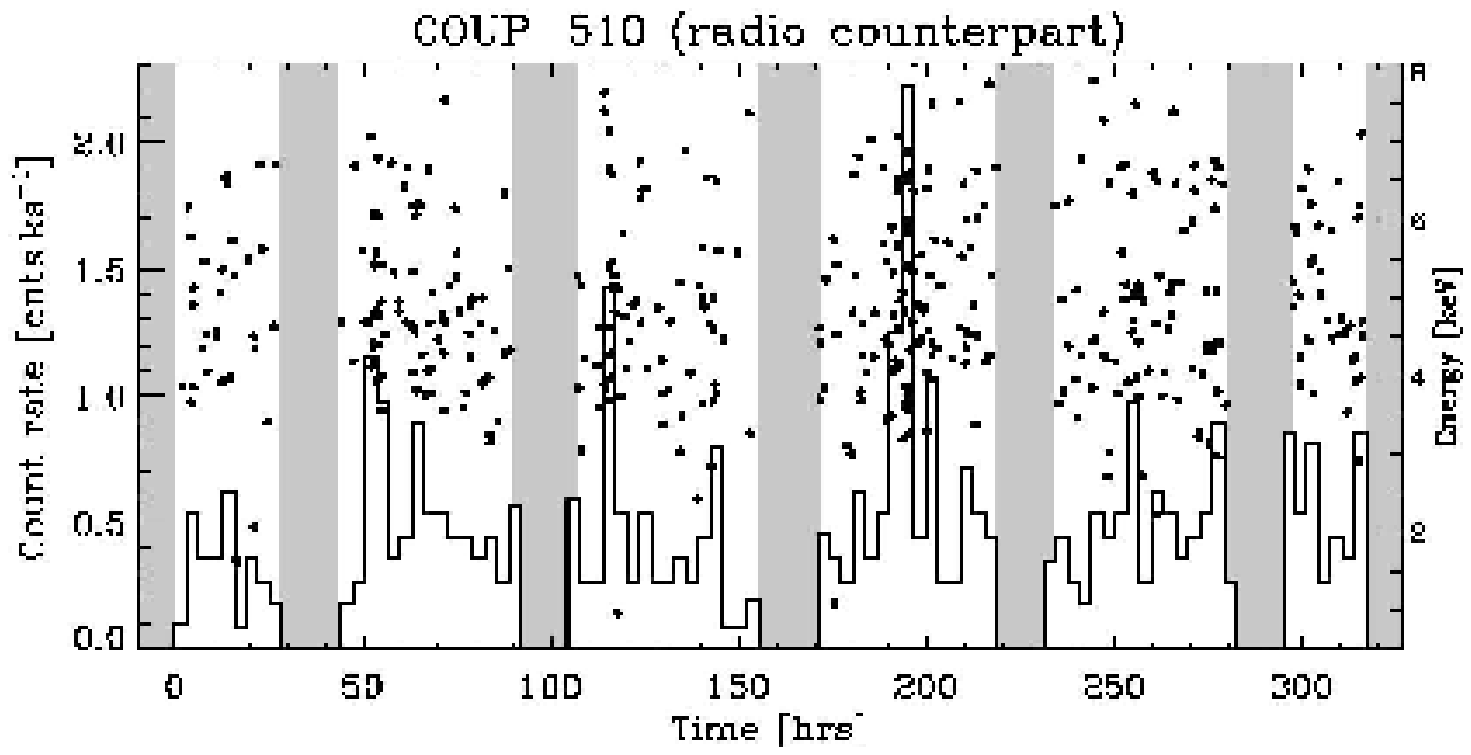} & \includegraphics[width=0.5\columnwidth]{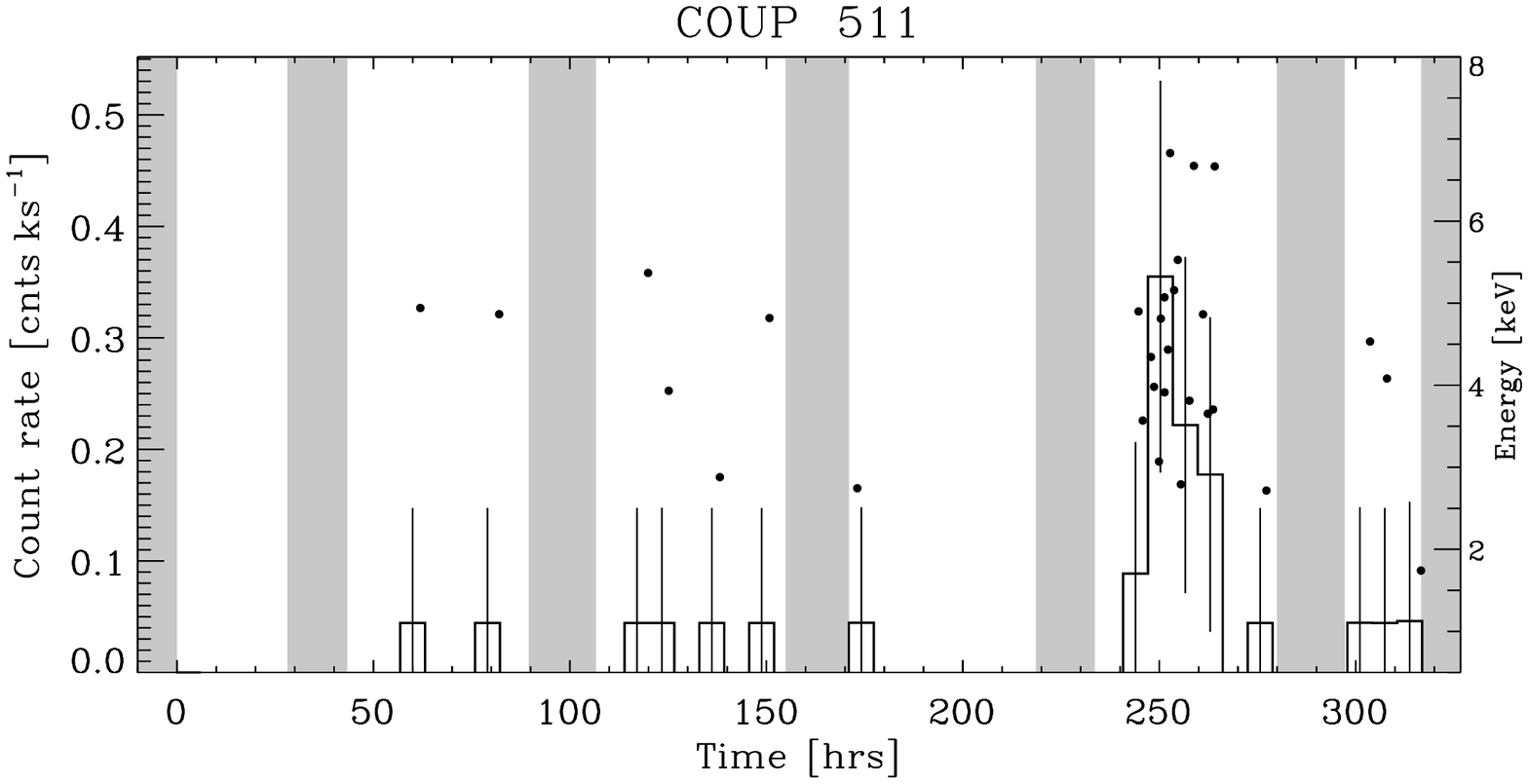}\\
\end{tabular}
 \caption{0.5--8.0\,keV light curves for COUP sources in OMC-1S without 
infrared counterparts. Vertical grey stripes indicate the five passages of \cxo{}
through the Van Allen belts where ACIS was taken out of the focal plane, 
and thus was not observing Orion. Dots mark the arrival times of 
individual X-ray photons with their corresponding energies given on the 
righthand axis. 
}
\end{figure*}

\begin{figure*}[!h]
\centering
\begin{tabular}{cc}
\includegraphics[width=0.5\columnwidth]{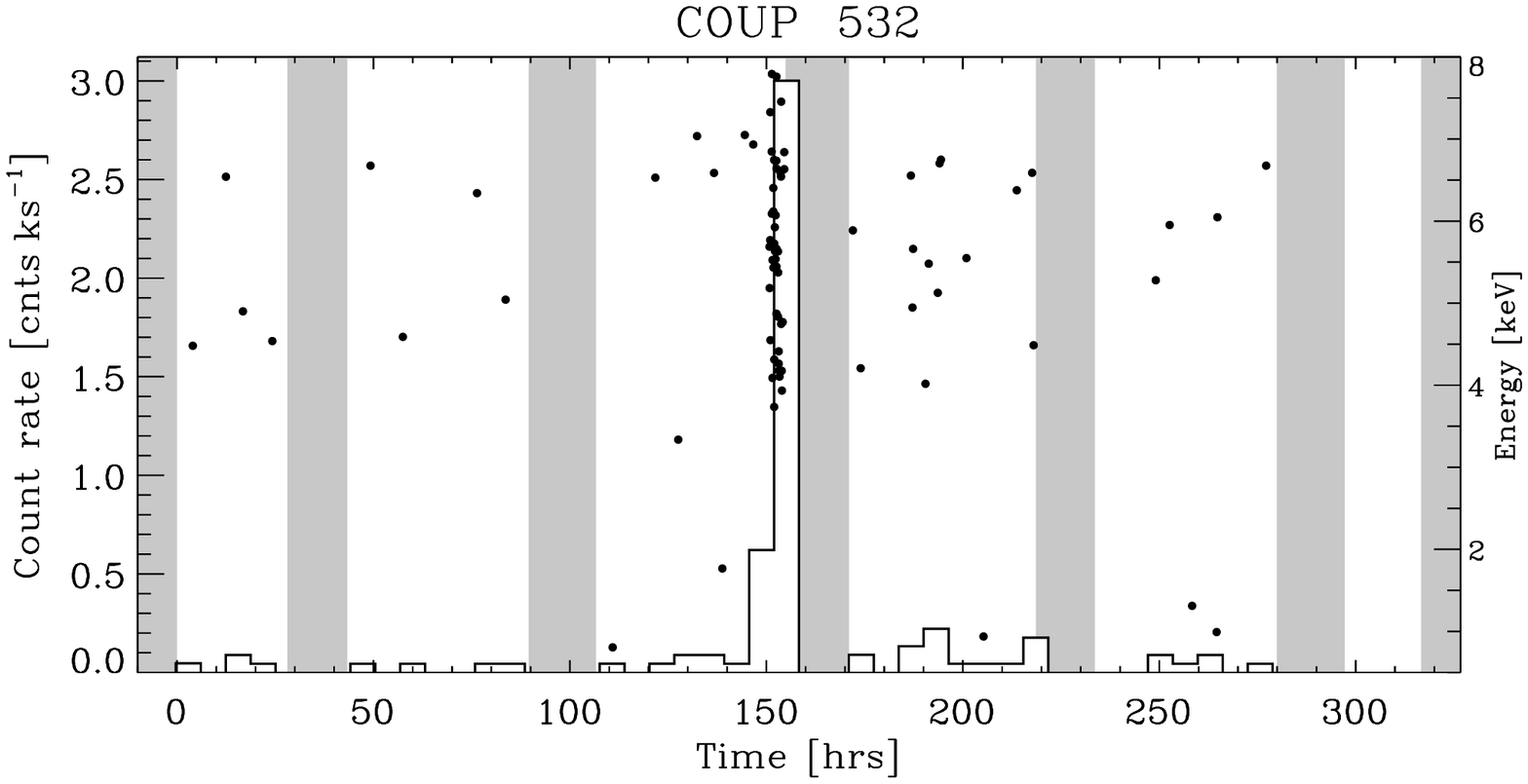} & \includegraphics[width=0.5\columnwidth]{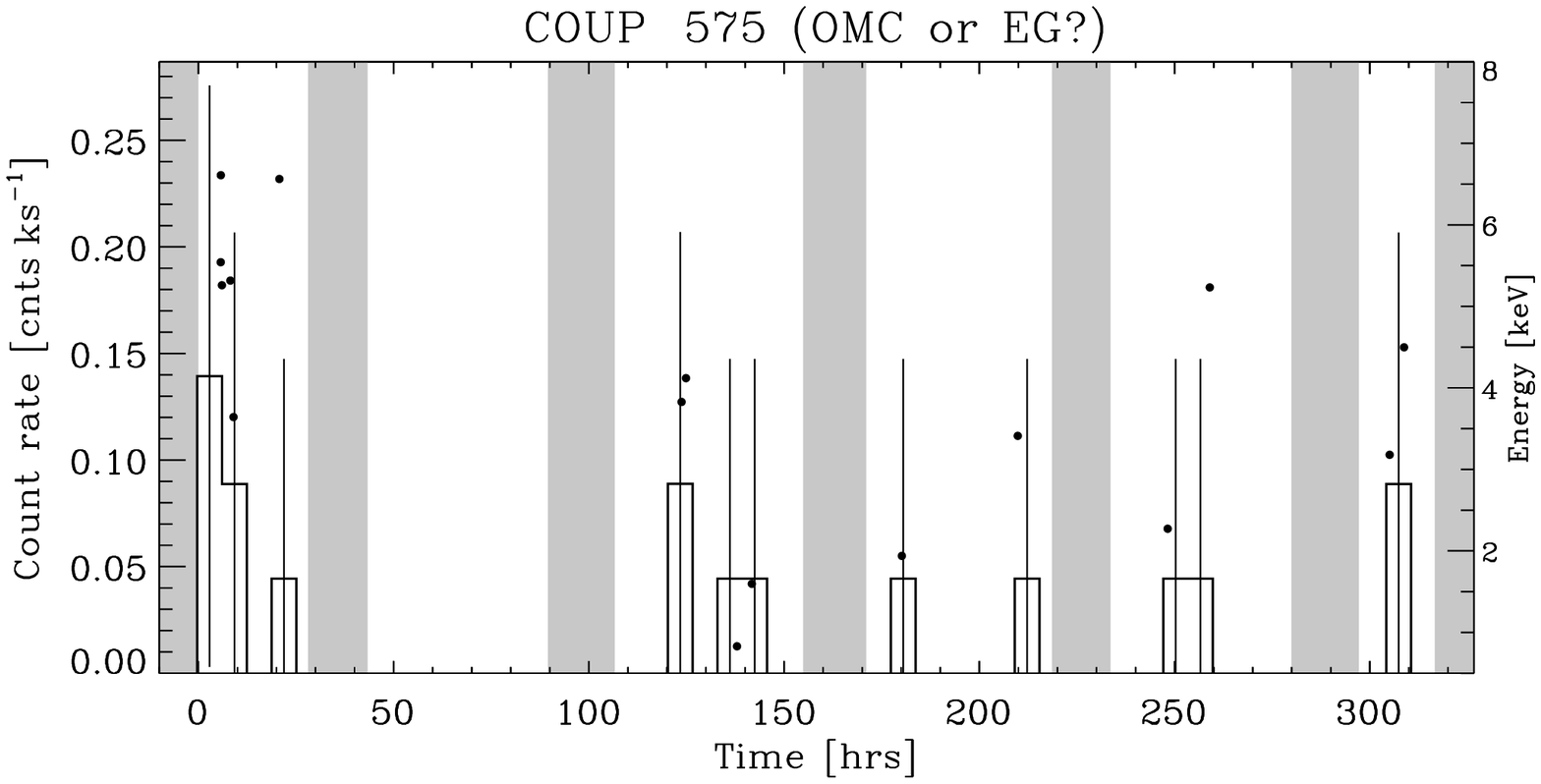}\\
\includegraphics[width=0.5\columnwidth]{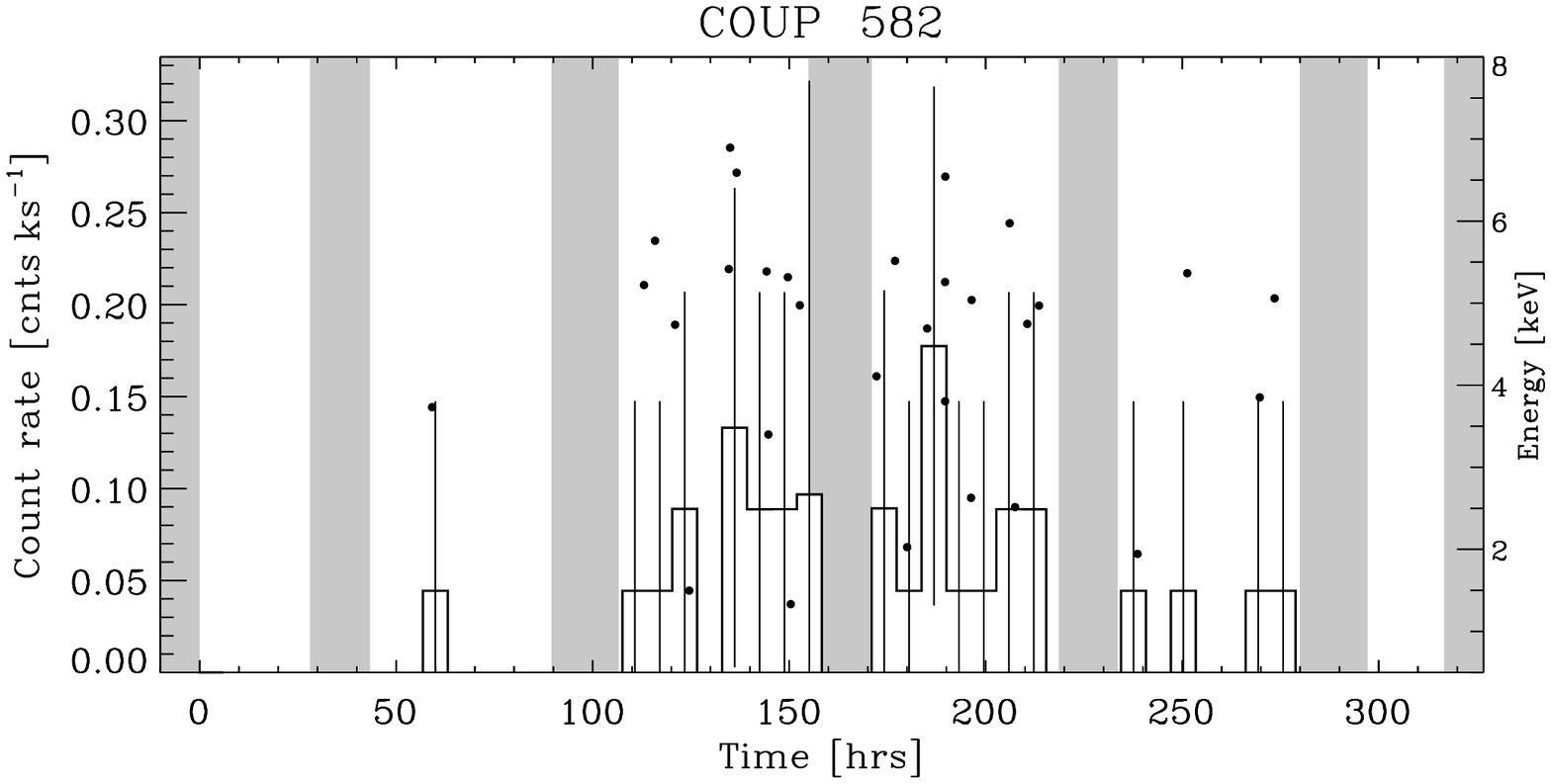} & \includegraphics[width=0.5\columnwidth]{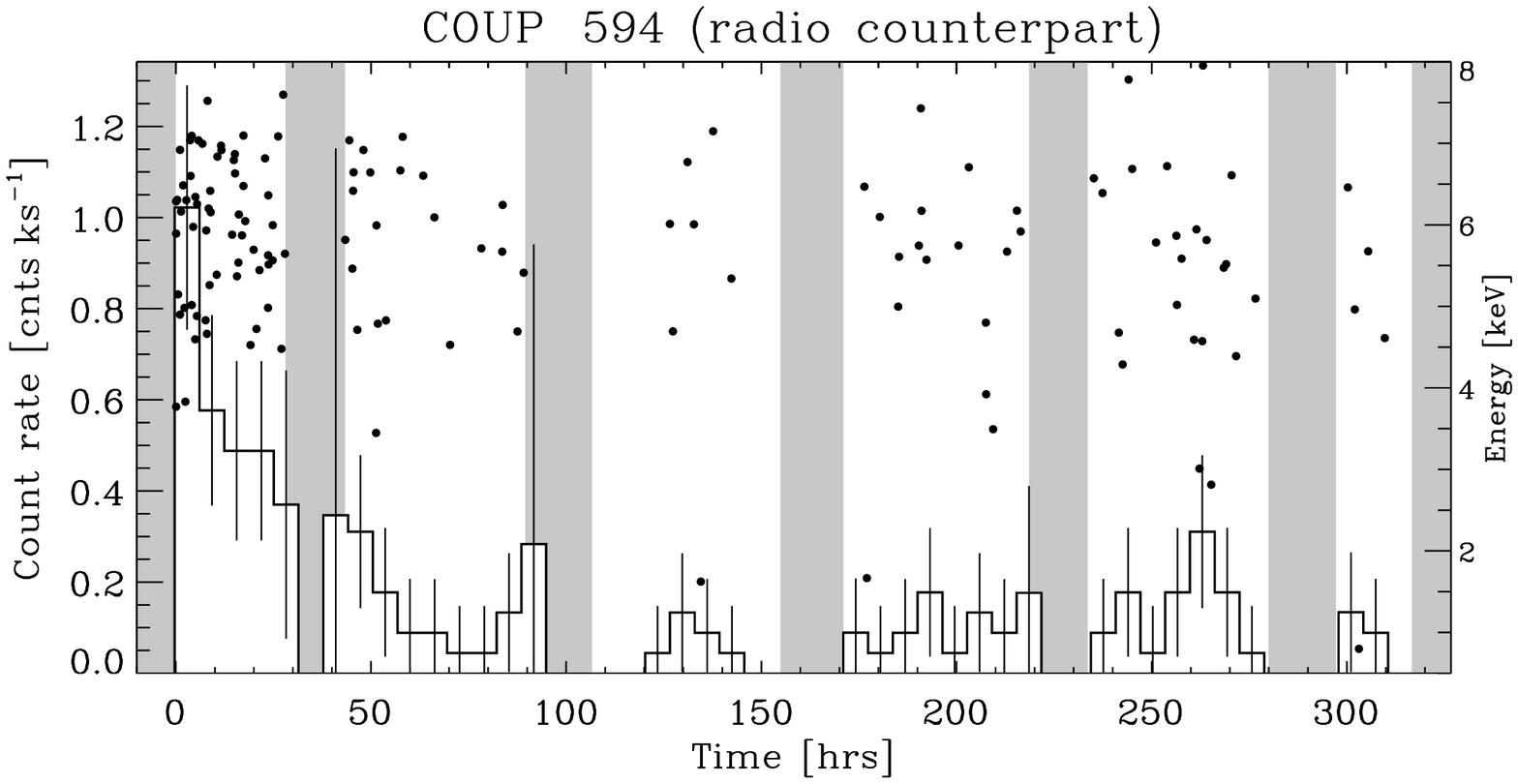}\\
\includegraphics[width=0.5\columnwidth]{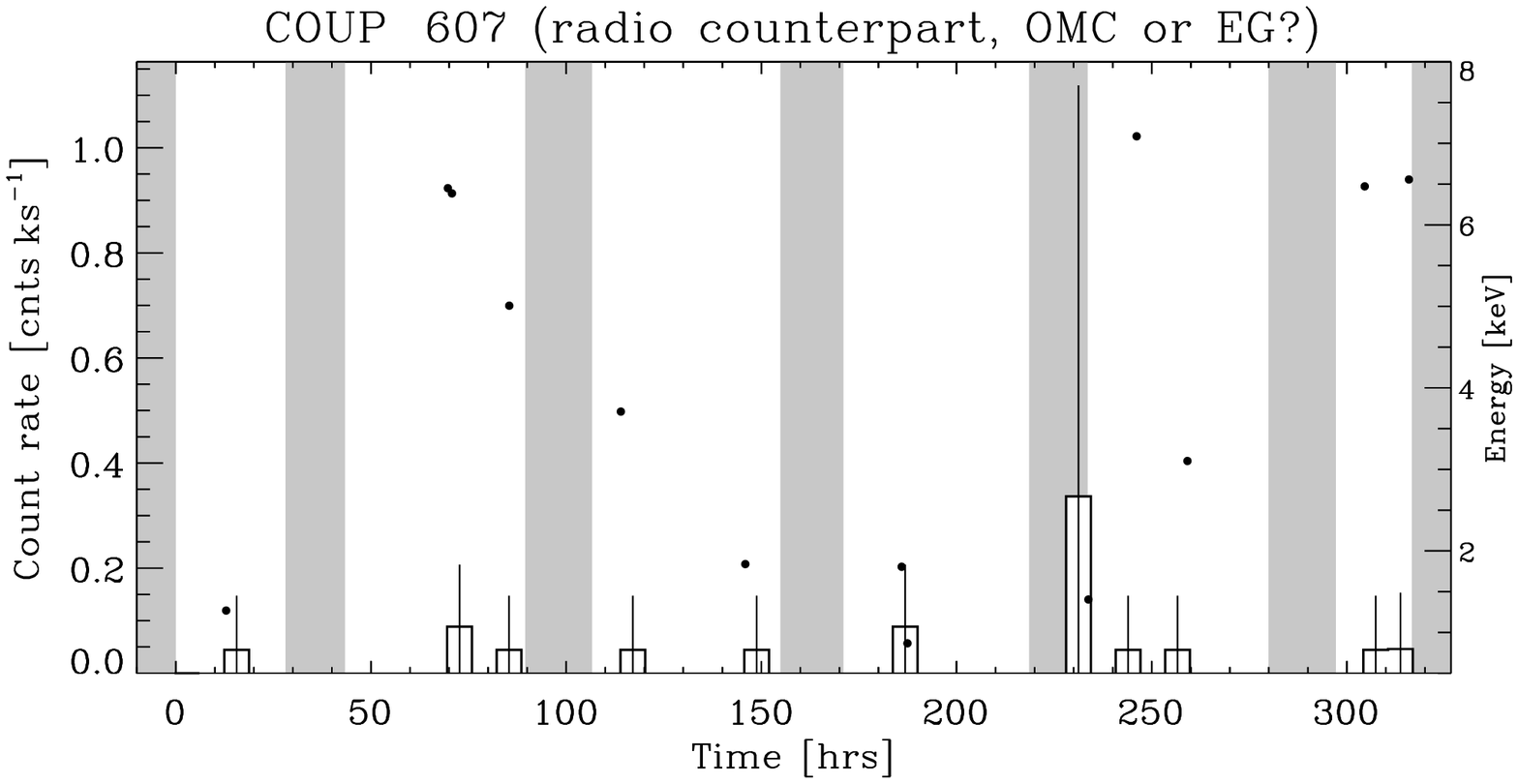} & \includegraphics[width=0.5\columnwidth]{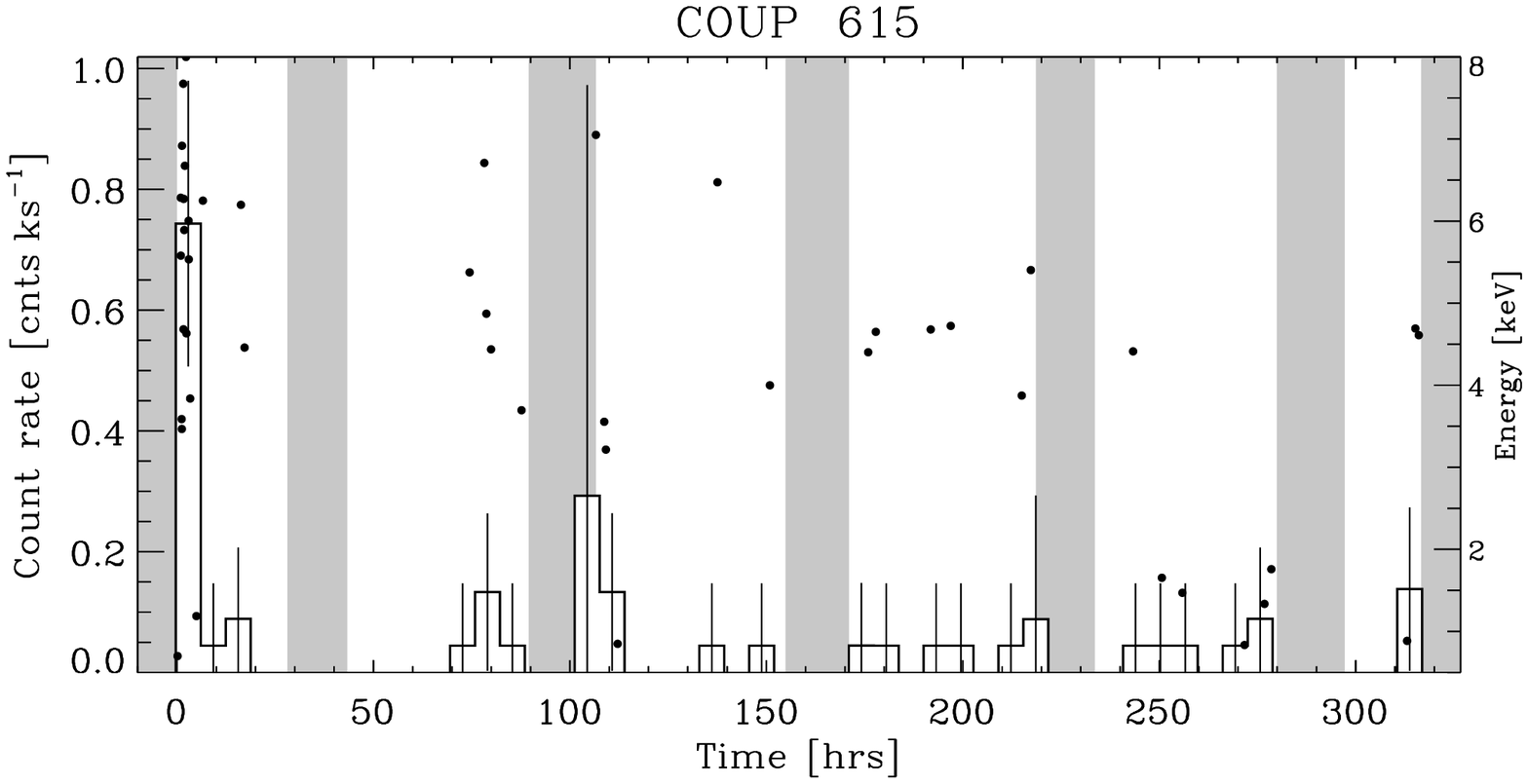}\\
\includegraphics[width=0.5\columnwidth]{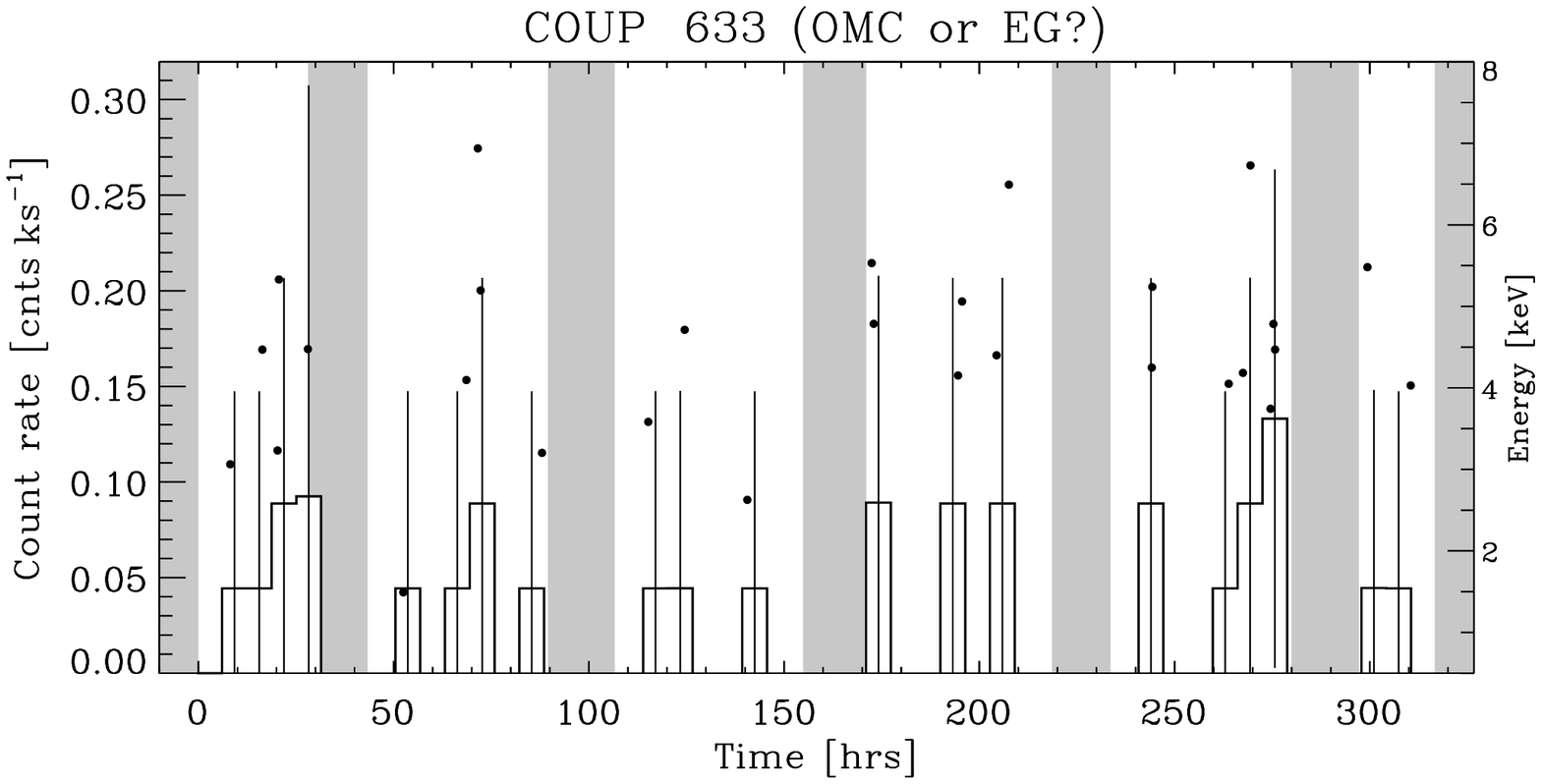} & \includegraphics[width=0.5\columnwidth]{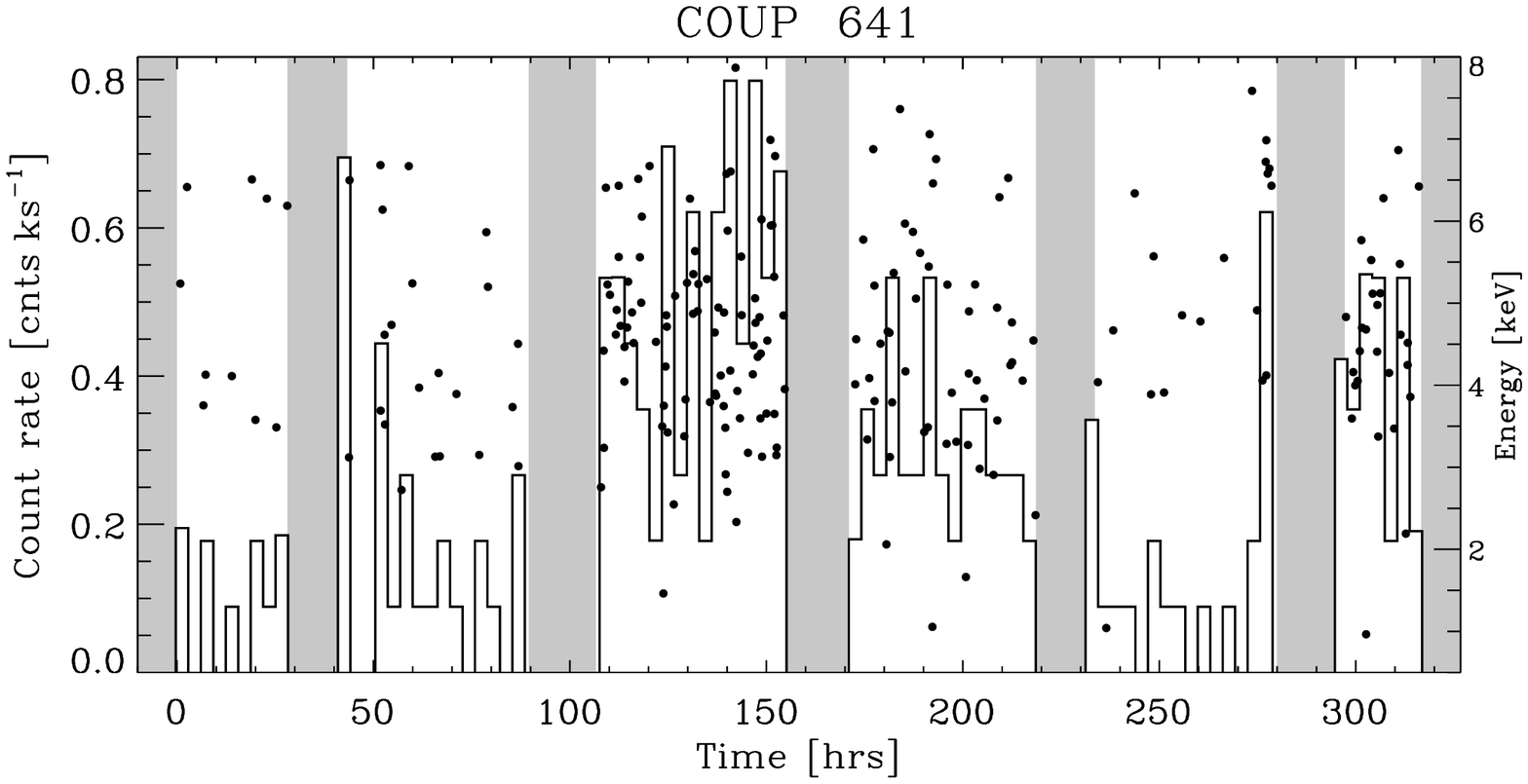}\\
\includegraphics[width=0.5\columnwidth]{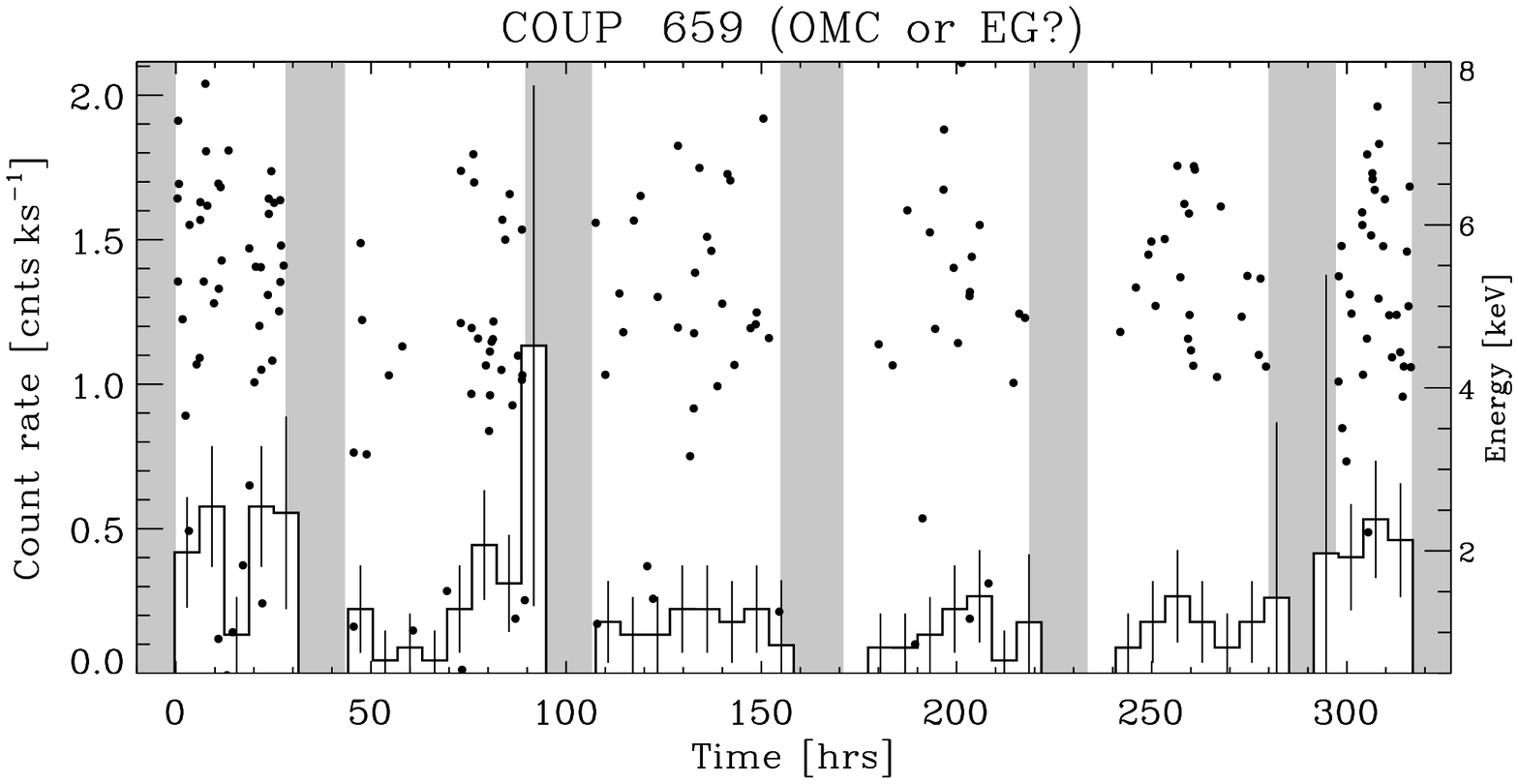} & \includegraphics[width=0.5\columnwidth]{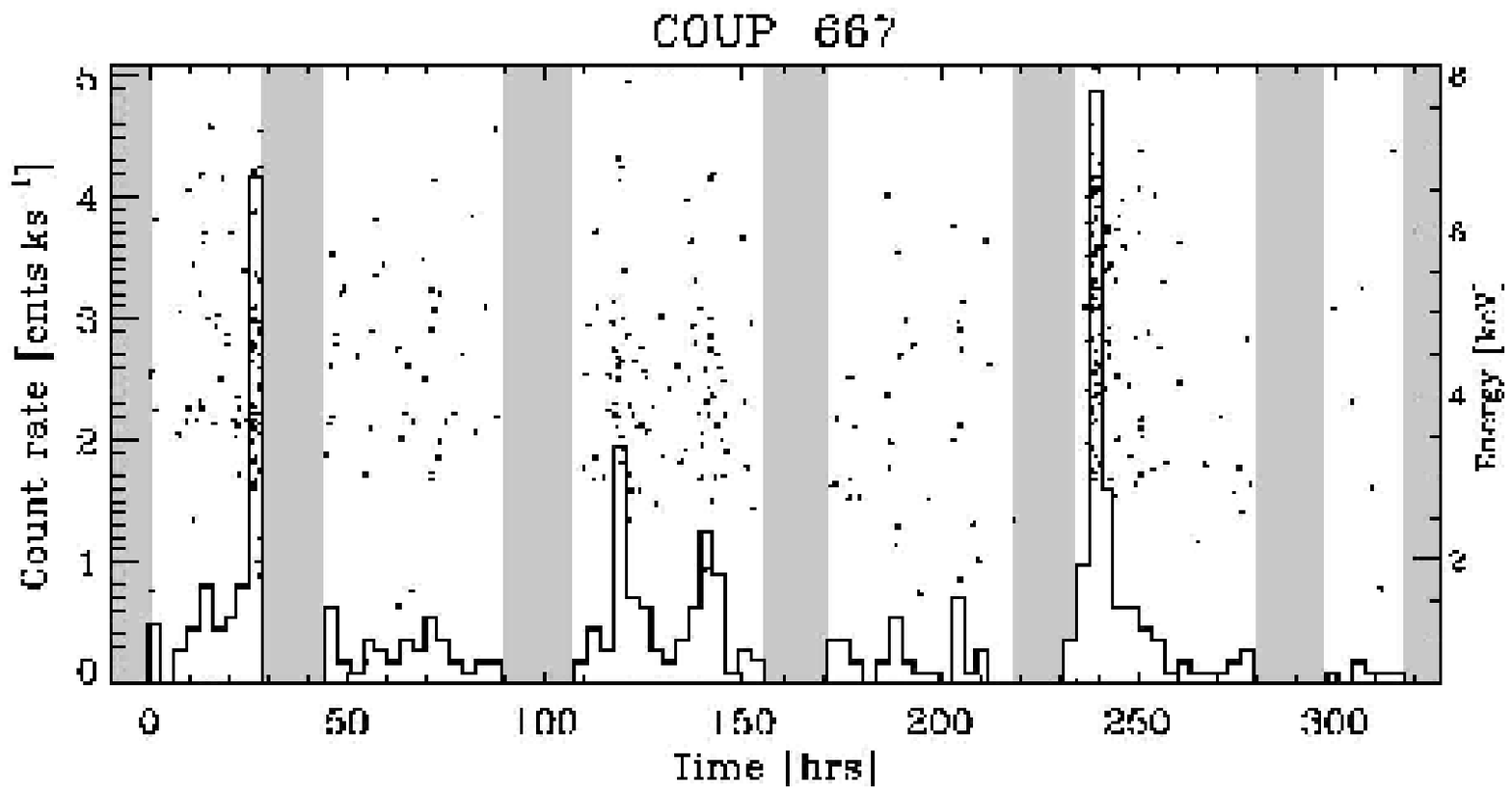}\\
\end{tabular}
\addtocounter{figure}{-1}
\caption{{\it Continued.}}
\label{lc_source_x_1s}
\end{figure*}

\begin{figure*}[h]
\begin{minipage}{0.5\columnwidth}
\centering
 \includegraphics[width=\columnwidth]{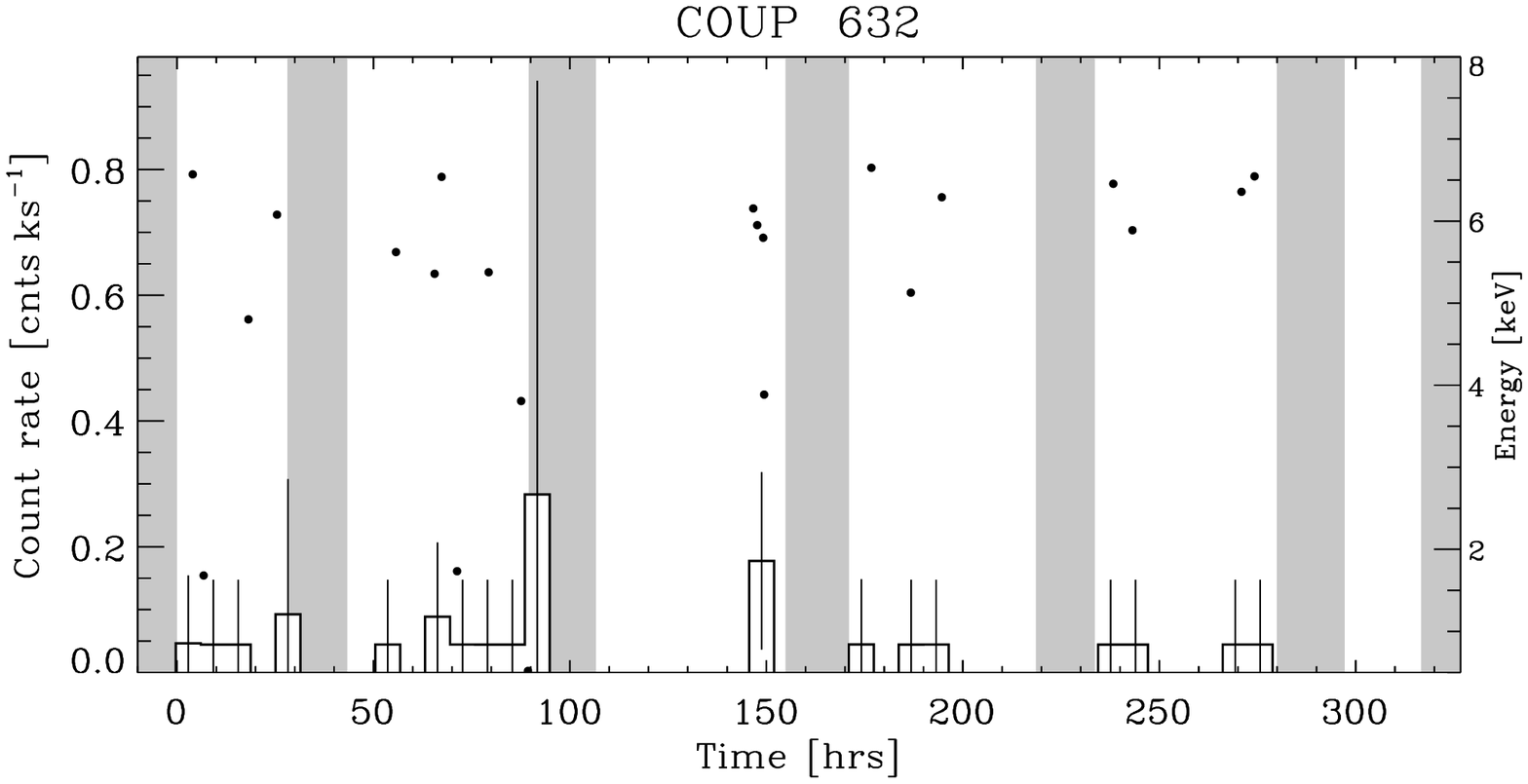}
\end{minipage}
\ \
\begin{minipage}{0.5\columnwidth}
\centering
\includegraphics[angle=-90,width=0.65\columnwidth]{f13b.ps}
\end{minipage}
\caption{0.5--8.0\,keV band X-ray light curve and X-ray spectrum for
COUP\,632 in OMC-1S\@. This X-ray source is the most embedded COUP source 
with $\log N_{\rm H}$=23.94 ($A_{\rm V}\sim500$\,mag) and has a counterpart only 
detected in mid-infrared \citep{smith04} and at 1.3\,cm \citep{zapata04b}.}
\label{most_embedded}
\end{figure*}

\begin{figure*}[!h]
\centering
\begin{tabular}{cc}
\includegraphics[width=0.43\columnwidth]{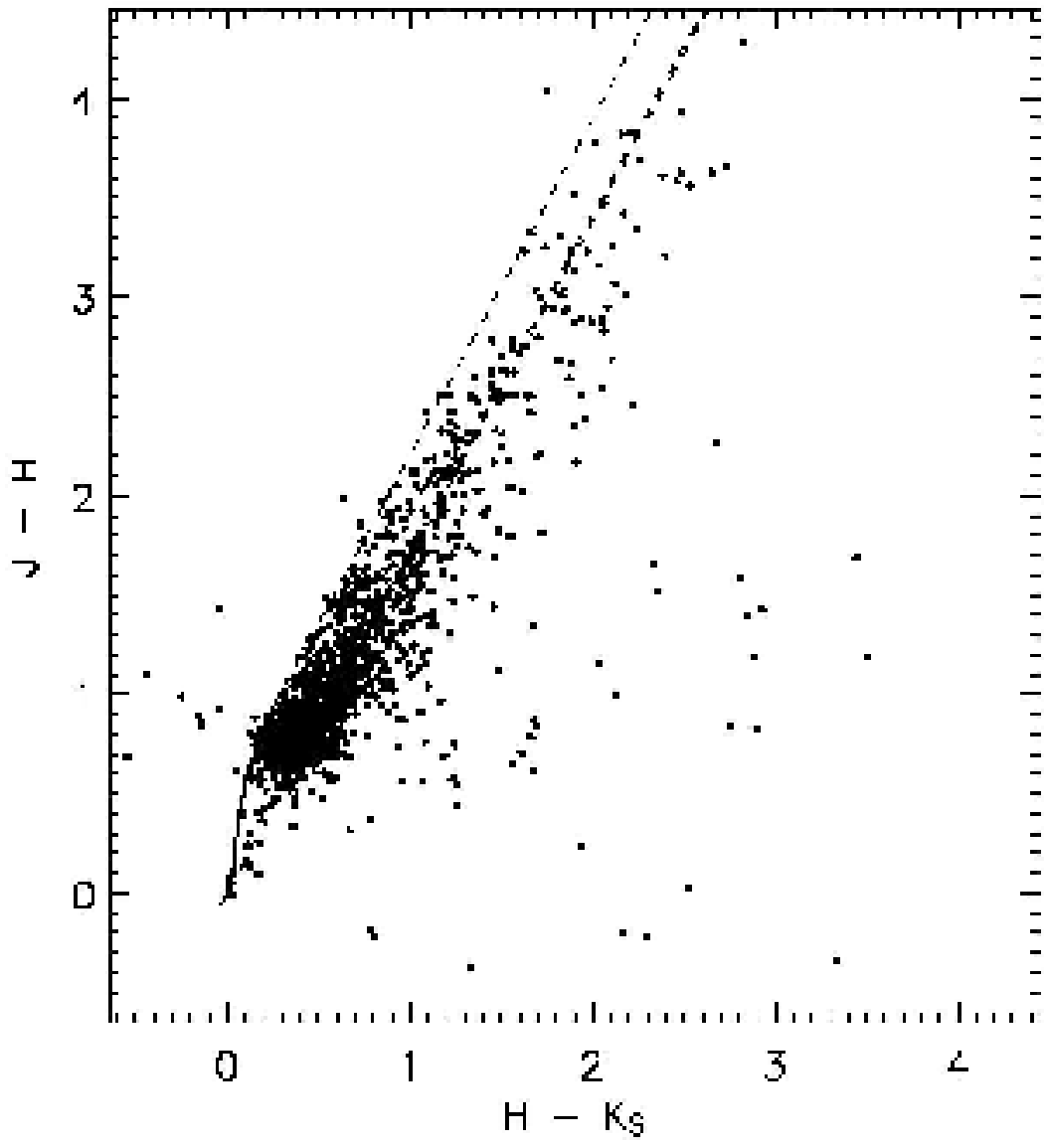} & \includegraphics[width=0.43\columnwidth]{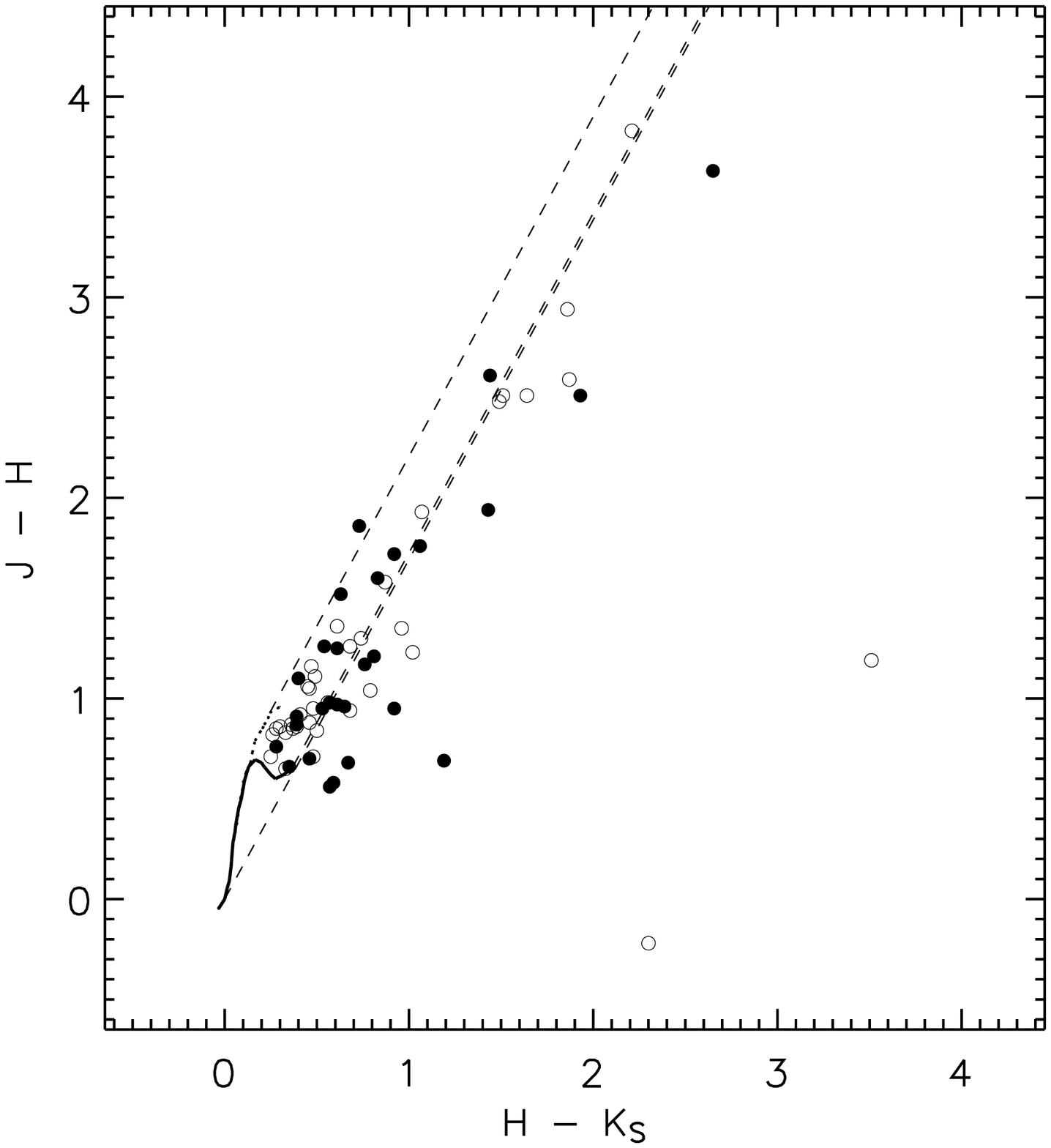}\\
\includegraphics[width=0.43\columnwidth]{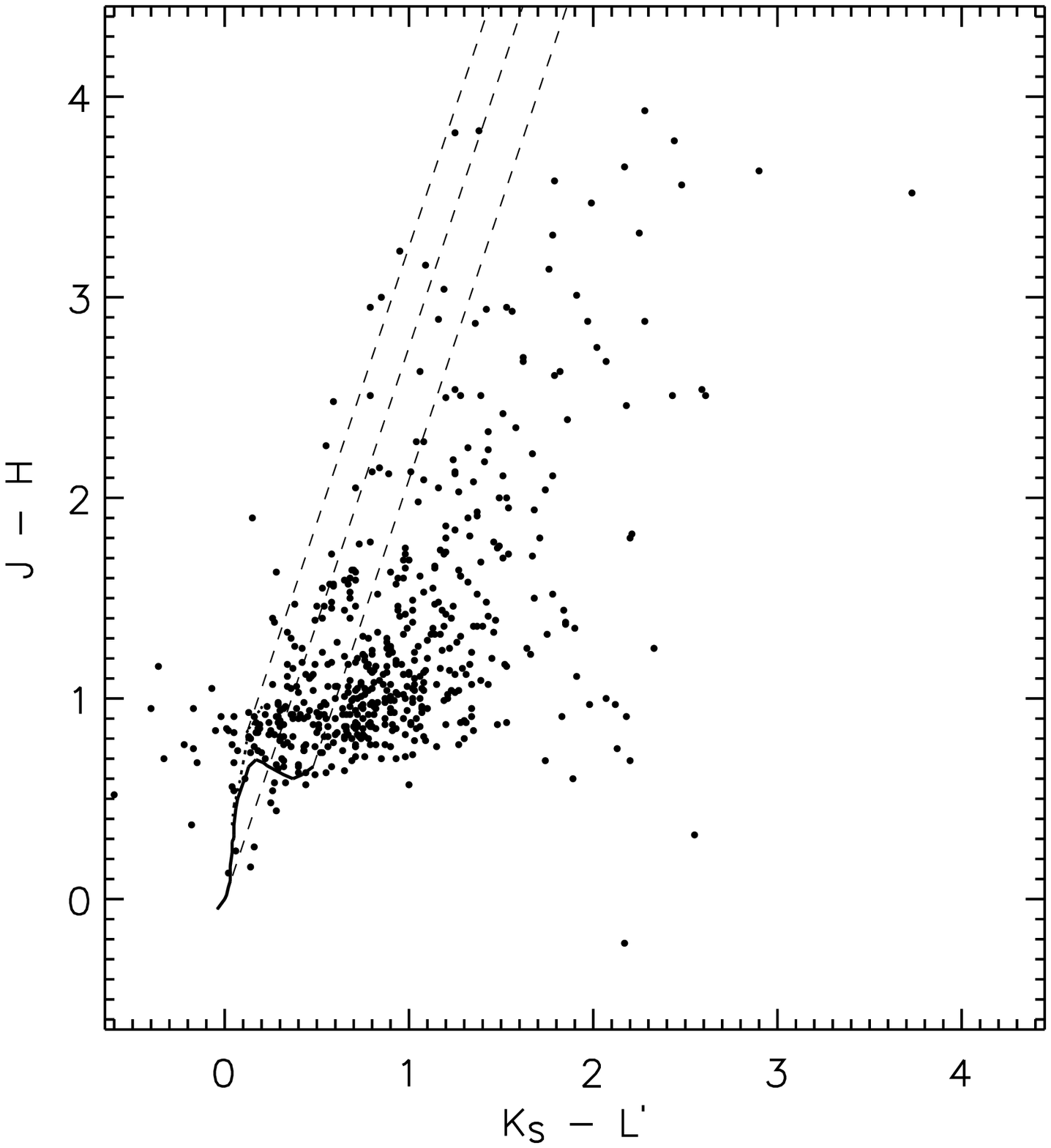} & \includegraphics[width=0.43\columnwidth]{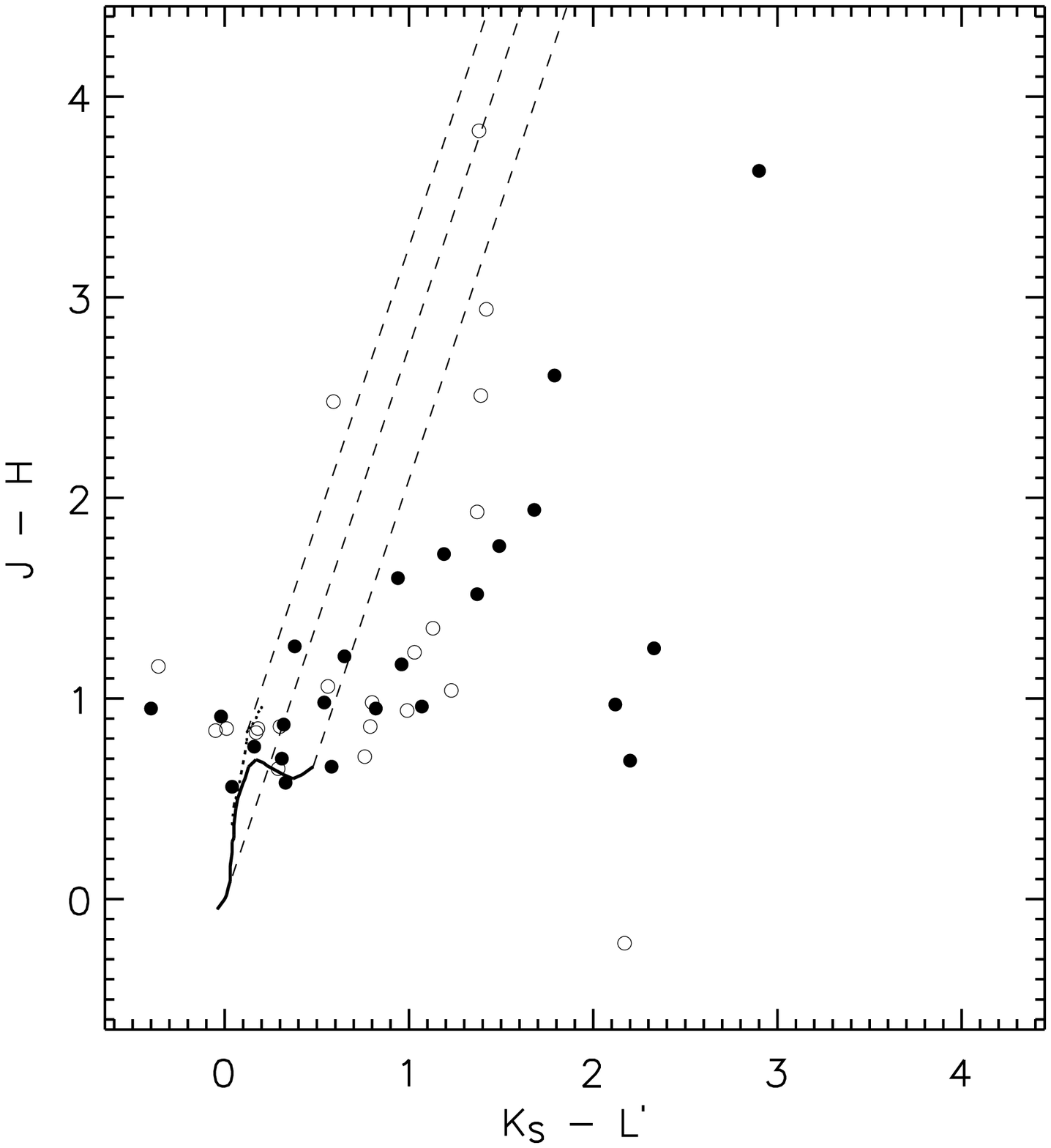}\\
\includegraphics[width=0.43\columnwidth]{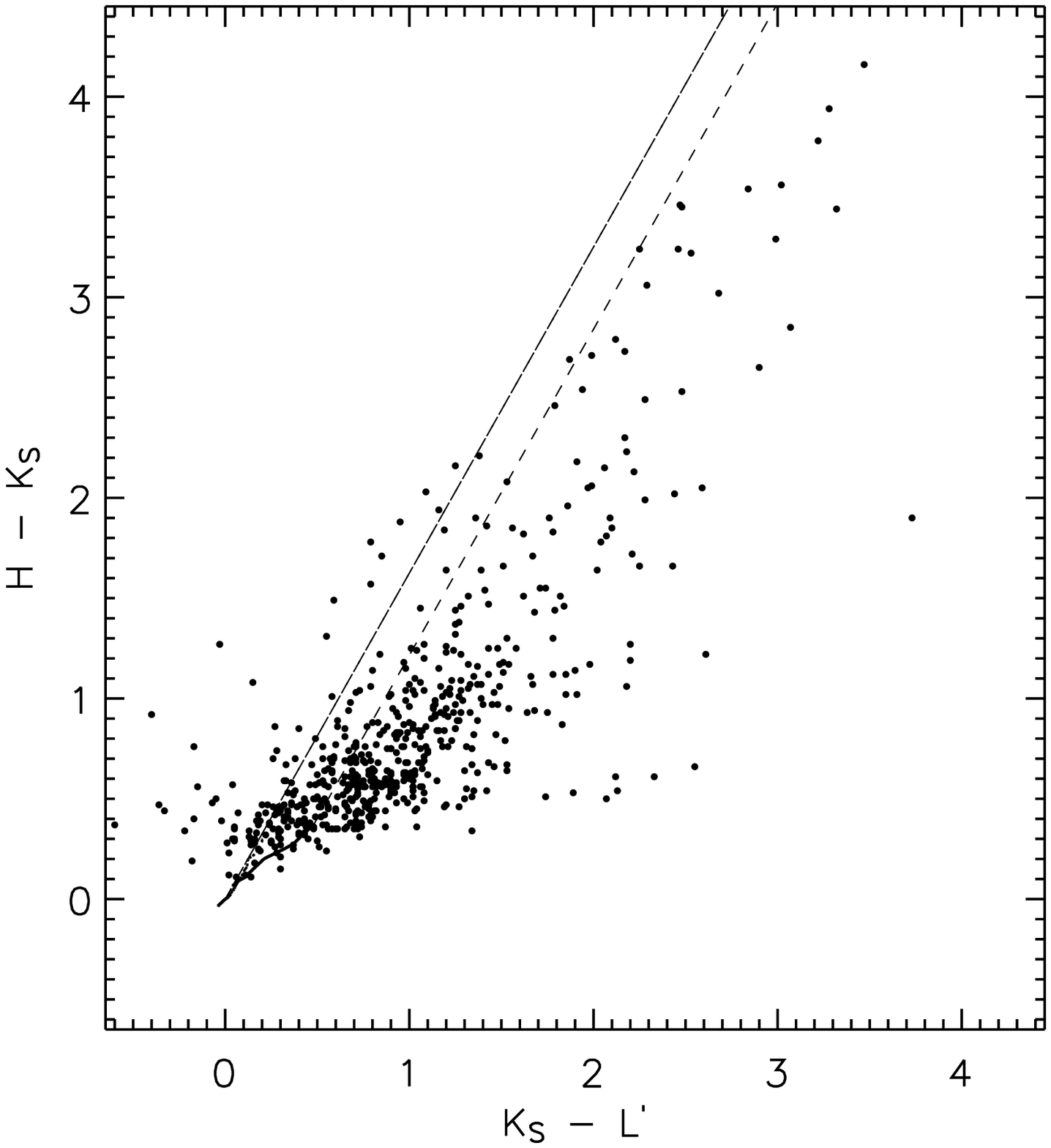} & \includegraphics[width=0.43\columnwidth]{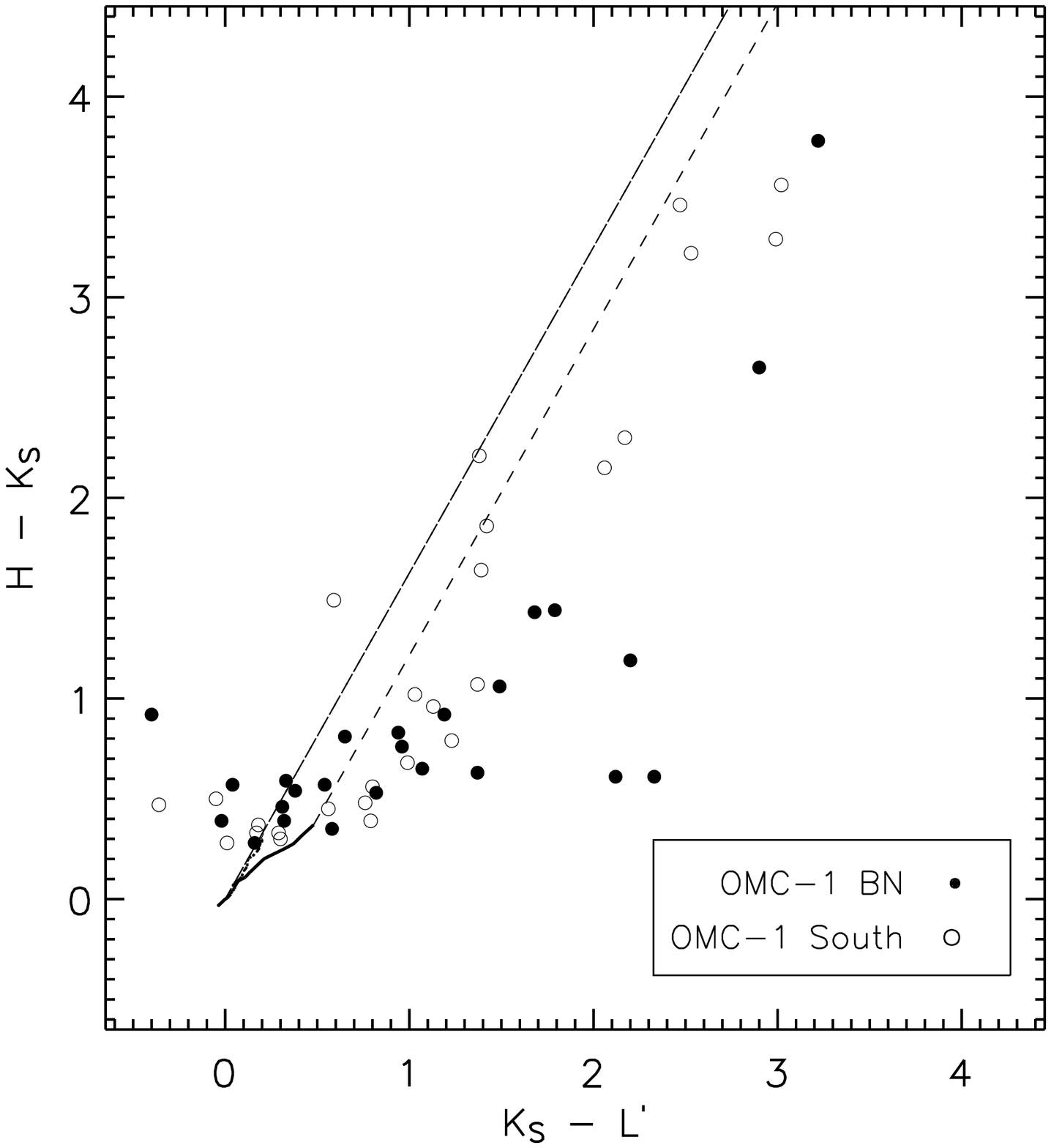}\\
\end{tabular}
\caption{Color-color diagrams for all COUP sources (left) and the embedded
OMC-1 (BN-KL and OMC-1S) sources (right).}
\label{color_color}
\end{figure*}

\begin{figure*}[!h]
\centering
\begin{tabular}{@{}ccc@{}}
\includegraphics[angle=0,width=0.33\columnwidth]{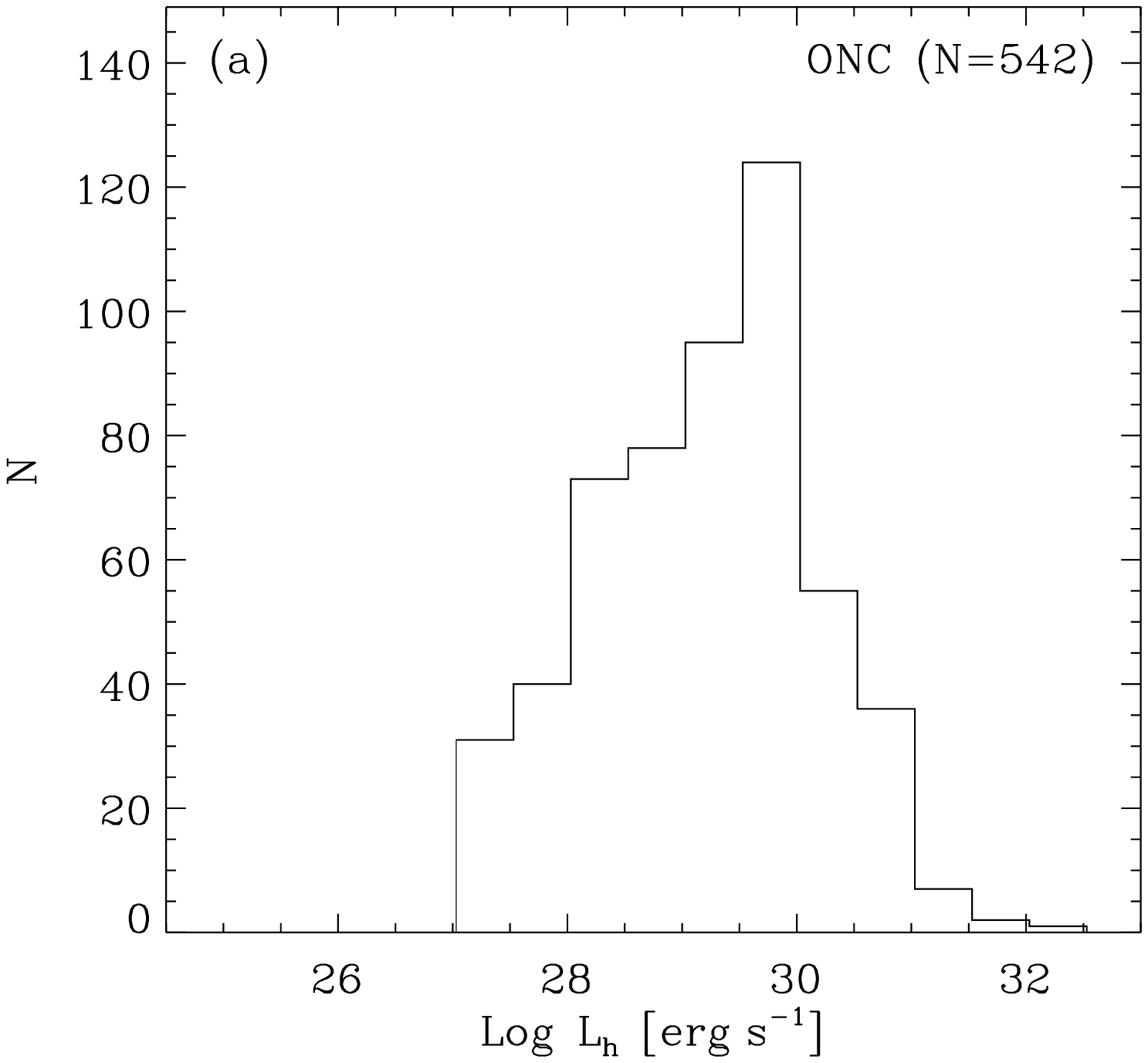}  & \includegraphics[angle=0,width=0.33\columnwidth]{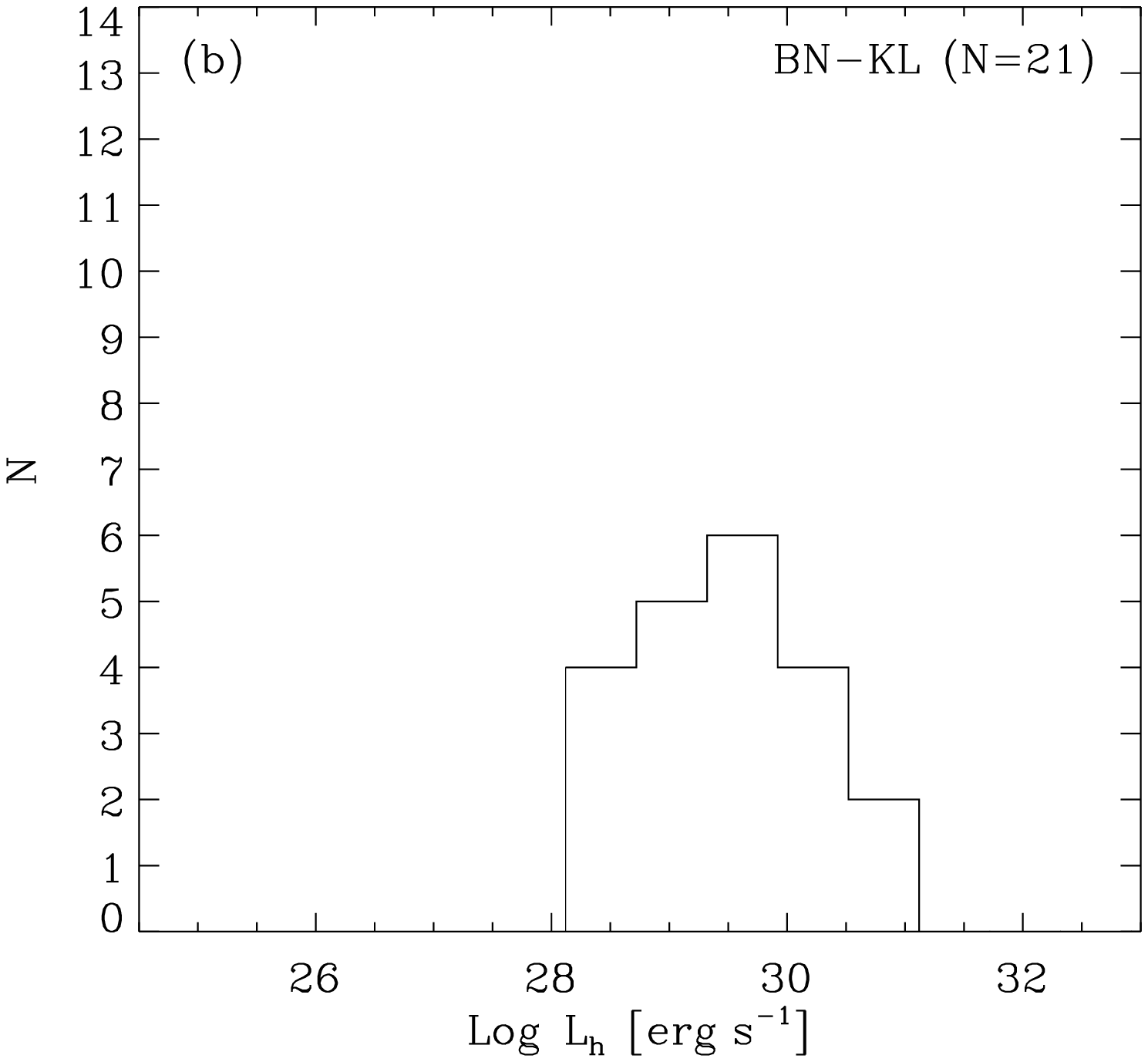}  & \includegraphics[angle=0,width=0.33\columnwidth]{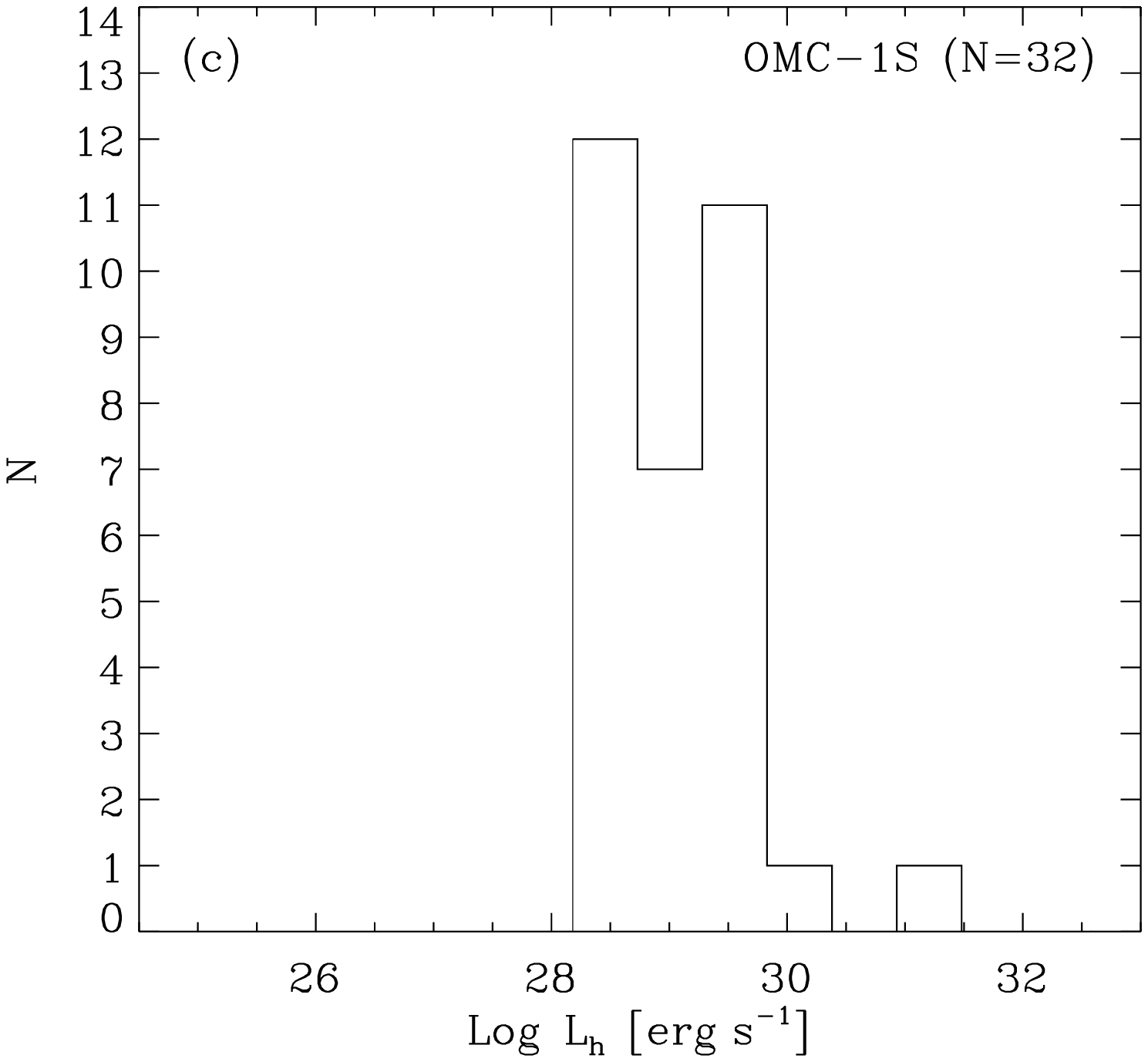} \\
\includegraphics[angle=0,width=0.33\columnwidth]{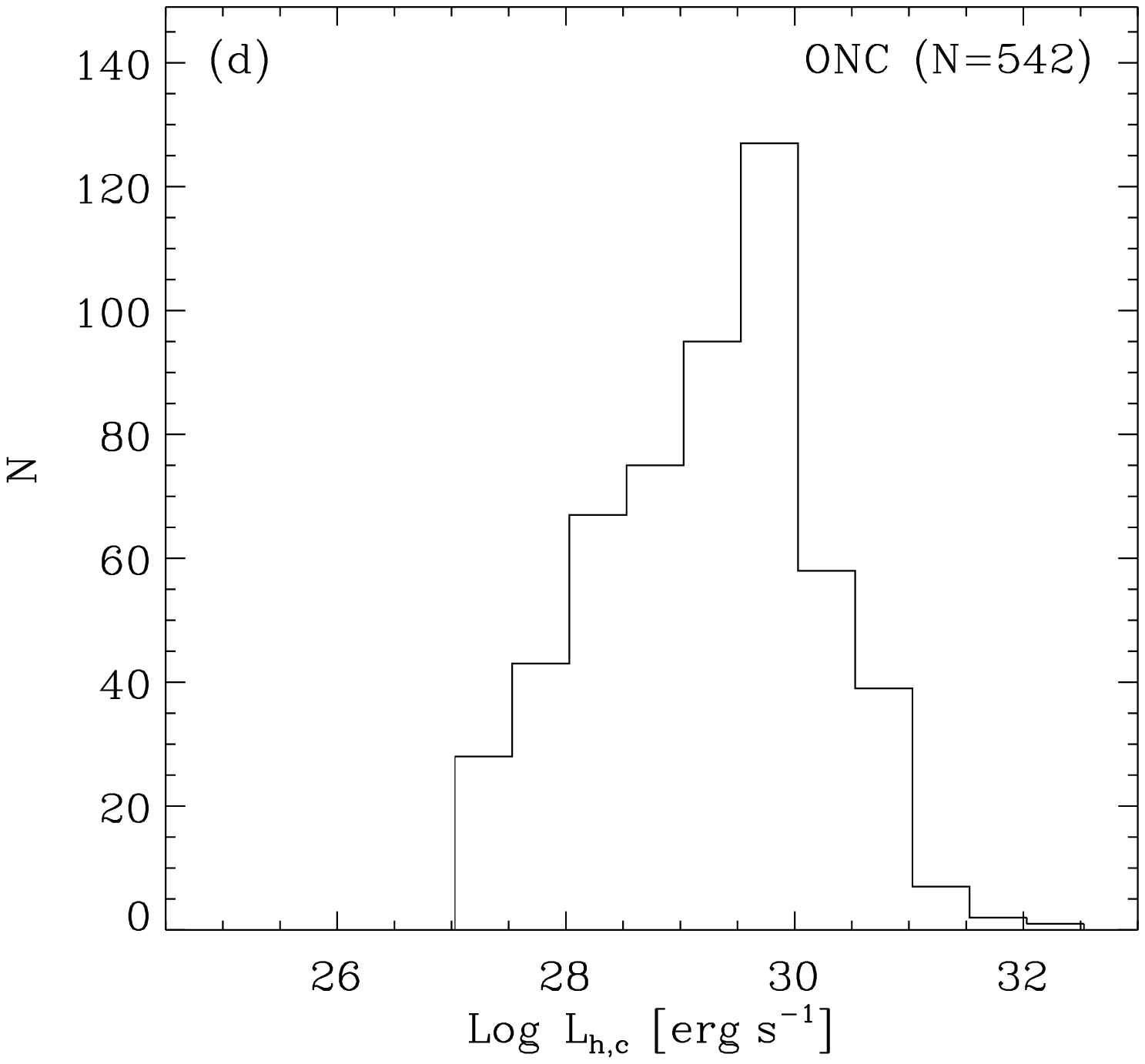} & \includegraphics[angle=0,width=0.33\columnwidth]{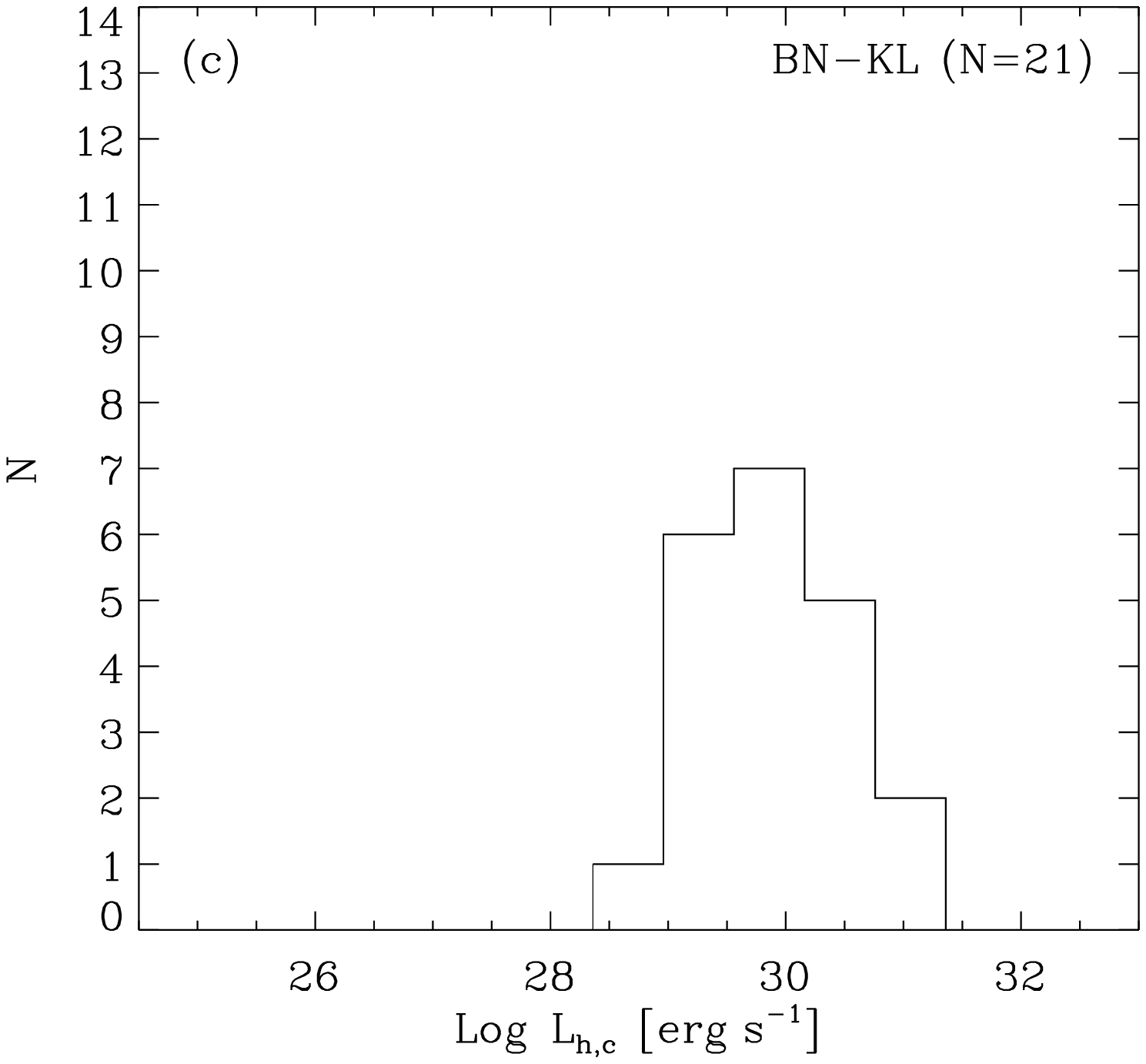} & \includegraphics[angle=0,width=0.33\columnwidth]{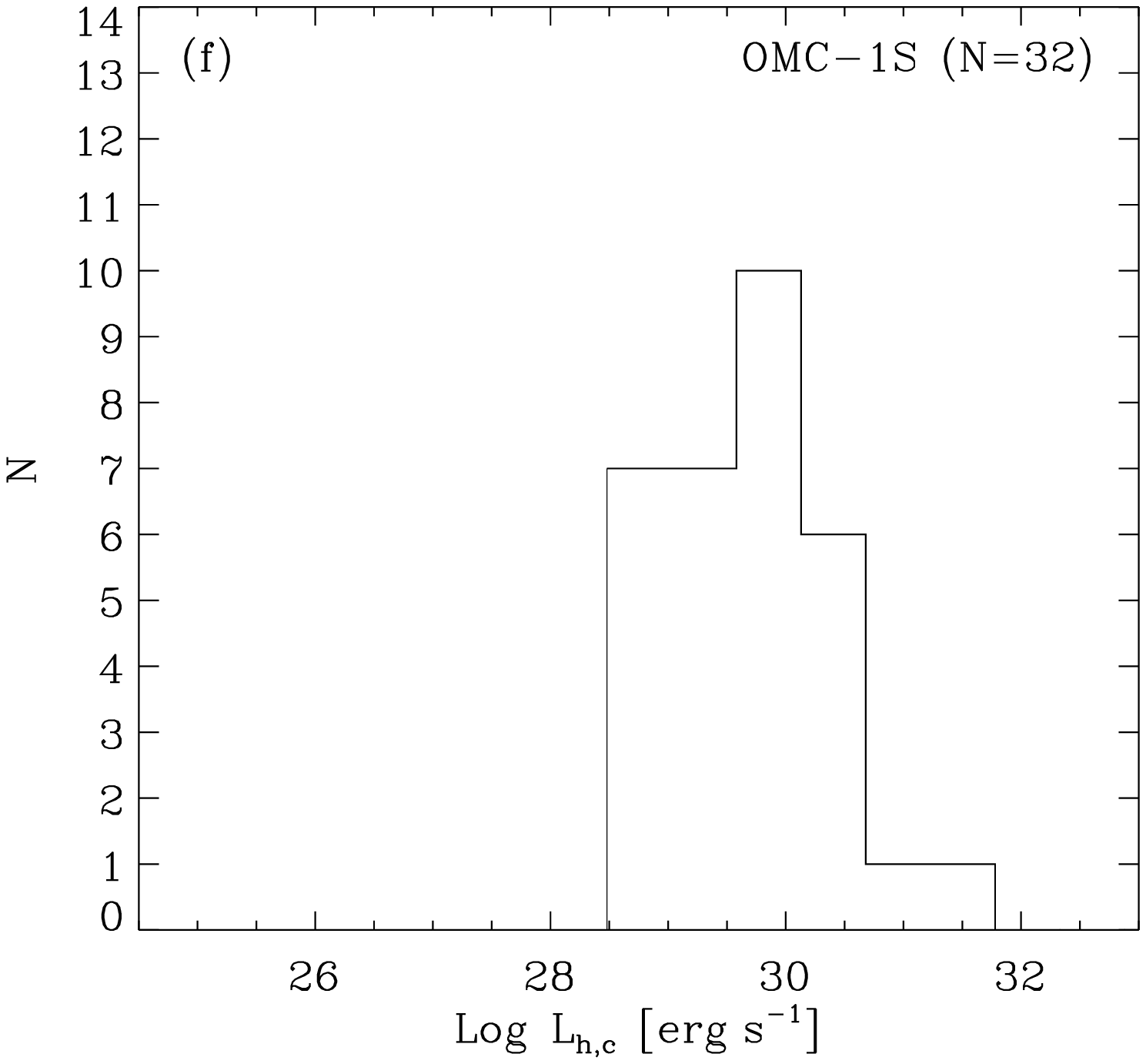}\\
\end{tabular}
 \caption{Distributions of the hard energy band (2.0--8.0\,keV) X-ray 
luminosities for the ONC lightly-absorbed optical sample (left), obscured 
COUP sources in BN-KL (middle) and OMC-1S (right). The top panels show the
distribution of observed luminosities, while the lower panels show the
luminosities after correction for absorption. The size of each sample is 
given in parentheses.
}
\label{lh}
\end{figure*}

\begin{figure*}[!h]
\centering
\begin{tabular}{@{}c@{}c@{}}
\includegraphics[width=0.5\columnwidth]{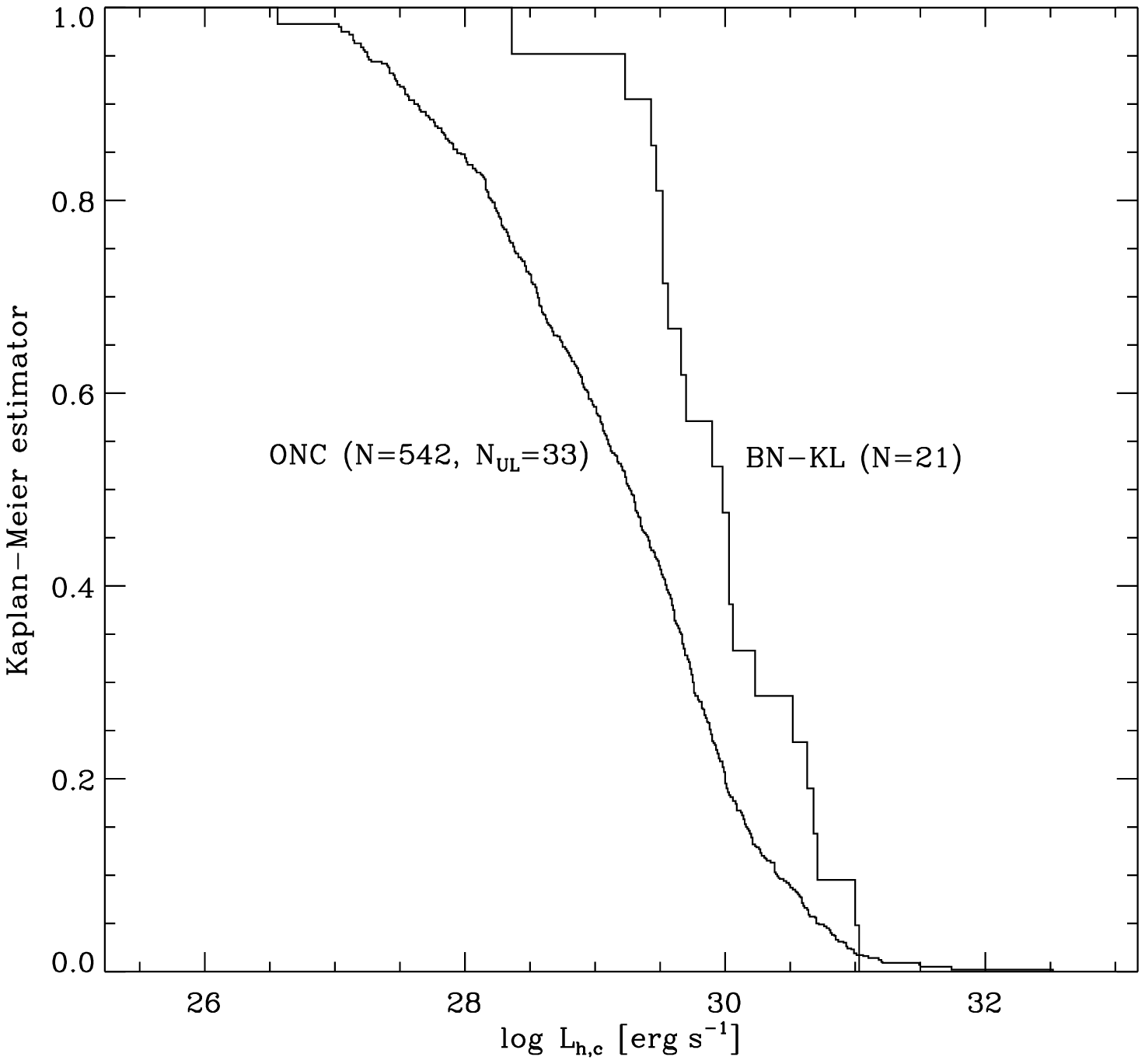} & \includegraphics[width=0.5\columnwidth]{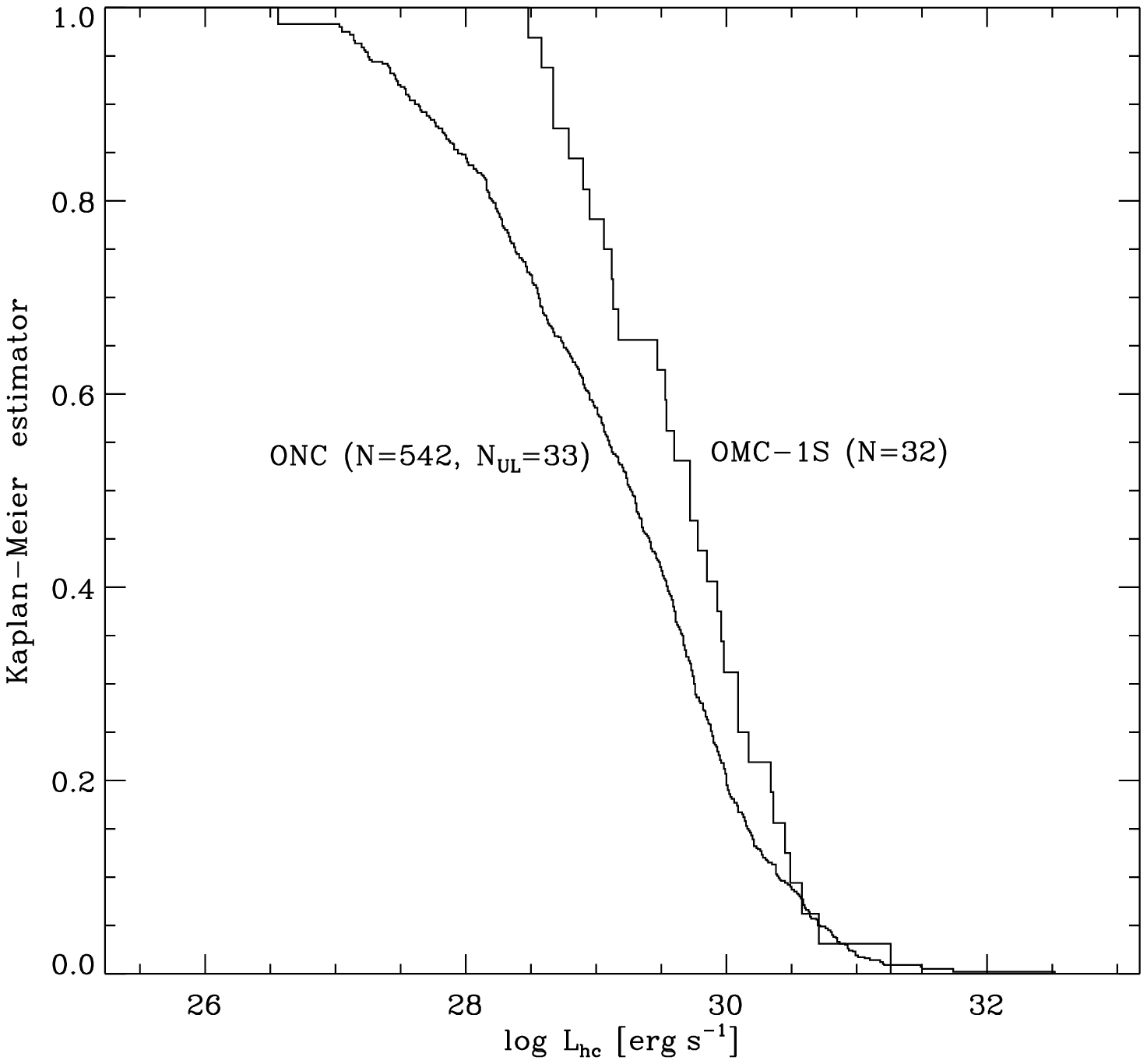} \\
\end{tabular}
 \caption{Comparison of the X-ray hard band intrinsic luminosity functions. 
Left: the ONC lightly-absorbed optical sample (mean=29.14$\pm$0.04, 
median=29.26; $N_{\rm ul}$ indicates the number of upper limits in the sample) 
compared with obscured COUP sources in BN-KL (mean=29.96$\pm$0.14, 
median=29.94). Right: the ONC lightly-absorbed optical sample compared with
obscured COUP sources in OMC-1S (mean=29.66$\pm$0.12, median=29.66).
}
\label{integral_xlf}
\end{figure*}

\begin{figure*}[!h]
\centering
\begin{tabular}{@{}ccc@{}}
\includegraphics[width=0.31\columnwidth]{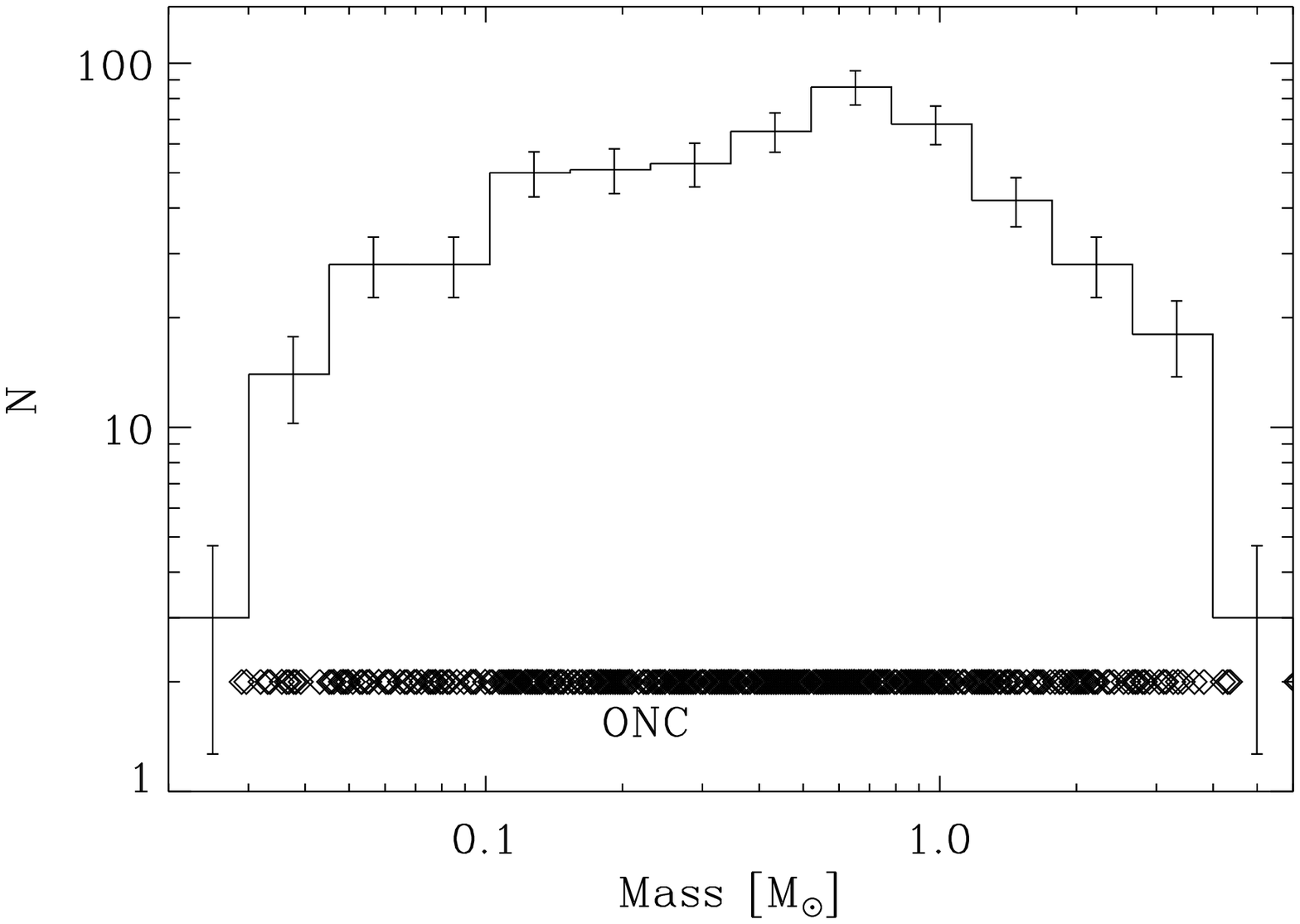} & \includegraphics[width=0.31\columnwidth]{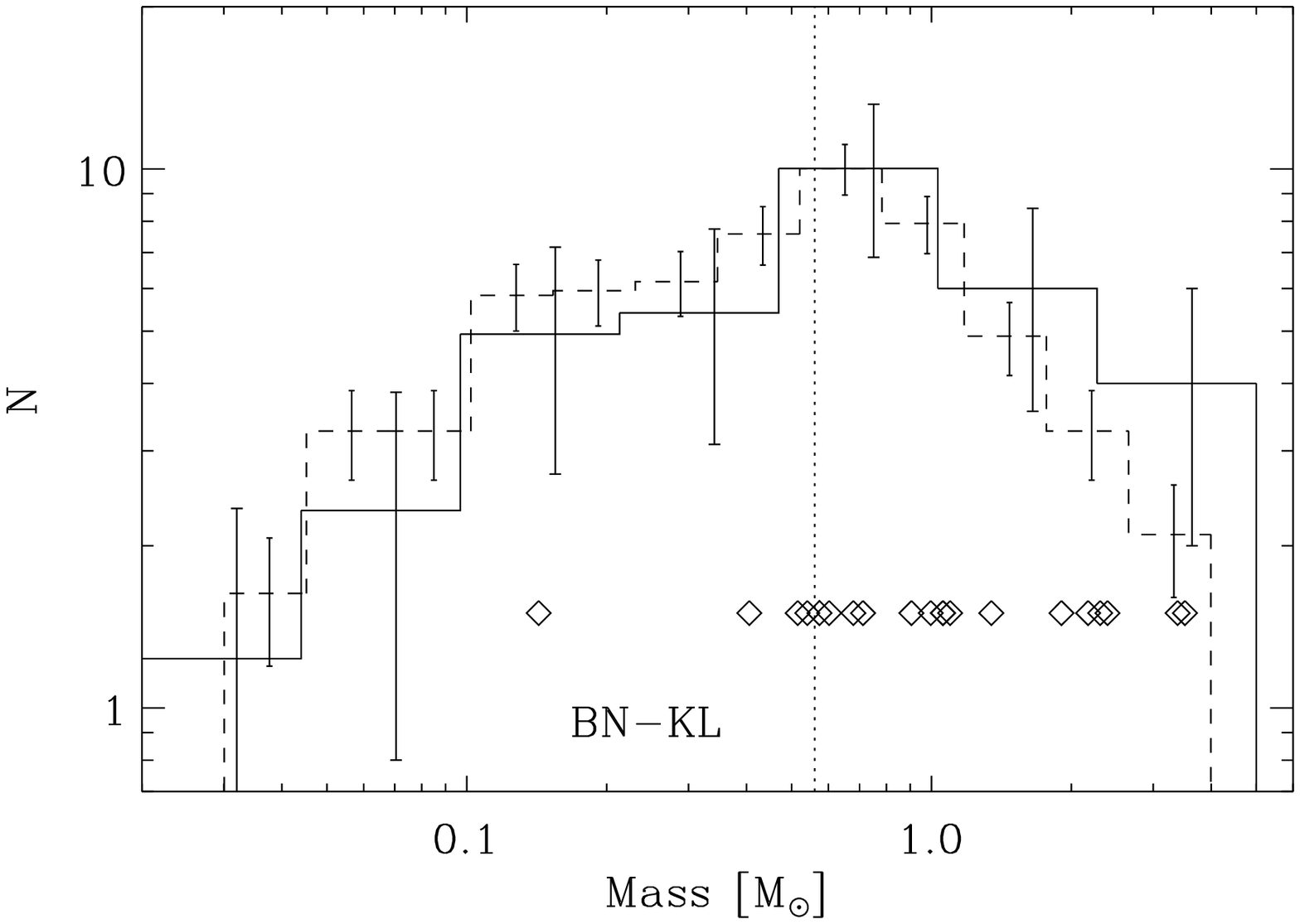} & \includegraphics[width=0.31\columnwidth]{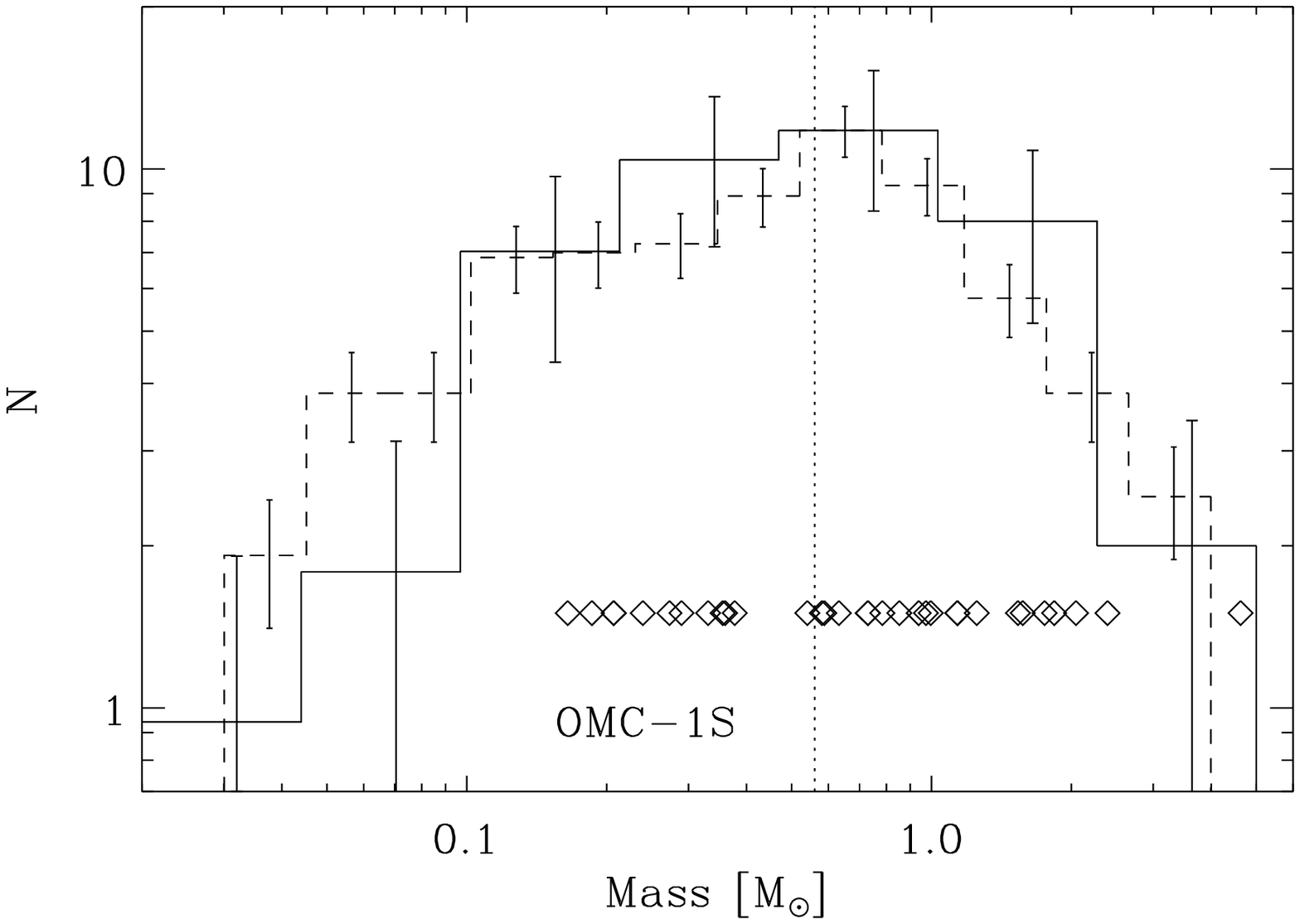}  \\
\end{tabular}
 \caption{The X-ray determined Initial Mass Function for the lightly-absorbed 
ONC sources, and for the obscured X-ray sources in BN-KL and OMC-1S\@. For
reference, the first panel shows the IMF of the ONC lightly-absorbed sample 
below 5\,M$_\odot$ built from the X-ray hard band luminosities corrected for 
absorption. Error bars show $1\sigma$ Poisson errors. Diamonds mark the masses 
of the COUP sources. The samples used to construct IMFs for BN-KL and OMC-1S have 
been supplemented in the mass interval 0.01--0.53\,M$_\odot$ (dotted line) 
with 13 and 10 stars, respectively, randomly distributed to match the ONC 
lightly-absorbed sample; the IMFs below $\sim$0.5\,M$_\odot$ are the averages 
of 10,000 trial sets. For comparison, the dashed line shows the IMF of the 
of the ONC lightly-absorbed sample normalized to the peak of the histogram.
}
\label{bn_imf}
\end{figure*}


\begin{thebibliography}{}
\bibitem[\protect\astroncite{Allen \& Burton}{1993}]{allen93} 
Allen, D.~A.~\& Burton, M.~G.\ 1993,  \nat,  363, 54
\bibitem[\protect\astroncite{Bally et al.}{1987}]{bally87} 
Bally, J., Stark, A.~A., Wilson, R.~W., \& Langer, W.~D.\ 1987, \apjl, 312, L45 
\bibitem[\protect\astroncite{Bally et al.}{2000}]{bally00} 
Bally, J., O'Dell, C.~R., \& McCaughrean, M.~J.\ 2000,  \aj,  119, 2919
\bibitem[\protect\astroncite{Becklin \& Neugebauer}{1967}]{becklin67} 
Becklin, E.~E.~\& Neugebauer, G.\ 1967,  \apj,  147, 799
\bibitem[\protect\astroncite{Bonnell \& Clarke}{1999}]{bonnell99} 
Bonnell, I.~A., \& Clarke, C.~J.\ 1999,  \mnras,  309, 461
\bibitem[\protect\astroncite{Carpenter et al.}{2000}]{carpenter00} 
Carpenter, J.~M., Heyer, M.~H., \& Snell, R.~L.\ 2000, \apjs,  130, 381
\bibitem[\protect\astroncite{Churchwell et al.}{1987}]{churchwell87} 
Churchwell, E., Wood, D.~O.~S., Felli, M., \& Massi, M.\ 1987,  \apj,  321, 516
\bibitem[\protect\astroncite{Dougados et al.}{1993}]{dougados93} 
Dougados, C., L{\'e}na, P., Ridgway, S.~T., Christou, J.~C., \& Probst, R.~G.\ 1993,  \apj,  406, 112
\bibitem[\protect\astroncite{Feigelson \& Nelson}{1985}]{feigelson85} 
Feigelson, E.~D.~\& Nelson, P.~I.\ 1985, \apj, 293, 192
\bibitem[\protect\astroncite{Feigelson et al.}{2002}]{feigelson02} 
Feigelson, E.~D., Broos, P., Gaffney, J.~A., Garmire, G., Hillenbrand, L.~A., Pravdo, S.~H., Townsley, L., \& Tsuboi, Y.\ 2002, \apj, 574, 258
\bibitem[\protect\astroncite{Feigelson et al.}{2005}]{feigelson05} 
Feigelson, E.~D., Getman, K.~V., Townsley, L., Garmire, G., Preibisch, T., Grosso, N., \& Montmerle T.\ 2005, \apjs{} (this issue)
\bibitem[\protect\astroncite{Feigelson \& Getman}{2005}]{feigelson05b} 
Feigelson, E.~D.\ \& Getman, K.~V.\ 2005, in The Initial Mass Function: 50 Years Later, E.\ Corbelli et al.\ (eds.), Kluwer, in press
\bibitem[\protect\astroncite{Flaccomio et al.}{1999}]{flaccomio99} 
Flaccomio, E., Micela, G., Sciortino, S., Favata, F., Corbally, C., \& Tomaney, A.\ 1999,  \aap,  345, 521
\bibitem[\protect\astroncite{Flaccomio et al.}{2003}]{flaccomio03} 
Flaccomio, E., Damiani, F., Micela, G., Sciortino, S., Harnden, F.~R., Murray, S.~S., \& Wolk, S.~J.\ 2003,  \apj,  582, 382
\bibitem[\protect\astroncite{Flaccomio et al.}{2005}]{flaccomio05} 
Flaccomio, E., Micela, G., Sciortino, S., Feigelson, E.~D., Herbst, W., Favata, F., Harnden, F.~R., \& Vrtilek, S.~D., 2005,  \apjs{} (this issue)
\bibitem[\protect\astroncite{Gagn{\'e} et al.}{1995}]{gagne95} 
Gagn{\'e}, M., Caillault, J., \& Stauffer, J.~R.\ 1995,  \apj,  445, 280
\bibitem[\protect\astroncite{Garmire et al.}{2000}]{garmire00} 
Garmire, G., Feigelson, E.~D., Broos, P., Hillenbrand, L.~A., Pravdo, S.~H., Townsley, L., \& Tsuboi, Y.\ 2000,  \aj,  120, 1426
\bibitem[\protect\astroncite{Garmire et al.}{2003}]{garmire03} 
Garmire, G.~P., Bautz, M.~W., Ford, P.~G., Nousek, J.~A., \& Ricker, G.~R.\ 
2003,  \procspie,  4851, 28
\bibitem[\protect\astroncite{Gaume et al.}{1998}]{gaume98} 
Gaume, R.~A., Wilson, T.~L., Vrba, F.~J., Johnston, K.~J., \& Schmid-Burgk, J.\ 1998,  \apj,  493, 940
\bibitem[\protect\astroncite{Giacconi et al.}{2001}]{giacconi01} 
Giacconi, R., Rosati, P., Tozzi, P., et al.\ 2001,  \apj,  551, 624
\bibitem[\protect\astroncite{Genzel \& Stutzki}{1989}]{genzel89} 
Genzel, R.~\& Stutzki, J.\ 1989,  \araa, 27, 41
\bibitem[\protect\astroncite{Getman et al.}{2005a}]{getman05a}
Getman, K.~V., Feigelson, E.~D., Grosso, N., McCaughrean, M.~J., Micela, G., Broos, P., Garmire, G., \& Townsley, L.\ 2005a, \apjs{} (this issue)
\bibitem[\protect\astroncite{Getman et al.}{2005b}]{getman05b}
Getman, K.~V., et al.\ 2005b, \apjs{} (this issue)
\bibitem[\protect\astroncite{Gezari et al.}{1998}]{gezari98} 
Gezari, D.~Y., Backman, D.~E., \& Werner, M.~W.\ 1998,  \apj,  509, 283
\bibitem[\protect\astroncite{Gomez et al.}{1993}]{gomez93} 
Gomez, M., Hartmann, L., Kenyon, S.~J., \& Hewett, R.\ 1993,  \aj,  105, 1927
\bibitem[\protect\astroncite{Greenhill et al.}{2004}]{greenhill04} 
Greenhill, L.~J., Gezari, D.~Y., Danchi, W.~C., Najita, J., Monnier, J.~D., \& Tuthill, P.~G.\ 2004,  \apjl,  605, L57
\bibitem[\protect\astroncite{Hillenbrand}{1997}]{hillenbrand97} 
Hillenbrand, L.~A.\ 1997,  \aj,  113, 1733
\bibitem[\protect\astroncite{Hillenbrand et al.}{1998}]{hillenbrand98} 
Hillenbrand, L.~A., Strom, S.~E., Calvet, N., Merrill, K.~M., Gatley, I., Makidon, R.~B., Meyer, M.~R., \& Skrutskie, M.~F.\ 1998,  \aj,  116, 1816
\bibitem[\protect\astroncite{Hillenbrand et al.}{2001}]{hillenbrand01} 
Hillenbrand, L.~A., Carpenter, J.~M., \& Skrutskie, M.~F.\ 2001,  \apjl,  547, L53
\bibitem[\protect\astroncite{Johnston et al.}{1992}]{johnston92} 
Johnston, K.~J., Gaume, R., Stolovy, S., Wilson, T.~L., Walmsley, C.~M., \& Menten, K.~M.\ 1992,  \apj,  385, 232
\bibitem[\protect\astroncite{Johnstone \& Bally}{1999}]{johnstone99} 
Johnstone, D.~\& Bally, J.\ 1999,  \apjl,  510, L49
\bibitem[\protect\astroncite{Kastner et al.}{2005}]{kastner05}
Kastner, J.~H., Franz, G., Grosso, N., Bally, J., McCaughrean, M., Getman, K.~V., Feigelson, E.~D., \& Schulz, N.~S.\ 2005, \apjs{} (this issue)
\bibitem[\protect\astroncite{Kleinmann \& Low}{1967}]{kleinmann67} 
Kleinmann, D.~E.~\& Low, F.~J.\ 1967,  \apjl,  149, L1
\bibitem[\protect\astroncite{Ku \& Chanan}{1979}]{ku79} 
Ku, W.~H.-M., \& Chanan, G.~A.\ 1979, \apjl, 234, L59 
\bibitem[\protect\astroncite{Lacombe et al.}{2004}]{lacombe04} 
Lacombe, F., et al.\ 2004,  \aap,  417, L5
\bibitem[\protect\astroncite{Lada et al.}{2000}]{lada00} 
Lada, C.~J., Muench, A.~A., Haisch, K.~E., Lada, E.~A., Alves, J.~F., Tollestrup, E.~V., \& Willner, S.~P.\ 2000,  \aj,  120, 3162
\bibitem[\protect\astroncite{Lada et al.}{2004}]{lada04} 
Lada, C.~J., Muench, A.~A., Lada, E.~A., \& Alves, J.~F.\ 2004, \aj, 128, 1254 
\bibitem[\protect\astroncite{Lagrange et al.}{2004}]{lagrange04} 
Lagrange, A.-M., et al.\ 2004,  \aap,  417, L11
\bibitem[\protect\astroncite{Lawson et al.}{1996}]{lawson96} 
Lawson, W.~A., Feigelson, E.~D., \& Huenemoerder, D.~P.\ 1996, \mnras, 280, 1071 
\bibitem[\protect\astroncite{Lomb}{1976}]{lomb76} 
Lomb, N.~R.\ 1976,  \apss,  39, 447
\bibitem[\protect\astroncite{Lonsdale et al.}{1982}]{lonsdale82} 
Lonsdale, C.~J., Becklin, E.~E., Lee, T.~J., \& Stewart, J.~M.\ 1982,  \aj,  87, 1819
\bibitem[\protect\astroncite{Lucy}{1974}]{lucy74} 
Lucy, L.~B.\ 1974, \aj, 79, 745
\bibitem[\protect\astroncite{McCaughrean}{1988}]{mccaughrean88}
McCaughrean, M.~J. 1988, PhD thesis, Univ.\ Edinburgh
\bibitem[\protect\astroncite{McCaughrean et al.}{1996}]{mccaughrean96}
McCaughrean, M.~J., Rayner, J.~T., Zinnecker, H., \& Stauffer, J.~R. 1996,
in S.~V.~W. Beckwith et al., eds., Disks and outflows around young stars,
Lecture Notes in Physics 465, 33
\bibitem[\protect\astroncite{Mezger et al.}{1990}]{mezger90} 
Mezger, P.~G., Zylka, R., \& Wink, J.~E.\ 1990,  \aap,  228, 95
\bibitem[\protect\astroncite{Menten \& Reid}{1995}]{menten95} 
Menten, K.~M.~\& Reid, M.~J.\ 1995,  \apjl,  445, L157
\bibitem[\protect\astroncite{Muench et al.}{2002}]{muench02} 
Muench, A.~A., Lada, E.~A., Lada, C.~J., \& Alves, J.\ 2002,  \apj,  573, 366
\bibitem[\protect\astroncite{Mundy et al.}{1986}]{mundy86} 
Mundy, L.~G., Scoville, N.~Z., Baath, L.~B., Masson, C.~R., \& Woody, D.~P.\ 1986, \apjl, 304, L51 
\bibitem[\protect\astroncite{O'Dell \& Wong}{1996}]{odell96} 
O'Dell, C.~R.~\& Wong, K.\ 1996,  \aj,  111, 846
\bibitem[\protect\astroncite{O'Dell}{2001}]{odell01} 
O'Dell, C.~R.\ 2001,  \araa,  39, 99
\bibitem[\protect\astroncite{O'Dell \& Doi}{2003}]{odell03} 
O'Dell, C.~R.~\& Doi, T.\ 2003,  \aj,  125, 277
\bibitem[\protect\astroncite{Preibisch et al.}{2005a}]{preibisch05a} 
Preibisch, T., et al.\ 2005a, \apjs{} (this issue)
\bibitem[\protect\astroncite{Preibisch et al.}{2005b}]{preibisch05b} 
Preibisch, T., et al.\ 2005b, \apjs{} (this issue)
\bibitem[\protect\astroncite{Richardson}{1972}]{richardson72} 
Richardson, W.~H.\ 1972, Optical Society of America Journal, 62, 55
\bibitem[\protect\astroncite{Rieke et al.}{1973}]{rieke73} 
Rieke, G.~H., Low, F.~J., \& Kleinmann, D.~E.\ 1973,  \apjl,  186, L7
\bibitem[\protect\astroncite{Robberto et al.}{2005}]{robberto05} 
Robberto, M., Beckwith, S.~V.~W., Panagia, N., Patel, S.~G., Herbst, T.~M., et al.\ 2005, AJ, in~press [astro-ph/0412665]
\bibitem[\protect\astroncite{Scargle}{1982}]{scargle82} 
Scargle, J.~D.\ 1982,  \apj,  263, 835
\bibitem[\protect\astroncite{Schmid-Burgk et al.}{1990}]{schmid-burgk90} 
Schmid-Burgk, J., Guesten, R., Mauersberger, R., Schulz, A., \& Wilson, T.~L.\ 1990,  \apjl,  362, L25
\bibitem[\protect\astroncite{Schulz et al.}{2003}]{schulz03} Schulz, N.~S., 
Canizares, C., Huenemoerder, D., \& Tibbets, K.\ 2003, \apj, 595, 365 
\bibitem[\protect\astroncite{Siess et al.}{2000}]{siess00} 
Siess, L., Dufour, E., \& Forestini, M.\ 2000, \aap,  358, 593
\bibitem[\protect\astroncite{Smith et al.}{2004}]{smith04} 
Smith, N., Bally, J., Shuping, R.~Y., Morris, M., \& Hayward, T.~L.\ 2004,  \apjl,  610, L117
\bibitem[\protect\astroncite{Smith et al.}{2005}]{smith05} 
Smith, N., Bally, J., et al. 2005, \apj, submitted
\bibitem[\protect\astroncite{Stelzer et al.}{2003}]{stelzer03} 
Stelzer, B., Hu{\'e}lamo, N., Hubrig, S., Zinnecker, H., \& Micela, G.\ 2003, \aap, 407, 1067 
\bibitem[\protect\astroncite{Stelzer et al.}{2005}]{stelzer05} 
Stelzer, B., Flaccomio, E., Montmerle, T., Micela, G., Sciortino, S., Favata, F., Preibisch, T., \& Feigelson, E.~D.\ 2005, \apjs{} (this issue)
\bibitem[\protect\astroncite{Stolovy et al.}{1998}]{stolovy98} 
Stolovy, S.~R., et al.\ 1998,  \apjl,  492, L151
\bibitem[\protect\astroncite{Tan}{2004}]{tan04} 
Tan, J.~C.\ 2004, \apj, L47
\bibitem[\protect\astroncite{Tsujimoto et al.}{2005}]{tsujimoto05}
Tsujimoto, M., Feigelson, E.~D., Grosso, N., Micela, G., Tsuboi, Y., Favata, F., \& Shang, S.\ 2005, \apjs{} (this issue)
\bibitem[\protect\astroncite{Lupton et al.}{2004}]{lupton04} 
Lupton, R., Blanton, M.~R., Fekete, G., Hogg, D.~W., O'Mullane, W., Szalay, A., \& Wherry, N.\ 2004,  \pasp,  116, 133
\bibitem[\protect\astroncite{Vuong et al.}{2003}]{vuong03} 
Vuong, M.~H., Montmerle, T., Grosso, N., Feigelson, E.~D., Verstraete, L., \& Ozawa, H.\ 2003, \aap, 408, 581 
\bibitem[\protect\astroncite{Weisskopf et al.}{2002}]{weisskopf02} 
Weisskopf, M.~C., Brinkman, B., Canizares, C., Garmire, G., Murray, S., \& Van Speybroeck, L.~P.\ 2002,  \pasp,  114, 1
\bibitem[\protect\astroncite{Werner et al.}{1976}]{werner76} 
Werner, M.~W., Gatley, I., Becklin, E.~E., Harper, D.~A., Loewenstein, R.~F., Telesco, C.~M., \& Thronson, H.~A.\ 1976,  \apj,  204, 420
\bibitem[\protect\astroncite{Wolk et al.}{2005}]{wolk05} 
Wolk, S., Harnden, F.~R., Flacommio, E., Micela, G., Favata, F., Glassgold, A.~E., Shang, S., \& Feigelson, E.~D.\ 2005, \apjs{} (this issue)
\bibitem[\protect\astroncite{Zapata et al.}{2004a}]{zapata04a} 
Zapata, L.~A., Rodr{\'\i}guez, L.~F., \& Kurtz, S.~E., \& O'Dell, C.~R.\ 2004a, \aj,  127, 2252
\bibitem[\protect\astroncite{Zapata et al.}{2004b}]{zapata04b} 
Zapata, L.~A., Rodr{\'\i}guez, L.~F., Kurtz, S.~E., O'Dell, C.~R., \& Ho, T.~P.\ 2004b, \apjl,  610, L121
\end{thebibliography}
\end{document}